\documentclass[fleqn,usenatbib,linenumbers]{mnras}
\usepackage{eso-pic}

\AddToShipoutPictureBG*{%
  \AtPageUpperLeft{%
    \hspace{0.75\paperwidth}%
    \raisebox{-3.5\baselineskip}{%
      \makebox[0pt][l]{\textnormal{DES-2021-0658}}
}}}%

\AddToShipoutPictureBG*{%
  \AtPageUpperLeft{%
    \hspace{0.75\paperwidth}%
    \raisebox{-4.5\baselineskip}{%
      \makebox[0pt][l]{\textnormal{FERMILAB-PUB-20-545-AE}}
}}}%

\usepackage{newtxtext,newtxmath}
\usepackage{mdframed}
\usepackage{graphicx}
\usepackage{subcaption}
	
\usepackage{lineno}

\usepackage{hyperref}
\usepackage[T1]{fontenc}
\usepackage{ae,aecompl}
\usepackage{todonotes}

\usepackage{graphicx}	
\usepackage{amsmath}	
\usepackage{adjustbox}

\newcommand{\snr}{S/N}
\newcommand{\mcal}{\textsc{metacalibration}}
\newcommand{\Mcal}{\textsc{Metacalibration}}
\newcommand{\MCAL}{\textsc{METACALIBRATION}}
\newcommand{\sx}{\textsc{SExtractor}}
\newcommand{\psfex}{\textsc{PSFEx}}
\newcommand{\ngmix}{\textsc{ngmix}}
\newcommand{\galsim}{\textsc{Galsim}}
\newcommand{\est}{e}
\newcommand{\pp}{\mbox{\boldmath $p$}}
\newcommand{\vesta}{\mbox{\boldmath $e_{\alpha}$}}
\newcommand{\vestb}{\mbox{\boldmath $e_{\beta}$}}
\newcommand{\vest}{\mbox{\boldmath $e$}}
\newcommand{\vwst}{\mbox{\boldmath $w$}}
\newcommand{\vqst}{\mbox{\boldmath $q$}}
\newcommand{\vecg}{\mbox{\boldmath $\gamma$}}
\newcommand{\vecc}{\mbox{\boldmath $c$}}
\newcommand{\vecgest}{\mbox{\boldmath $\gamma^{\mathrm{est}}$}}
\newcommand{\mcalR}{\mbox{\boldmath $R$}}

\newcommand{\mcalRg}{\mbox{\boldmath $R_\gamma$}}
\newcommand{\mcalRs}{\mbox{\boldmath $R_s$}}

\newcommand{\pixmappy}{\textsc{pixmappy}}

\newcommand{\treecorr}{\textsc{TreeCorr}}

\defcitealias{ZuntzY1}{ZS18}
\newcommand{\Zuntz}{\citetalias{ZuntzY1}}

\title[DES Y3 Shape Catalogue]{Dark Energy Survey Year 3 Results: Weak Lensing Shape Catalogue}

\author[Gatti \& Sheldon et~al.]{
\parbox{\textwidth}{
\Large 
M.~Gatti$^{1\star}$,  
E.~Sheldon$^{2\dag}$, 
A.~Amon$^{3}$,
M.~Becker$^{4}$, 
M.~Troxel$^{5}$, 
A.~Choi$^{6}$, 
C.~Doux$^{7}$, 
N.~MacCrann$^{6,8}$, 
A.~Navarro-Alsina$^{9}$,
I.~Harrison$^{10}$
% ...
\begin{center} (DES Collaboration) \end{center}
}
}

\vspace{0.2cm}

\date{Accepted XXX. Received YYY; in original form ZZZ}

\pubyear{2020}

\begin{document}
\label{firstpage}
\pagerange{\pageref{firstpage}--\pageref{lastpage}}
\maketitle
\definecolor{pink}{rgb}{0.848, 0.188, 0.478}
\begin{abstract}
We present and characterise the galaxy shape catalogue from the first 3 years of Dark Energy Survey (DES) observations, over an effective area of ~4143 deg$^2$ of the southern sky. We describe our data analysis process and our self-calibrating shear measurement pipeline \mcal , which builds and improves upon the pipeline used in the DES Year 1 analysis in several aspects. The DES Year 3 weak-lensing shape catalogue consists of 100,204,026 galaxies, measured in the \textit{riz} bands, resulting in a weighted source number density of {$n_{\rm eff} = 5.59$} gal/arcmin$
^{2}$ and corresponding shape noise {$\sigma_e = 0.261$}. We perform a battery of internal null tests on the catalogue, including tests on systematics related to the point-spread function (PSF) modelling, spurious catalogue B-mode signals, catalogue contamination, and galaxy properties.

\end{abstract}


\makeatletter
\def \blfootnote{\xdef\@thefnmark{}\@footnotetext}
\makeatother

\blfootnote{$^{\star}$ E-mail: mgatti@ifae.es}
\blfootnote{$^{\dag}$ E-mail: erin.sheldon@gmail.com}


\section{Introduction} \label{sec:intro}

The measurement of weak gravitational lensing is an important component for constraining dark energy with current and planned imaging surveys \citep[e.g.][]{KIDS2015,TakadaHSC2010,DESWhitePaper,IvezicLSST08,Euclid2011,SpergelWFIRST2015}. For the Dark Energy Survey \citep{DESWhitePaper, Flaugher2015,DES_book_2020}, weak lensing is one of four ``key probes'', the others being galaxy angular clustering, galaxy cluster abundances, and type IA supernovae distances.  With these combined probes, DES will constrain cosmological parameters such as the dark energy equation of state parameter $w$ with high precision.  The goal of this work is to present empirical tests of the weak lensing shear measurements performed on the DES first 3 years (Y3) data set (DES Y3) in order to assess systematic errors that may degrade this precision.

Weak gravitational lensing is the deflection of light as it passes by mass concentrations in the universe \citep{SchneiderBook92}.  The distant objects observed in our images appear in a different location than they would had their light passed through a homogeneous universe.  This deflection can be inferred only in the rare cases that the unperturbed light path is known, for example in strong lens systems with multiple images of the source \citep{Walsh0957}.  There is a higher order effect that can be inferred without such knowledge:  the light deflections differ slightly across the galaxy image, resulting in a small distortion of its shape. This distortion induces an ellipticity that is directly related to the mass concentrations that caused the deflections.  This weaker ``shear'' effect results in a departure from isotropy in the orientations of galaxies that is spatially coherent: the ellipticities of galaxies become correlated on the sky \citep[see, e.g.][and references therein]{BartelmannSchneider2001}.

Because the shear is directly related to the lensing mass, the effect can be cleanly predicted given an accurate model of the mass concentrations.  In turn, the distribution of matter in the universe inferred by modelling the shear signal depends sensitively on the cosmological parameters, such as the mass density $\Omega_{\rm m}$ and the equation of state of dark energy
\citep{HoekstraJainReview2008}.



{In the past decades a large variety of methods to infer the value of the shear field have been developed. Many of them use galaxy ellipticies as a proxy of the shear field, which usually involves assigning a set of numbers to each galaxy describing the observed galaxy light profile, once having assumed a galaxy model.} In order to infer the shear from measured ellipticities, one must therefore understand how the intrinsic ellipticities of galaxies are modified by gravitational shear, as well as other effects such as the point spread function of the atmosphere, telescope, and detector \citep{BJ02}. In addition, there are often biases present in the determination of the shape itself due to noise rectification or model misspecification \citep{HirataAlign04,Refreg12,Melchior12,Bernstein2010}. {We note that there exist methods to infer the shear field that do not require a per-galaxy shape estimate, which allows to avoid model biases (e.g., \citealt{Schneider2015}, or the Bayesian Fourier Domain (BFD) algorithm proposed by \citealt{BernBFD2016}). None of these methods are considered in this work, but we are planning to implement BFD in future DES releases.}



 

We can generally divide biases in the shear determination into two broad categories:  additive and multiplicative biases. Following standard notation \citep{Heymans2006,great3}, we can write an estimate of the two-component shear as:
\begin{align}
\label{eq:1}
	\vecgest = m \vecg + \vecc,
\end{align}
where \vecgest\ is a biased estimate of the true shear \vecg.  We call $m$ the multiplicative and \vecc\ the additive, or shear independent bias.

These biases can arise from a number of different causes.  PSF-misestimation can contribute to both multiplicative and additive biases:  if the size of the PSF is misestimated, a multiplicative bias will occur.  If the ellipticity of the PSF is misestimated, an additive bias will occur that is related to the PSF orientation. Another cause of multiplicative bias is calibration errors in the shear estimation algorithm itself, the method for converting an ensemble of ellipticity measurements into an estimate of a shear signal.  This can occur for a number of reasons, for example if the shear is not accurately inferred from the observed shapes due to aforementioned modelling errors or noise effects, or if any applied empirical or simulation based corrections have limited accuracy. In addition, selection and detection effects can induce significant
shear-dependent or PSF-dependent biases \citep{Kaiser2000,BJ02,BernBFD2016,Hoekstra2017,SheldonMcal2017,Fenech2017}. One noticeable example is the case of blended galaxies, where one single detection is actually associated to multiple, unresolved galaxies \citep{SheldonMetadetect2019,MacCrannSims2019}: in this case, the image pixels contain light from multiple sources, and if the shear estimation pipeline is not able to account for this effect, the shear measurement will be biased.



In this paper we present the weak lensing shape catalogue measured in DES Y3 imaging data, and perform empirical tests of the catalogue in order to assess potential biases. Our primary tool is the ``null test'':  we generate measurements that should yield zero signal in the absence of biases in the shape catalogue.  For example, if our PSF modelling is accurate we should detect no correlation between object ellipticities and PSF ellipticities.  Similarly, we should see no correlation between object ellipticities and unrelated quantities such as the location of an object's image within the focal plane or the observing conditions.

{This work is complemented by two other papers. The first one describes in more depth the PSF modeling used in the DES Y3 analysis \citep{Jarvis19}, and presents a number of diagnostic tests that are independent of the shape catalogue. The second work \citep{MacCrannSims2019} describes the suite of image simulations used to provide the overall calibration of the catalogue.} Indeed, some biases are difficult to test empirically due to the lack of an absolute calibration source for shear. Comparing subsamples of the data can reveal relative calibration biases between subsamples \citep{Becker2016,Troxel2018}, although the different selection biases affecting the subsamples can severely hamper the interpretation of the tests (e.g., \citealt{Mandelbaum2018}; but see \citealt{Amon2018}). 
For tests of the absolute calibration we therefore rely on simulations \citep{MacCrannSims2019}.

The specific method we employ for shear estimation in DES Y3 is \mcal\footnote{In particular, we used the following packages:
\begin{itemize}
    \item ngmix: v1.0.0, https://github.com/esheldon/ngmix
    \item ngmixer: v0.9.6, https://github.com/esheldon/ngmixer
\end{itemize}}  \citep{HuffMcal2017,SheldonMcal2017}.  This method is known to be unbiased for isolated galaxy images in the limit of weak shear and in the case of perfect knowledge of the PSF. \Mcal\ empirically corrects for noise, modelling, and selection biases \citep{SheldonMcal2017}.  However, \mcal\ can suffer a bias due to some of the effects mentioned above, for example PSF misestimation, and we test such biases in this work. The blending of galaxy images produces a calibration bias that is not addressed by the \mcal\ implementation used for DES Y3, {and which is large enough that cannot be ignored for the DES Y3 analysis}. In future releases we will apply empirical corrections using the \textsc{metadetection} method presented in \cite{SheldonMetadetect2019}.  For DES Y3, we instead rely on the aforementioned simulations to derive a correction.

{Contrary to the DES Y1 analysis where two different shape catalogues were produced with two different pipelines (\citealt*{ZuntzY1}, hereafter Z18), we only rely on one shape catalogue in the DES Y3 analysis. Despite the fact that having two different catalogues in the DES Y1 analysis increased our confidence in the robustness of the catalogues calibration, the DES Y3 shape catalogue is backed up by a much more powerful and accurate suite of image simulations \citep{MacCrannSims2019} compared to those used in the DES Y1 analysis. These image simulations replicate with high fidelity the features and properties of the DES Y3 shape catalogue, making us confident of the catalogue calibration.}

The outline of the paper is as follows: in \S~\ref{sect:data} we outline the new observations used in the DES Y3 analysis, and present improvements compared to DES Y1 observations. Updates concerning PSF modelling and PSF estimation are presented in \S~\ref{sect:PSF}.  In \S~\ref{sec:mcal} we discuss a few technical aspects of the \mcal\ algorithm implemented in the DES Y3 analysis.  In \S~\ref{sect:psf_diagnostic} we discuss systematic tests associated to the PSF modelling, and in \S~\ref{sect:sherartests} we present null tests of the shape catalogue, including shear variations in focal plane coordinates (\S~\ref{sec:shearvarfocal}), tangential shear around field centers (\S~\ref{sec:field_centers}), stellar contamination of the catalogue (\S~\ref{sec:starcontam}), B-modes tests (\S~\ref{sec:bmodes}), galaxy properties and observing conditions tests (\S~\ref{sec:additiveother}). In \S~\ref{sec:summary} we summarise our results. Last, Appendix~\ref{sect:2pt_response} lays out a generalisation of the \mcal\ calibration for shear two-point correlation functions, and Appendix~\ref{sec:stargal_appendix} provides more details on the star-galaxy separation algorithm implemented for the stellar contamination test.

\section{Data}\label{sect:data}
\subsection{New observations and footprint} \label{sec:newobs}

\begin{figure}
\begin{center}
    \includegraphics[width=0.4 \textwidth]{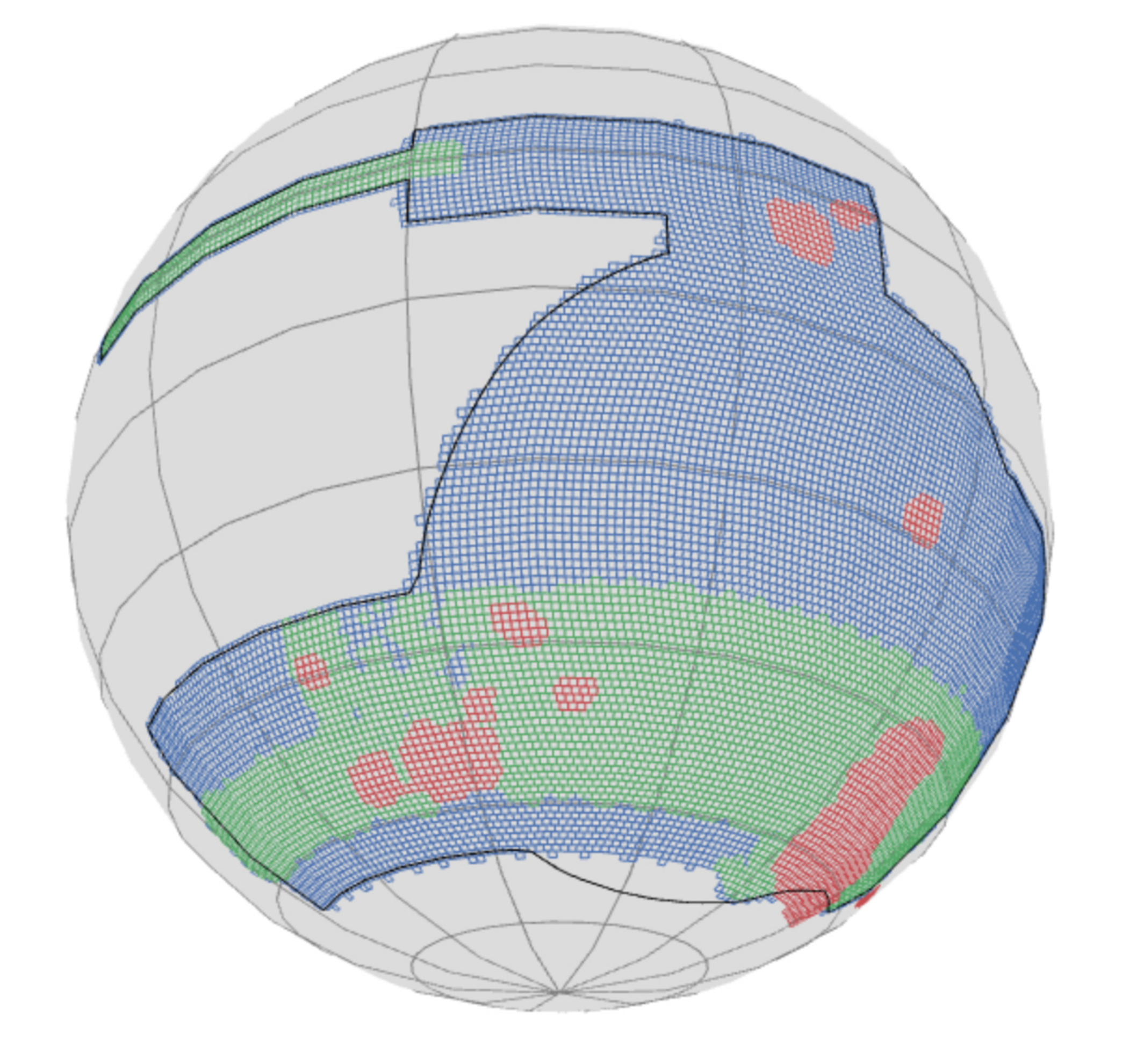}
\end{center}
\caption{Footprint of the DES Y3 shape catalogue.  The Y3 catalogue is shown in blue.  For comparison the SV and Y1 footprints, which
are nearly subsets of the Y3 are
overplotted in red and green respectively.}
\label{fig:footprint}
\end{figure}

The DES Y3 data represent a significant increase in total area compared to the Y1 data, with a similar depth. Slightly modified settings of the DES pipeline processing have lowered the threshold for detection \citep{Y3Gold}, {enabling an increase in the number of objects, more than expected from the increased area and depth alone}. The effective area of the wide survey with observations in the $griz$ bands, after masking for foregrounds and other problematic regions\footnote{\textsc{FLAGS\_FOREGROUND}=0 and \textsc{FLAGS\_BADREGIONS}<2}, is $\sim$4143 deg$^2$, compared to the $\sim$1321 deg$^2$ for the Y1 cosmic shear results \citep{Troxel2018}.  The area coverage is shown in Fig.~\ref{fig:footprint}. Object selection additionally required that the object belonged to the Gold catalogue \citep{Y3Gold}, that it was not marked as `anomalous'\footnote{\textsc{FLAGS\_GOLD}<8. {This rejects mainly objects with problems at the pixel level, such as saturation or truncation of the object at boundaries.}} and that it was successfully measured and where necessary, deblended by the multi-object fitting code, which simultaneously fits blended groups or isolated objects in the full multi-epoch, multi-band dataset \citep{Y3Gold}. 
This resulted in a final catalogue of 326,049,983 objects, a net improvement with respect to the $\sim$137 million objects detected in the Y1 catalogue \citep{Alex2018}.  For weak lensing, further cuts were performed using quantities measured as part of the \mcal\ procedure.  For details see \S~\ref{sec:mcalselect}.

The Y3 dataset includes other improvements, such as $\sim 0.003$ magnitude photometric accuracy, a better catalog for astrometric calibration \citep[2MASS,][]{2MASS}, better object flagging in the coadd catalogues using the \textsc{IMAFLAGS\_ISO} flag as described in \cite{Morganson2018}, and a more complete array of survey property maps \citep[see ][for details]{Y3Gold}.



\subsection{Astrometry}
\label{sec:astrom}

The pixels comprising each image of each source are assigned sky coordinates using the local first derivative (i.e. linearization) of the image-wide astrometric solution, using methods similar to those of the Y1 reductions \citep{Y3Gold}.  In brief, astrometric solutions are derived for all images in the survey by least-squares minimization of the residuals between different DES measurements of the same star, and between DES and an external astrometric reference catalog.  The astrometric model contains degrees of freedom representing the exposure pointings, atmospheric refraction, optical distortions (including chromatic terms for these two), and positioning of the CCDs in the focal plane.
Substantially improved solutions over Y1 are available for Y3 using the characterization of the The Dark Energy Camera (DECam, \citealt{Flaugher2015}) astrometric distortions derived in \citet{Bernstein2017}. The new solutions incorporated small-scale distortions due to stray electric fields in the detectors, and were registered to the Gaia DR1 catalogue \citep{gaiadr1}.  The dominant sources of astrometric calibration error are the 5--10~milliarcsec distortions induced by atmospheric turbulence in a typical exposure, with a smaller contribution from proper motions of the reference stars during the $\approx2$-year span of the imaging.  The improved astrometric solutions were applied to both the PSF determination and to the \mcal\ input postage stamps.

\subsection{Blinding of the catalogue}

A two-stage blinding procedure was adopted in the DES Y3 analysis to mitigate confirmation bias and avoid that experimenters (intentionally or unintentionally) tune the analysis to match expectations. A good blinding scheme must be capable of altering the output of the analysis without compromising the performance of systematic tests and pipeline validation. In particular, for the DES Y3 analysis we adopted a blinding scheme both at the shape catalogue level and at the summary statistics level. 

The blinding of the shape catalogue was performed in a similar fashion to the DES Y1 analysis (\Zuntz). The ellipticities \textbf{e} of the catalogue were transformed via |{\mbox{\boldmath $\eta$}}| $\equiv 2 {\rm arctanh} |\textbf{e}| \rightarrow f$|{\mbox{\boldmath $\eta$}}$ |$, with a hidden value 0.9 < f < 1.1. This mapping preserved the confinement of the $\textbf{e}$ values to the unit disc while rescaling all inferred shears. We made sure that the \mcal\ procedure did not accidentally absorb the blinding transformation.

\section{PSF modelling and estimation}\label{sect:PSF}
%

\begin{figure}
\begin{center}
\includegraphics[width=0.45\textwidth]{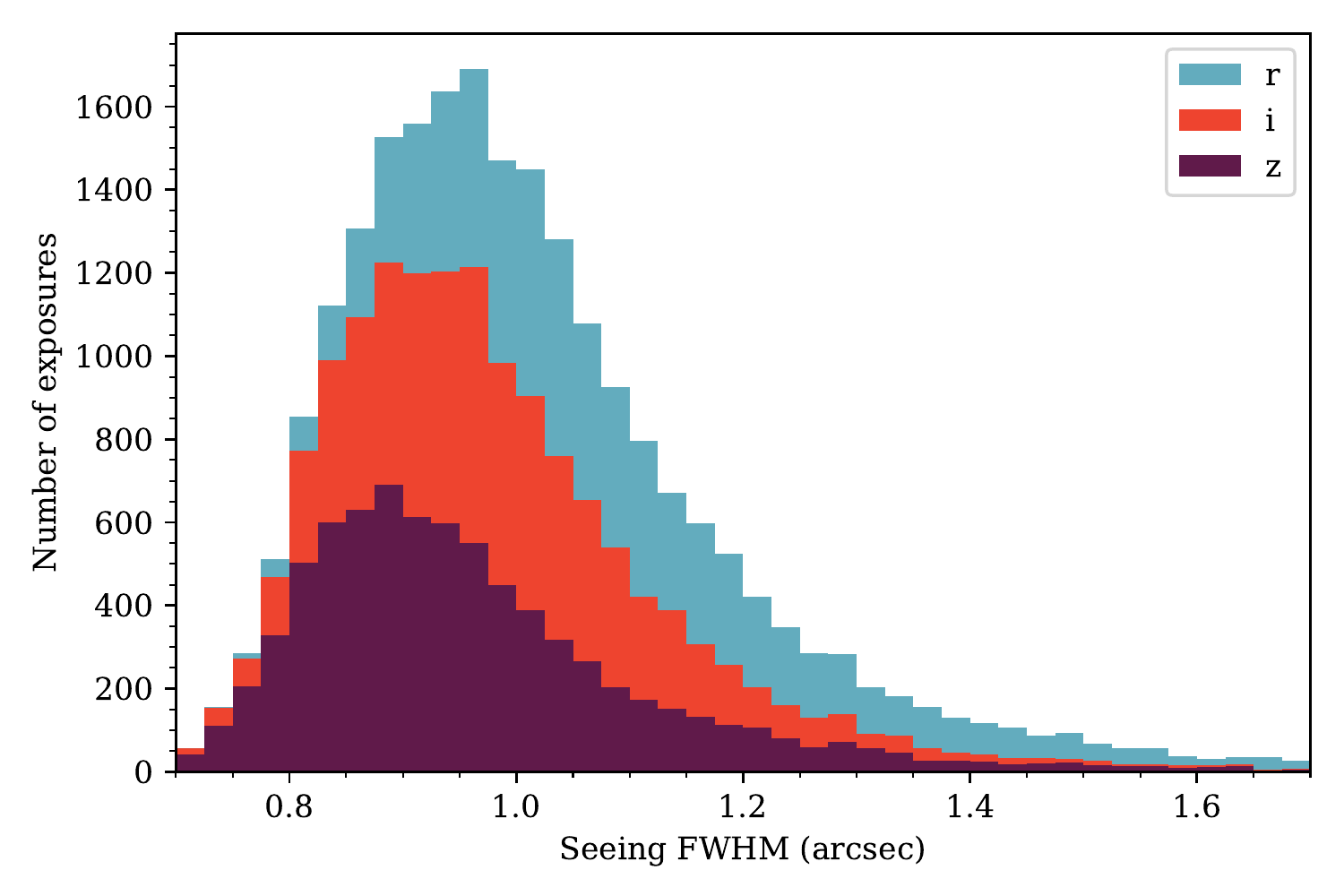}
\end{center}
\caption{The distribution of the median seeing FWHM of the stars used to model the PSF in the $riz$-bands. The median seeing for the distributions shown is 1.05" in the $r$-band, 0.97" in the $i$-band and 0.93" in the $z$-band. The overall median seeing is 0.98".}
\label{fig:fwhm}
\end{figure}
\subsection{PSF Measurement and Interpolation using PIFF}
\label{sec:piff}

For modeling the point-spread function (PSF), a new software package,  \textsc{Piff} (PSFs In the Full FOV)\footnote{
Specifically, the PSF modeling used release version 0.2.2. http://rmjarvis.github.io/Piff/ .} was used.
The full details of this software are described in \cite{Jarvis19}, but here we give an overview
of some salient features used in the DES Y3 analysis.

\textsc{Piff} has a number of available models it can use to describe the PSF at any given location, 
as well as a number of possible interpolation schemes to calculate the coefficients of the model at different locations.
For DES Y3, the \textsc{PixelGrid} model was used, which involves a grid of pixels, each with an independent amplitude at their centers.  
The amplitudes of the PSF between pixel centers were found using Lanczos interpolation (in particular, we used Lanczos interpolation kernels up to 3rd order). For DES Y3, we used for the model pixels of 0.3 arcsec on a side, slightly larger than the native image pixels (0.27 arcsec).  It was  found that this significantly increased the stability of the fits and reduced the prevalence of noise artifacts in the solutions.

To interpolate the PSF model at other locations besides the locations of the stars, a 3rd order \textsc{BasisPolynomial} (a class implemented in \textsc{Piff}) was used, which delayed the solution of the model coefficients for each star until also solving for the interpolation coefficients.  This helped handle moderately degenerate solutions for some stars (e.g. stars with masks that cover one or more of the model pixels),  as it allowed for all of the other stars help to constrain the overall fit. {The interpolation was performed over the CCD chip area}. 

\textsc{Piff} models the PSF in sky coordinates, rather than image coordinates.   We used the \textsc{pixmappy} \footnote{https://github.com/gbernstein/pixmappy} astrometric solutions to map from image coordinates to sky coordinates.  
This was a particularly important improvement over the Y1 PSF models, since the DES images have significant ``tree rings'' \citep{Estrada2010,Plazas2014a,Plazas2014b}, where the Jacobian of the astrometric solution changes significantly across regions with only a few stars.  {The modeling in sky coordinates helps reducing these spurious patterns through the use of accurate astrometric solutions, although some residual tree-ring features still remain, especially in the PSF size \citep{Jarvis19}}.

\subsection{Selection of PSF stars}
\label{sect:selection_stars}
Similar to \Zuntz, the initial selection of candidate PSF stars used a size-magnitude diagram of all the objects detected per image. For the magnitude, we used the \textsc{SEXTRACTOR} \citep{Bertin96} measurement \textsc{MAGAUTO}. For the size, we used the scale size as measured with \textsc{Ngmix} \citep{Sheldon2015}. The stars were easily identified {in each exposure (and for different bands)} at bright magnitudes as a locus of points with size nearly independent of magnitude. On the other hand, the galaxies have a range of sizes, all larger than the PSF size. The candidate PSF stars were taken to be this locus of objects from about $m\approx16$, where the objects begin to saturate, down to $m\approx22$, where the stellar locus merges with the locus of faint, small galaxies (the stars magnitudes are always relative to the exposure/band where the stars have been detected). {Binary stars are implicitly removed by \textsc{Piff},  as objects with ellipticity high enough to be recognised as having a different PSF are deemed not suitable for training the PSF model.} {The stellar density varies across the DES footprint; it tends to be higher in those exposures closer to the galactic plane and in the presence of stellar streams (see \citealt{Shipp2018} for more details).} 

From the list of candidate stars, we removed objects that were not suitable to use as models of the PSF. In Y1, we removed all objects within 3 magnitudes of the faintest saturated star in the same CCD exposure. This was done to avoid the interaction of charges in CCDs with the already accumulated charge distribution, which can cause an increase of observed size with flux, an effect also known as the ``brighter-fatter effect" (\citealt{Antilogus2014,Guyonnet2015,Gruen2015,Coulton2018,Lage2019,Astier2019}; see \S~\ref{brighterfatter}). {For DES Y3 we exploited the correction described in Antilogus et al. and implemented for DECam/DES in \cite{Gruen2015} as part of the initial image processing (but see \citealt{Coulton2018} for an alternative correction method)}. The correction has been applied after flat fielding and before the sky correction, which reduced the level of this effect seen on the images and allowed the selection of brighter stars. In particular, we imposed a lower magnitude limit which varies between CCD exposures and band considered, but it is typically of magnitude $\sim 16.5$ (to be compared to the Y1 cutoff at $\sim 18.5$). More details are provided in \cite{Jarvis19}. In the final star catalogue, each star has different entries for each exposure (and therefore band), as the DES Y3 PSF model is different for each exposure. Out of all the stars passing these selection cuts, we employ $\sim 80$ per cent out of all the stars passing these selection cuts to model the PSF, and reserve the remaining $\sim 20$ per cent of them for diagnostic tests (\S~\ref{sect:psf_diagnostic}). The stars reserved for diagnostic tests are selected randomly. 


In Fig.~\ref{fig:fwhm}, we show the distribution of the median measured full-width half-maximum (FWHM) for the PSF stars used in our study, restricted to the exposures used for shear measurements. The overall median seeing is 0.98".

\section{The \MCAL\ shape catalogue} \label{sec:mcal}

The shape catalogue was created using the \mcal\ algorithm presented in \cite{HuffMcal2017} and \cite{SheldonMcal2017}.  The implementation, code and configuration for DES Y3, was the same as that used for the DES Y1 catalogue and we refer the reader to  \Zuntz\ for details.  Here, we briefly describe the basic features of the algorithm and resulting differences between the Y3 and Y1 catalogues.

Consider a noisy, biased measurement, \vest\ , such as a two-component ellipticity estimated from pixel data in CCD images, from which we wish to calibrate a measurement of the gravitational shear, \vecg.  For small shear, we can Taylor expand this estimator as
\begin{align} \label{eq:Eexpand}
    \vest &= \vest|_{\gamma=0} + \frac{ \partial \vest }{ \partial \vecg}\bigg|_{\gamma=0} \vecg  + ... \nonumber \\
          &\equiv \vest|_{\gamma=0} + \mbox{\mcalRg}\vecg  + ... \, ,
\end{align}
where we have defined the {\em shear response matrix} \mcalRg.  In what follows we will drop higher order terms\footnote{{The next order term is $\propto \gamma^3$ \citep{SheldonMcal2017}; for large shears, such as in the case of tangential shear measurements near the centers of galaxy clusters (e.g., \citealt{McClintock2019}), it can introduce a $\sim$ per cent bias, but it can be safely neglected here.}}, and assume that the ellipticities in the absence of lensing $\vest|_{\gamma=0}$ average to zero.  

Given an ensemble of measurements \{\vest$_i$\} and responses \{{\mbox{\boldmath $R_{\gamma_i}$}}\}, we can form unbiased statistics of the shear \vecg.  For example, to measure an estimated mean shear \vecgest\ we can write 
\begin{align} \label{eq:rcorr}
    \langle \vecg^{\rm est} \rangle &\approx \langle \mbox{\mcalRg} \rangle^{-1}  \langle \vest \rangle \approx \langle \mbox{\mcalRg} \rangle^{-1} \langle \mcalRg \vecg \rangle,
\end{align}
where the averages for \vest\ and \mcalRg\ are taken over the ensemble of measurements.

The shear estimate \vecgest\ is a {\em weighted} mean of the measured \vest, with weights  $\langle\mcalRg \rangle$.  This weighting must be accounted for when calculating secondary statistics, such as the calculation of the effective redshift distribution for the ensemble.  Responses can also be derived for other statistics of the shear, such as two-point correlation functions (see Appendix~\ref{sect:2pt_response}).

For \mcal, the response matrix \mcalRg\ for each galaxy was measured using finite difference derivatives.  The derivative was calculated by producing versions of the image that had been sheared by small amounts $\pm \gamma \sim 0.01$, and repeating the measurement \vest\ on those sheared images. We used a central finite difference estimate:
\begin{equation} \label{eq:Rnum}
    R_{\gamma_{i,j}} = \frac{\est_i^+ - \est_i^-}{\Delta \gamma_j},
\end{equation}
{where $\est_i^+$ and $\est_i^-$ are the $i$-th component of the ellipticities measured on images sheared by an artificial shear with $j$-th component equal to ${\pm \Delta \gamma}$.} In order to perform this shearing, the image must be deconvolved by the PSF, sheared, and reconvolved by the PSF. {Before reconvolution, the PSF is further symmetrized (\citealt{SheldonMcal2017}; \Zuntz) in order to correct for PSF anisotropy; then, the symmetrized PSF is slightly dilated in order to suppress amplified noise due to the deconvolution. The dilation depends on the PSF ellipticity: if the PSF is round it corresponds to a $\sim$2 per cent dilation, if not the dilation is slightly larger.} 
Because the reconvolution results in a different PSF, the basic ellipticity measurement used as the shear estimator must be performed on a similarly reconvolved but {\em unsheared} image. To optimise computational efficiency, the DES Y3 implementation of \textsc{METACALIBRATION} deconvolves the original image by the complete PSF solution, but then uses a simplified single Gaussian model and Gaussian PSF to fit the detected objects in the sheared images. \textsc{METACALIBRATION} has been shown to calibrate also biases introduced by this simplified model \citep{SheldonMcal2017}. We performed all calculations using the \ngmix\ package\footnote{The basic \mcal\ measurements were performed using the \ngmix\ code, which is publicly available as free software: \url{https://github.com/esheldon/ngmix}}. The image manipulations are part of \textsc{ngmix.metacal}, which in turn makes use of the \galsim\ \citep{GALSIM2015} software for the convolution operations. Ellipticities were calculated using a maximum likelihood fit of a single Gaussian to the multi-epoch, multi-band observations for each object (\Zuntz). 

{The typical values of the diagonal elements of the shear response} $\langle \mcalRg \rangle$ are of order $\approx 0.6$ for galaxies in DES, although the value depends on the details of the measurements such as object signal-to-noise ratio (\snr) and size relative to the PSF (\Zuntz). In addition to the shear response matrix described above, the response of the estimator to the selections that define the science sample under study must also be taken into account (see \S~\ref{sec:mcalselect} for a summary of the DES Y3 shape catalogue selections).  Selection effects are typically a few percent for selections made on DES catalogues (\Zuntz). This effect can be calculated by selecting on sheared measurements and calculating a new ensemble response $\langle \mcalRs \rangle$ \citep{SheldonMcal2017}.  Then the total ensemble response is then given by
\begin{align}
	\langle \mcalR \rangle = \langle \mcalRg \rangle + \langle \mcalRs \rangle
\end{align}
and averages are then performed using the total response:
\begin{align} \label{eq:rcorr}
    \langle \vecg^{\rm est} \rangle &= \langle \mbox{\mcalR} \rangle^{-1}  \langle \vest \rangle
\end{align}
We split into $\langle \mcalRg\rangle$ and $\langle \mcalRs\rangle$ because $\langle \mcalRg\rangle$ can be calculated for each object separately and then used as a weight in other averages, for example to calculate the redshift distribution of the catalogue.  However, the total ensemble averaged response $\langle \mcalR \rangle$ can also be calculated directly by treating the measurements from the sheared images as completely
separate catalogues, and performing the selections and ensemble averages on each catalogue separately \citep{SheldonMcal2017}.
In that case, mean weights can be derived by binning the catalogue, e.g.  by redshift, and calculating the $\langle \mcalR \rangle$ in each bin.

As noted in \cite{SheldonMcal2017}, the total ensemble response matrix $\langle \mcalR \rangle$ is, to good approximation, diagonal: as a consequence, the response correction reduces to element-wise division. 

\subsection{Differences Between the Y3 and Y1 Catalogues}
\label{sec:catdiff}

The \mcal\ shape catalogue differs from DES Y1 in the
following ways:
\begin{itemize}

    \item \textsc{Piff} PSF solutions were used for the \mcal\ deconvolutions rather
        than the \psfex\ solutions that were used for Y1 (See \S
        \ref{sec:piff}).
        
    \item The Jacobian of the world coordinate systems (WCS) WCS transformation was taken from the \pixmappy\ astrometry solutions (see \S \ref{sec:astrom}).
    
       \item We altered the weak-lensing selection criteria (see \S \ref{sec:mcalselect}).

    \item We applied an inverse variance weight to galaxies (see \S \ref{sect:weight}).

\end{itemize}

In addition to these differences, we also applied a calibration correction (2-3\%) to the catalogue based on simulations (see \S~\ref{image_sims_summary}). This correction mostly calibrates a {\em shear-dependent detection bias} which affects the shear estimates when objects are blended. We do not expect the aforementioned detection related biases to be addressed by the tests in this paper, as the tests presented herein are mainly sensitive to additive shear biases.  

\newpage
\subsection{Object Selection from the \MCAL\ Catalogue} \label{sec:mcalselect}
\begin{figure*}
\begin{center}
\includegraphics[width=\textwidth]{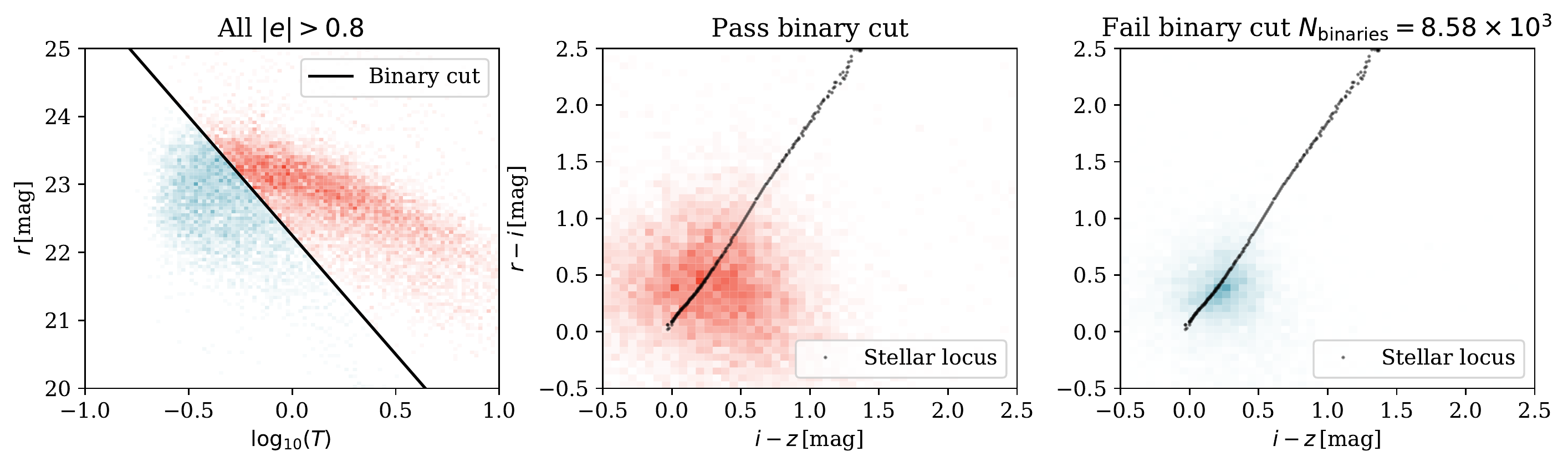}
\end{center}
\caption{Shear bias from contamination by unresolved binary stars. \emph{Left} shows the cut applied to isolate unresolved binaries from the population of objects in in our catalogue with measured $|e| > 0.8$, \emph{centre} shows objects from above the cut, which have galaxy colours, and \emph{right} shows objects below the cut, which have stellar colours.}
\label{fig:stellar:binaries}
\end{figure*}
Here we discuss the standard weak lensing selection employed in all the DES Y3 shear analyses. If additional selections are included, such as tomographic binning, these can induce further selection biases that must be accounted for by an appropriate selection response, \mcalRs\ .

{We performed \mcal\ measurements on all objects detected by \sx\ in the DES coadds, using the $riz$ bands. We excluded the $g$ band measurements due to known issues in the estimation of the PSF (see \citealt{Jarvis19} for a discussion).}  The Y3 detections are significantly different from those in Y1 due to changes to the \sx\ configuration that resulted in a more pure and complete catalogue \citep{Y3Gold}.  
A small subset of objects (less than a percent) were not measured due to lack of data in one or more bands, typically near the survey boundaries.


For objects processed with \mcal, we made the following further selections:
\begin{itemize}

    \item The object measurements had to belong to the unmasked regions of the DES Y3 Gold catalogue after problematic regions had been removed and had not to be marked as ``anomalous'' \citep{Y3Gold}. These selections should be nearly shear independent.

    \item We selected objects with $ 10 <$ \snr $< 1000$, as determined by the Gaussian fit to the unsheared image. The \snr\ definition is the same as used in \Zuntz\footnote{In particular, we define \snr $= \left(\sum_{\rm p} m_{\rm p} I_{\rm p}/\sigma^2_{\rm p}\right)/\left(\sum_{\rm p} m^2_{\rm p}/\sigma^2_{\rm p}\right)^{1/2}$, where the sum runs over the pixel ${\rm p}$, $m_{\rm p}$ is the best fit model for the galaxy, $I_{\rm p}$ the measured pixel value and $\sigma_{\rm p}$ the estimated pixel variance.}, {and it is computed combining information from all filters and exposures.} The low cut limited faint objects impacted by detection biases. The high cut removed very bright objects, for which Poisson noise could create fluctuations larger than the typical background noise, erroneously flagging the detections as problematic\footnote{In the implementation used for the DES Y3, matched pixels from different single epoch postage stamps of a detected object were compared, and if some of the values were too far from the median, the object was rejected. This ``outlier rejection'' algorithm was implemented mostly to remove problematic pixels, e.g., those affected by cosmic rays and correctly masked.}. 

    \item We selected the objects with galaxy to PSF size ratio $T / T_{\rm PSF} > 0.5$, as in DES Y1, to reduce the impact of PSF modeling errors.  $T$ is a measure of the size squared of the object, and it is defined following \Zuntz\ as $T = I_{xx}+I_{yy}$, with $I_{\mu \nu} =\left[\int dx dy I(x,y)(\mu-\bar{\mu})(\nu-\bar{\nu})\right]/\left[\int dx dy I(x,y)\right]$. To compute $T$, the galaxy Gaussian best fit model $I(x,y)$ is used.  The $T_{\rm PSF}$, determined by \mcal\ , is the size squared of the PSF, also from the Gaussian fit of the PSF. {The $T$ and $T_{\rm PSF}$ used for the selection are the average $T$ and $T_{\rm PSF}$ over all the exposures and bands, with weight according to the weight maps from each epoch. The selection, therefore, is not applied on a per exposure basis, but on each single galaxy.}
        
         \item We imposed the selection $T < 10$ arcsec$^2$, which removed the largest objects.  By visual inspection, many of these detections are not large objects, but their size estimate is affected by the light emitted by close, large neighbours. 
         
         \item We excluded the objects characterised {simultaneously} by $ T > 2$ arcsec$^2$ and ${\rm S/N} < 30$. These relatively large, faint objects are mostly blends upon visual inspection, and their inclusion could potentially introduce biases in the catalogue. 
         
          \item We limited the objects to those for which the most reliable photometric redshifts could be obtained: $18 < i < 23.5$, $ 15 < r,z < 26$ and fine-tuning against any outlier colours with $-1.5 < (r-i,z-i) < 4$ \citep{Myles2020}. 
         
         \item We imposed a selection to limit the binary star contamination of the galaxy catalogue.  For high-ellipticity objects of the shape catalogue, unresolved binary stars could contribute significantly and are difficult to distinguish from galaxies. Following \cite{Hildebrandt2017}, we cut our high ellipticity ($|\vest|>0.8$) shape catalogue in $r$ magnitude -- size ($T$) space according to: $\log_{10} (T/{\rm arcsec}^2) < (22.5 - r)/2.5$ (see left panel of Fig.~\ref{fig:stellar:binaries}). Colour-colour plots of these objects tend to follow better a stellar locus than the remainder of the catalogue (central and right hand panels of Fig.~\ref{fig:stellar:binaries}), although the difference is not conspicuous as stars and galaxies are not well separated in the $r-i$, $i-z$ plane. Hence, we inferred that these were indeed unresolved binary stars and removed them from the shape catalogue. These objects constituted $20\%$ of the $|\vest|>0.8$ objects in the shape catalogue before their removal. {We note that the removal might not be perfect, and some binary stars could be still contaminating our catalogue. The impact of stars contamination is further discussed in \S~\ref{sec:starcontam}.}

\end{itemize}

{All the selections described here are combined using logical conjunction to obtain our final weak lensing selection. Except for the first selection, all the others are shear dependent and can induce a selection bias that has to be corrected for using the selection response term \mcalRs . We recall that in the current implementation of \mcal\, detection effects are not corrected for by any selection response terms, and need to be calibrated for using image simulations}. The selection discussed here constitutes a reliable weak lensing selection and is applied for all tests detailed in this paper, as well as further studies. The number of objects passing this selection is 100,204,026.
\bigskip

\subsection{Inverse variance weight}\label{sect:weight}

An estimator of a shear signal is usually a linear combination of individual galaxy shapes. In that linear combination,  one can assign equal weight to each galaxy or alternatively, a different weight $w_i$ to each galaxy. A dependence of that weight on shear could introduce selection biases which, however, can be corrected by the \mcal\ formalism if the weight is determined from quantities also measured on artificially sheared versions of the galaxy image. For  minimising the variance of the measured shear signal, it can be shown that the weight should be proportional to the  inverse of the variance of the shear estimated from each galaxy.

The variance of mean shear estimated from a sample of galaxies as in Eq.~\ref{eq:rcorr} is
\begin{align} \label{eqn:shearvar}
\sigma_{\gamma}^2=\sigma_{e}^2 \langle \mcalRg \rangle^{-2} \; , 
\end{align}
where $\sigma_{e}^2$ is the variance of \vest including intrinsic and measurement-related shape noise. While for any individual galaxy it is difficult to evaluate Eq.~\ref{eqn:shearvar}, e.g. due to the noise in $\mcalRg$, for a large ensemble of galaxies it is straightforward to estimate both $\sigma_{e}^2$ and $\langle \mcalRg \rangle^{-2}$. We therefore chose to estimate $\sigma_{\gamma}^2$ and thus assigned a piecewise-constant weight for 
ensembles of galaxies binned by the quantities S/N and $T/T_{\rm PSF}$:
\begin{multline}
w^{i}\left(T/T_{\rm PSF},\:{\rm \snr} \right)= \sigma_{\gamma}^{-2} \left(T/T_{\rm PSF},\:{\rm \snr} \right) = \\\left[\sigma_{e}^{-2}\langle \mcalRg \rangle^{2}\right]\left(T/T_{\rm PSF},\:{\rm \snr} \right),
\end{multline}
with
\begin{equation}
 \sigma_{e}^2 \left(T/T_{\rm PSF},\:{\rm \snr} \right) = \frac{1}{2}\left[\frac{\sum (e_{i,1})^2}{n_{\rm gal}^2}+\frac{\sum (e_{i,2})^2}{n_{\rm gal}^2}\right].
\end{equation}
In the above equations,$\langle \mcalRg \rangle^{2}\left(T/T_{\rm PSF},\:{\rm \snr} \right)$ is the response and $n_{\rm gal}$ is the raw number count of galaxies in a given bin of $\left(T/T_{\rm PSF},\:\snr \right)$, and the sum over the ellipticities squared runs only over the galaxies belonging to that bin. Similarly, We used the size ratio and \snr\ because they are main proxies for measurement-related shape noise and variations of  response. 

\begin{figure*}
\begin{center}
\includegraphics[width=0.8 \textwidth]{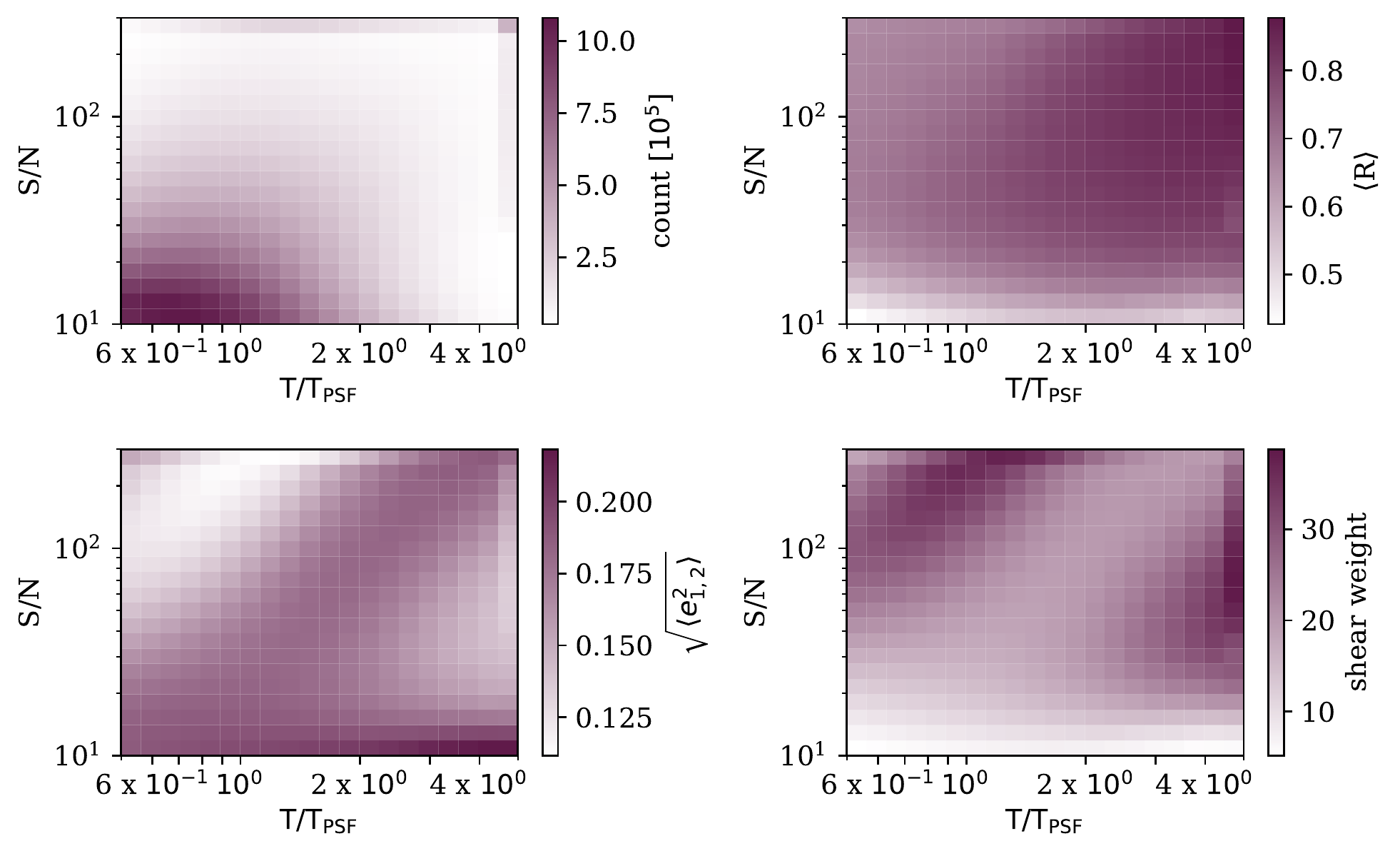}
\end{center}
\caption{Properties of the Y3 \mcal\ catalogue as a function of galaxy S/N and size ratio $T / T_{\rm PSF}$ (defined as the ratio between galaxy size and PSF size). \textit{Top left:} galaxy number counts; \textit{top right:} \mcal\ response, as defined in \S~\ref{sec:mcal};  \textit{bottom left:} root-mean-square of components of galaxy ellipticity; \textit{bottom right:} shear weights, as defined in \S~\ref{sect:weight}.}
\label{fig:weights}
\end{figure*}

Fig.~\ref{fig:weights} shows the counts, $\langle \mcalRg \rangle$, and $\sqrt{\sigma_{e}^2}$ of galaxies in $20\times20$ logarithmically scaled bins of ${\rm S/N}=10\ldots300$ and $T/T_{\rm PSF}=0.5\ldots5$. The upper limit of each range is chosen such that more than $97.5\%$ of the sample lies below it. Remaining galaxies with large S/N or $T/T_{\rm PSF}$ are subsumed into the respective last bin.

While shear response is a mostly monotonic function of S/N and a weak function of size, we found the scatter in the measured ellipticity to have a more complex behavior. Visual inspection of samples of galaxies with small and large size ratio at high S/N, and with small and large S/N at large size ratio indicated that this is a result of how galaxy morphology maps to this space of observed properties: the large scatter in ellipticity of galaxies with large S/N and size ratio results from the incidence of highly elliptical, nearly edge-on disk galaxies. We also note that a redshift dependence is hidden in the four plots of Fig.~\ref{fig:weights}: generally, the high \snr\ bins are characterised by a low mean redshift, whereas low \snr\ bins have a high mean redshift. 

The inverse-variance weighting significantly increases the statistical power of the \mcal\ catalog; 
without weighting of galaxies, the fiducial sample triples the statistical power of DES Y1. Inverse-variance weighting increases this further by $\sim 25\%$. The relative gain in statistical power is only a weak function of the S/N cut-off chosen. However, we note that for even lower S/N than the minimum of 10 usable here, the statistical power of the unweighted catalogue has a maximum in the cut-off S/N due to the noise introduced by faint galaxies.


\subsection{Number Density}\label{sect:number_density}
\begin{figure*}
\begin{center}
\includegraphics[width=0.48\textwidth]{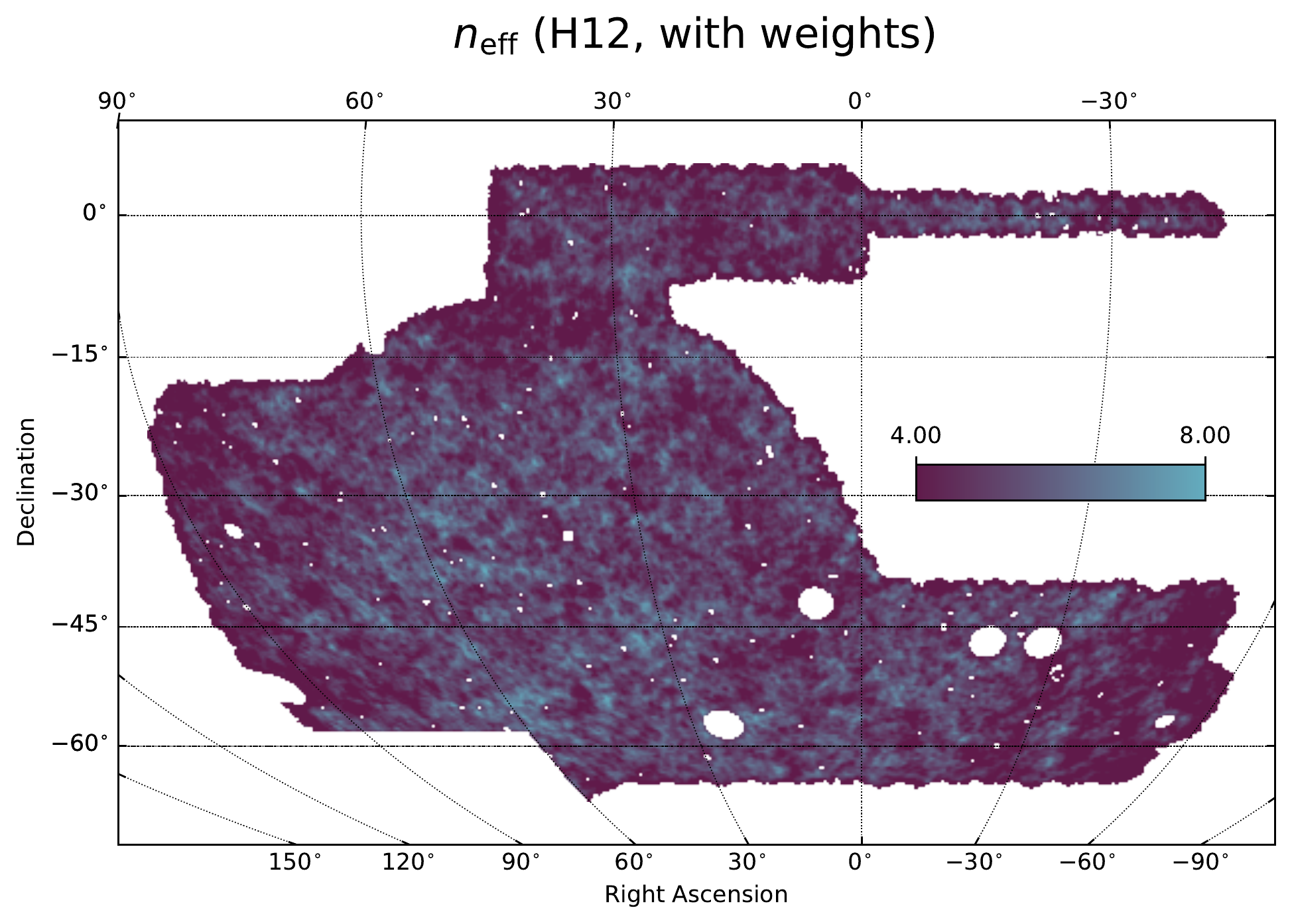}
\includegraphics[width=0.48\textwidth]{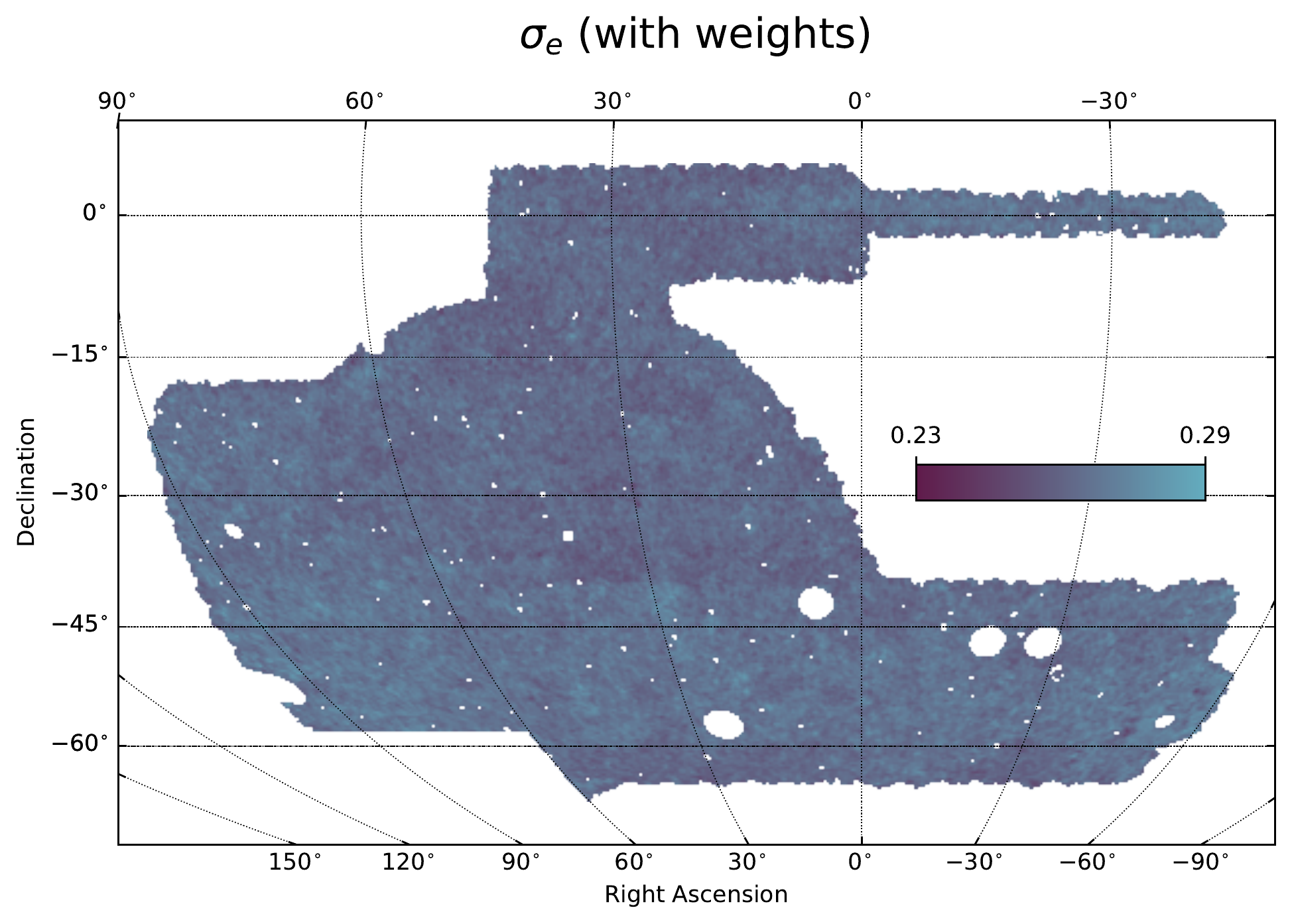}
\end{center}
\caption{Weighted effective number density, $n_{\rm eff}$, and shape variance, $\sigma_e$, of sources in the weak lensing selection across the survey footprint. }
\label{fig:number_density_map}
\end{figure*}




\begin{table}
\centering
\caption{Number density values and noise per component using the different definitions described in \S~\ref{sect:number_density}. The last two columns, $c_1$ and $c_2$, indicate the per-component mean shear measured in the catalogue.}
\begin{adjustbox}{width=0.45\textwidth}
\begin{tabular}{|c|c|c|c|c|}
 \hline
{Definition} & $n_{\rm eff}$ &{$\sigma_e$} & $c_1$ & $c_2$ \\
 \hline
     Chang+13     &  {5.320}      & {0.255} & {0.00035} & {0.00006} \\
      Heymans+12      &   {5.592}     & {0.261} & {0.00035} & {0.00006}\\
 \hline
\label{table:n_eff}
\end{tabular}
\end{adjustbox}
\end{table}
After applying the appropriate selections, the effective number density, $n_{\rm eff}$, and shape variance, $\sigma_{\rm e}$, are computed and reported in Table \ref{table:n_eff}, using the definitions from \cite{Chang2013} and \cite{Heymans2012}. These quantities, together, typically quantify the overall constraining power of a shape catalogue as the variance of the estimated shear, $\sigma_{\gamma}^2=\sigma_e^2/n_{\rm eff}$.  

The definition for the shape variance given by \cite{Chang2013} reads as follows:
\begin{equation}
\sigma_{e,\rm  C13}^{2} = \frac{1}{2}\frac{\sum w_i^2 \left(e_{i,1}^2+e_{i,2}^2-\sigma_{\rm m,i}^2\right)}{\sum w_i^2},
\end{equation}
where $\sigma_{\rm m,i}$ is the per-galaxy measurement noise as provided by \mcal\ . The effective number density is given by \cite{Chang2013} in terms of the area of the survey, $A$, as:
\begin{equation}
n_{\rm eff, C13} = \frac{1}{A} \frac{\sigma_{e,\rm  C13}^2 \sum w_i^2}{\sum w_i^2 \left(\sigma_{e,\rm  C13}^2 +\sigma_{\rm m,i}^2/2\right)}\, .
\end{equation}
%


Alternatively, the definition by \cite{Heymans2012} is given in terms of the shear weight, $w_i$, for each galaxy as:
\begin{equation}
n_{\rm eff, H12} = \frac{1}{A} \frac{(\sum w_i)^2}{\sum w_i^2} \, .
\end{equation}
The shape variance is given by, 
\begin{equation}
\label{eq:s_eff}
\sigma^2_{\rm eff, H12} = \frac{1}{2}\left[\frac{\sum (w_{i} e_{i,1})^2}{(\sum w_{i})^2}+\frac{\sum (w_{i} e_{i,2})^2}{(\sum w_{i})^2}\right]\left[\frac{(\sum w_i)^2}{\sum w_i^2}\right] \, .
\end{equation}
{We use the H12 definition to compute the analytical covariances needed for the cosmic shear cosmological analysis \citep{Amon2020,Secco2020}.} We note that in Eq.~\ref{eq:s_eff} we already assumed the ellipticities to be corrected by the response matrix, so the above equations provide the correct $\sigma^2_{\rm eff}/n_{\rm eff}$ needed to estimate the analytical covariance matrix. In principle, if a further calibration of the shear multiplicative bias $m$ is applied to the catalogue (beyond the \mcal\ response matrix), this has to be included in the estimate of Eq.~\ref{eq:s_eff} \citep{J20}. Since in our case this additional calibration factor is only a few percent and it will be partially self-calibrated by our data, we decided not to include it in our estimate of $\sigma_{\rm eff}$.

In Table \ref{table:n_eff} we further report the per component mean ellipticity measured in the catalogue ($c_1$ and $c_2$). The mean ellipticity is defined as the weighted sum of the galaxy ellipticities, corrected by the mean response. In particular, the mean shear measured for the first component is larger than the expected mean shear from cosmic variance ($\sim 0.5$ $10^{-5}$). The mean ellipticity needs to be subtracted before any science application. More details about the potential origin of such mean shear are provided in \S~\ref{sec:additiveother}.

Fig.~\ref{fig:number_density_map} shows the spatial pattern of the weighted effective number density of the survey, as well as the shape variance. 

\subsection{Absolute calibration from image simulations}\label{image_sims_summary}

In \cite{MacCrannSims2019} we tested the performance of the \mcal\ methodology described above using image simulations. We generated a suite of simulated multi-band DES-like images in which known shear signals were applied. {The simulations are generated following closely the real DES Y3 data. We first simulate complete sets of single-epoch images forming 400 DES Y3 tiles (selected at random among all the available ones) in all four photometric bands. The single epoch images have the same pixel geometry of the real data images; the noise and weight-maps are estimated from the corresponding images in real data. For every single epoch image, the same WCS used in the real images is implemented, and pixels are masked based on the data bad pixel mask. Parametric models for stars and galaxies are injected in the images using  \galsim\ \citep{GALSIM2015}; models from the COSMOS field \citep{deepfields} are used for galaxies, whereas models for stars are taken from \cite{trilegal2019}. Galaxies and stars are convolved with smoothed versions of the PSF estimated from real data. The simulated images include realistic levels of seeing; small anisotropies in the data PSF are also included in the simulations.} The images were then analysed with much of the same machinery as the real DES data, e.g., detection using \textsc{SExtractor} was performed on a \textsc{SWarp}-generated coadd, and \mcal\ was used to estimate the mean shear, which can be compared to the known input shear to estimate shear calibration biases.

We expect to observe biases at some level since the \mcal\ methodology described above does not account for possible shear dependence in the detection and  deblending of sources performed by \textsc{SExtractor}. Indeed in \cite{MacCrannSims2019} we find an average multiplicative biases of $m=(-2.08\pm0.12)$ per cent and additive biases of $c_1 = (-1.0 \pm 1.4)10^{-4}$,  $c_2 = (-1.2 \pm 1.4)10^{-4}$ for the full shear catalogue (i.e. after the standard weak lensing selection described in \S \ref{sec:mcalselect}). We describe in detail the source of these biases. We note that in image simulations we do not measure any statistically significant $\langle e_1 \rangle$, in contrast to what we measured on data, meaning that the root cause of that positive mean shear is not modelled in image simulations. The multiplicative bias quoted above is strictly only applicable as a correction to a constant shear signal. In the presence of blending and redshift dependent shear signals a more general approach to correcting theoretical predictions of cosmological lensing signals is required; we again refer the reader to \cite{MacCrannSims2019} where we describe and implement such an approach. 

\subsection{Y3 Shear Catalogue Public Release} \label{sec:release}

The usage of the Y3 \mcal\ catalogue is identical to the usage of the year 1 catalogue.  Please see \Zuntz\ for details.  The full \mcal\ catalogue will be made publicly available following publication, at the URL \url{https://des.ncsa.illinois.edu/releases}. {We remind that in order to correctly use the catalogue for any scientific purposes, the calibration based on image simulations \citep{MacCrannSims2019} needs to be applied.}

\section{PSF Diagnostics}\label{sect:psf_diagnostic}
In this section we detail the systematic effects that are connected to the DES Y3 PSF model and residuals. In particular, we discuss the tests we performed on:
\begin{itemize}
    \item the impact of the \textbf{brighter-fatter effect} (\S~\ref{brighterfatter}) in the stars used for the PSF modeling;
    \item dependencies of the \textbf{PSF model residuals on stars and galaxy colours} (\S~\ref{sect:chromatic});
    \item \textbf{additive biases due to PSF misestimation} (\S~\ref{sec:PSF_modeling_error}) using $\rho$ statistics \citep{Rowe2010}, both in sky coordinates and focal plane coordinates;
    \item \textbf{tangential shear around stars} (\S~\ref{sec:gammat_stars}). 
\end{itemize}
{These tests aim at empirically detecting biases in the shape catalogue due to PSF modeling errors. Additional tests of the DES Y3 PSF modeling that are independent of the shape catalogue can be found in the DES Y3 PSF model paper \citep{Jarvis19}.}

 \subsection{Brighter-fatter effect}\label{brighterfatter} 
\begin{figure}
    \begin{center}
        \includegraphics[width=0.4 \textwidth]{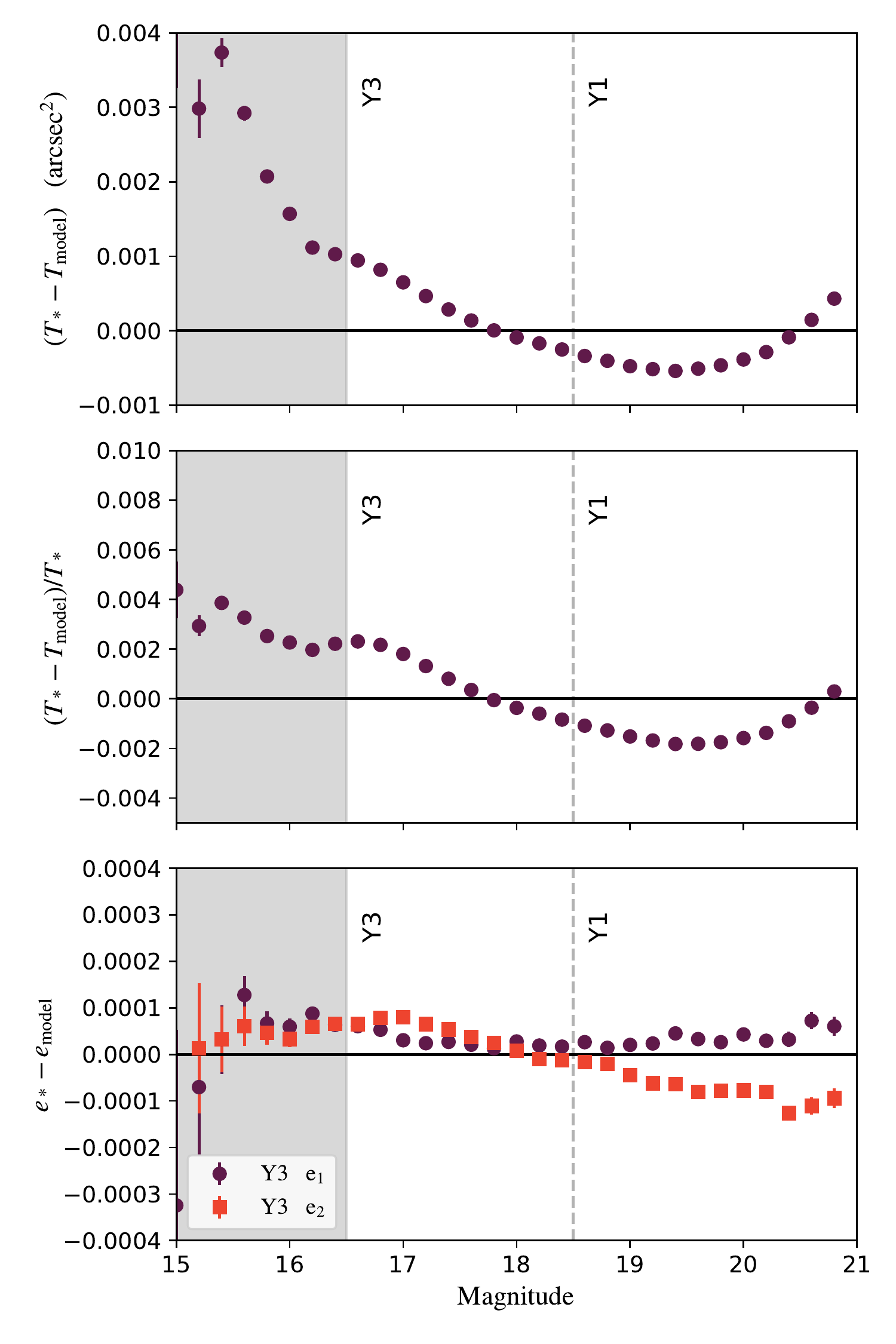}
    \end{center}
    \caption{The PSF residual size (top), fractional size (middle) and shape (bottom) of stars as a function of their magnitude {(relative to the band where the star has been detected)}. {The brighter-fatter effect can be noted as an increase in the PSF size residual at bright magnitudes}. To reduce the impact of the brighter-fatter effect, bright stars are excluded from our PSF models; the cut-off varies between CCD exposures but the shaded grey region shows a typical example. {For the stars passing the cut, the fractional size residuals are below 0.5 percent, at all magnitudes.}} 
    \label{fig:BF}
\end{figure}

The interaction of charges in CCDs with the already accumulated charge distribution causes an increase of observed size with flux, also known as the brighter/fatter effect \citep{Antilogus2014,Guyonnet2015,Gruen2015}. In Fig.~\ref{fig:BF} we show size residuals (upper panel), fractional size residuals (second panel), and $e_1$ and $e_2$ shape residuals (lower panel) {of the \textsc{PIFF} model for the reserved stars catalogue, relative to the actual PSF measurements, as a function of their magnitude}. The impact of the brighter-fatter effect observed in DES Y1 was reduced by the exclusion of the bright stars from the PSF modeling procedure, with the cut-off varying between CCD exposures, but typically at magnitude $\sim$18.5. For DES Y3, we implemented a correction of the effect, which is applied directly to the pixel values early in the data reduction process
\citep{Gruen2015,Morganson2018}, which allowed for the utilisation of stars down to magnitude $\sim$16.5. The gain of stars two magnitudes brighter than those considered in the Y1 analysis contributed to improving the PSF solutions for DES Y3. For stars brighter than  $\sim$16.5, an upturn in the size residuals can still be seen, indicating that the correction implemented was not enough to remove the brighter-fatter effect for the brightest objects. We note, however, that this upturn is a bit milder in the central panel, which shows the fractional size residual as a function of magnitude (which is the key quantity here, as biases in the PSF size should lead, at first order, to a multiplicative bias that scales a $\Delta T/T$). The trend at fainter magnitudes might be related to potential galaxy contamination or noise biases, although deeper investigation is needed to confirm the nature of these trends. Shape residuals show no significant trend with magnitude within all the magnitude range considered here. We also produced a per-chip version of Fig.~\ref{fig:BF}, following \cite{Giblin2020}; most of the chips followed the expected patterns, except chip 6, which exhibited a mild flux dependence of $e_{*,1}-e_{\rm model,1}$. The origin of this mild flux dependence is unknown (but see \citealt{Giblin2020} for a list of potential causes), although we do not think it could cause any problem to the DES Y3 analysis. Moreover, our dithering strategy puts each galaxy on a different chip each exposure, further mitigating this effect.


\subsection{PSF residual with colour}\label{sect:chromatic}

We investigate the dependence of the PSF residuals on the colour of the stars, as compared to the colour of galaxies, in order to ensure that the PSF is well matched to the galaxies. In general, different effects cause the PSF to be wavelength-dependent, with potential consequences on the robustness of cosmic shear measurements \citep{Cypriano2010,Voigt2012,Plazas2012,Semboloni2013,Meyers2015}; no chromatic correction is included in the DES Y3 PSF model, so if the typical colours of the stars used to model the PSF are different from those of the galaxy sample, this can induce a bias. {We anticipate that the Y6 PSF model will include a chromatic correction in the form of a single colour parameter to be included during the PSF fit (see \citealt{Jarvis19} for more details), but this has not been included in the DES Y3 PSF model as it was deemed not necessary.}

\begin{figure*}
\begin{center}
\includegraphics[width=0.9 \textwidth]{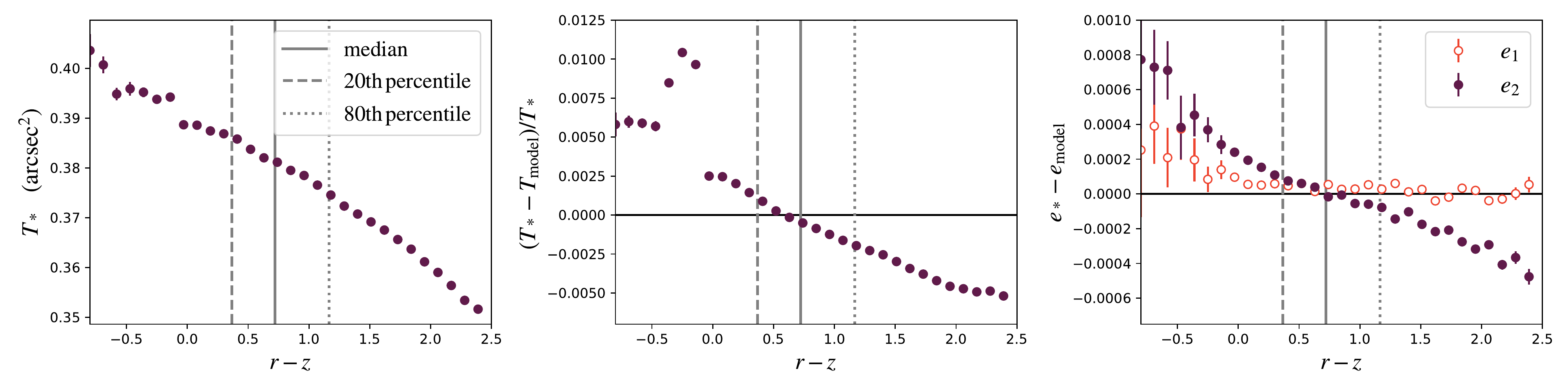}
\end{center}
\caption{The PSF  size (left), fractional size (middle) and shape (right) of stars as a function of their $r-z$ colour. {The PSF  size and shapes are relative to the exposure/band where the star has been detected. The colour for a given entry has been computed by matching by RA/DEC the stars observed in different band/exposures.}For our galaxy sample, the median is $(r-z)=0.75$ (vertical solid line). Most of the DES Y3 galaxy sample (the vertical dashed lines indicate the 20th and 80th percentiles) is within $dT/T<0.002$ and  $\Delta e<0.0001.$}
\label{fig:psfres_colour}
\end{figure*}

Fig.~\ref{fig:psfres_colour} shows the PSF size, the fractional PSF size and shape residuals as function of colour. A noticeable dependence on colour can be seen. It is possible that part of the mean shear trend can be explained by differential chromatic diffraction, while the PSF size trend is probably dominated by Kolmogorov seeing \citep{Jarvis19}, but further investigation is needed to fully clarify the nature of these trends. In each panel, the median colour of the DES Y3 galaxy sample is over-plotted, which corresponds to $(r-z)=0.75$, as well as the 20th and 80th percentile colour of the sample. This indicates that most of the sample is within $dT/T<0.002$ and  $\Delta e<0.0001$, {deemed acceptable as it would roughly correspond to an additive bias of the same order of magnitude of the expected cosmic variance on $\langle e \rangle$ ($\sim 0.5$ $10^{-4}$). Even if we do not directly correct for this, we stress that biases due to unaccounted chromatic effects should ultimately be captured by the $\rho$ statistics test, described in the next section. } {Last, we note that in the central panel of Fig.~\ref{fig:psfres_colour} a few points at $r-z\sim-0.3$ seem to not follow the main fractional PSF size - colour relation. This is probably caused by a few AGN/quasars contaminating our PSF stars catalogue (see Fig. 6 of \citealt{Jarvis19}), as the size of these objects is larger than the one predicted by our PSF model. We did not consider this contamination problematic, as the number of objects with $r-z<0$ in our PSF catalogue is less than 0.5 per cent.}



\subsection{Additive biases from PSF Modeling: $\rho$ statistics}\label{sec:PSF_modeling_error}
\begin{figure*}
\begin{center}
\includegraphics[width=0.9 \textwidth]{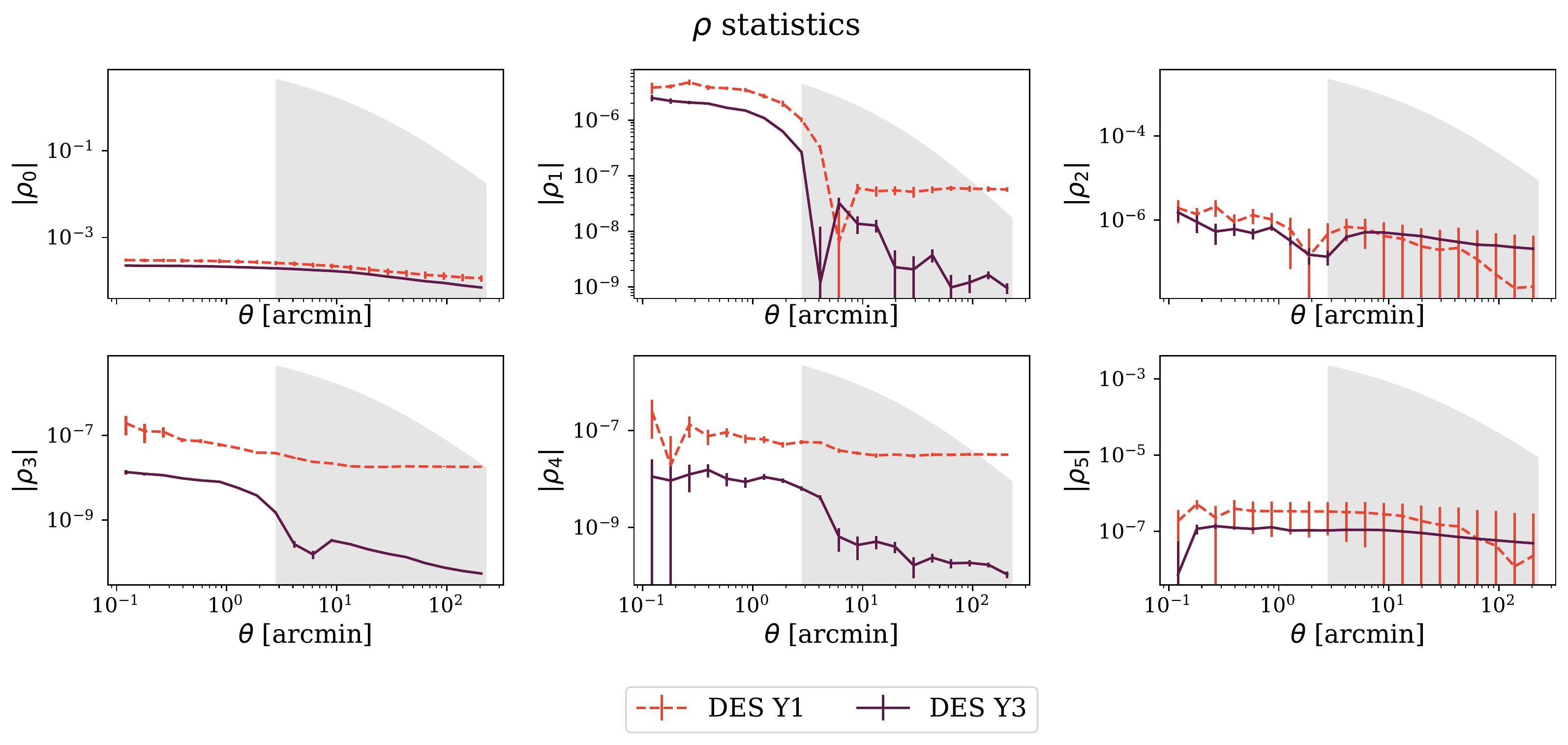}
\end{center}
\caption{$\rho$ statistics as measured for the catalogue of reserved stars. Only the $\rho_+$ components are shown. DES Y3 values are compared to DES Y1 values, showing a substantial improvement owing to a better PSF modeling. As an order of magnitude comparison, we show as grey regions 10 per cent of the weakest expected cosmic shear $\xi_+$ signal, which is from the lowest redshift tomographic bin. In order to effectively compare the cosmic shear signal to each single $\rho$ statistics, we divided it by $\alpha^2$, $\beta^2$, $2 \beta \alpha$, $\eta^2$, $2 \beta \eta$, $2 \alpha \eta$, depending on whether we compare to $\rho_0$, $\rho_1$, etc. We furthermore assumed the following realistic values: $\alpha = 0.001 $, $\beta = 1 $, $\eta = 1 $. We recall this only serves as an order of magnitude comparison, the impact of PSF residuals on the cosmic shear analysis has been quantified in \citealt{Amon2020} and deemed negligible.}
\label{fig:rho_stats}
\end{figure*}
\begin{figure*}
\begin{center}
\includegraphics[width=0.9 \textwidth]{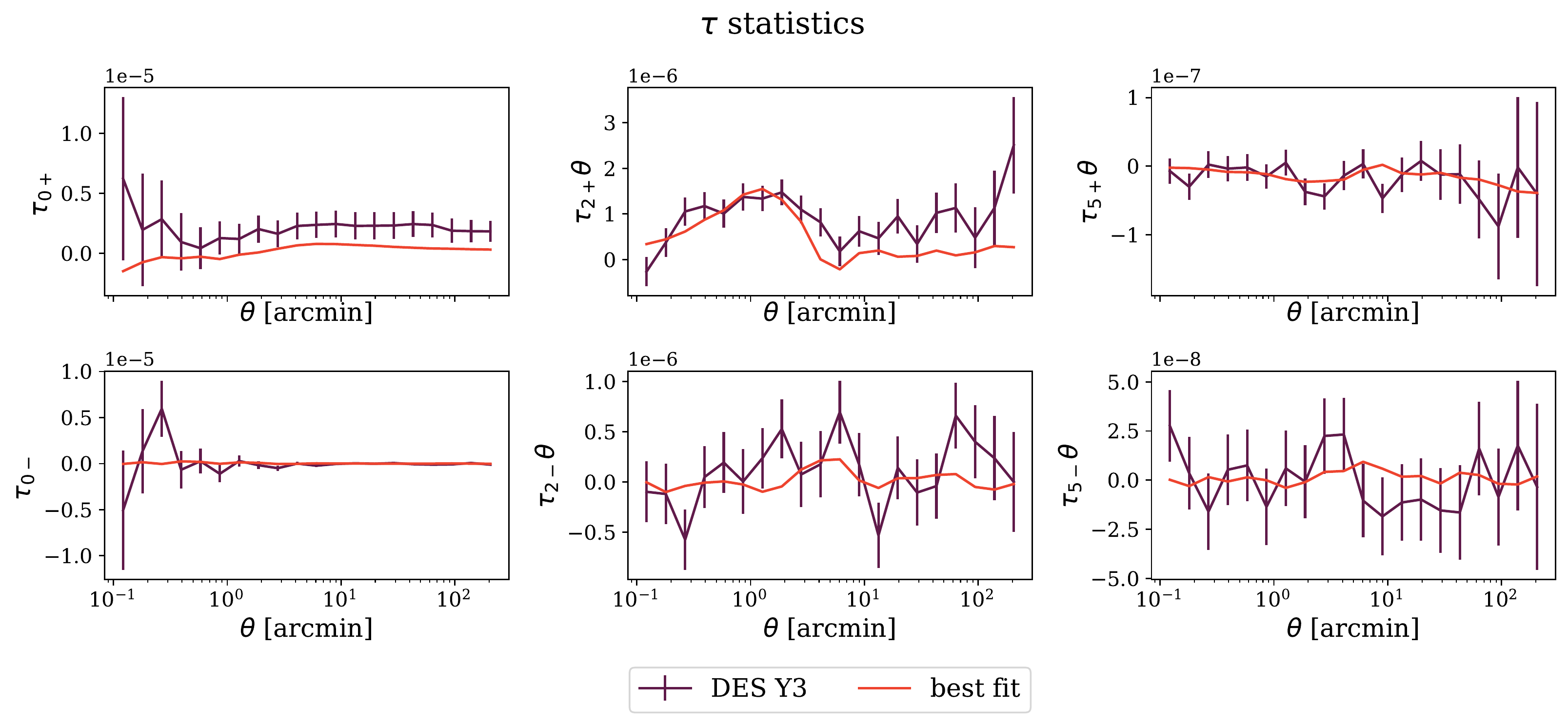}
\end{center}
\caption{Measured $\tau_{+}$ and $\tau_{-}$ together with the best fit models.}
\label{fig:rhoandtaustats}
\end{figure*}
%
%
In this section, the propagation of additive systematic errors due to PSF-misestimation to measurements of the ellipticity of galaxy images is quantified. It is assumed that the observed shape of a galaxy inherits additional contributions due to additive systematic errors and noise:
\begin{equation}
\label{eq: observed}
\vecgest=\vecg +\delta \vest^{\textrm{sys}}_{\textrm{PSF}}+\delta \vest^{\textrm{noise}}
\end{equation} 
Specifically, $\delta \vest^{\textrm{sys}}_{\textrm{PSF}}$ quantifies additive systematic biases from PSF modelling errors. Other sources of additive systematic biases are  explored in \S~\ref{sec:additiveother}. Note that, in contrast to Eq.~\ref{eq:1}, Eq.~\ref{eq: observed} does not include any source of multiplicative biases, which are instead discussed in \cite{MacCrannSims2019}.

While we expect that $\langle \delta \vest^{\textrm{noise}} \rangle = 0$, detection of a signal for the PSF residual, $\langle \delta \vest^{\textrm{sys}}_{\textrm{PSF}} \rangle$, would point to a problem. Following \cite{Paulin-Henriksson2008,Jarvis2016}, we describe PSF modelling errors as:
\begin{equation}
\label{eq:new}
\delta \vest^{\textrm{sys}}_{\textrm{model}}=\alpha \textbf{e}_{\rm model}+\beta\left(\textbf{e}_{\rm *}-\textbf{e}_{\rm model}\right)+\eta\left(\textbf{e}_{\rm *}\frac{T_{\textrm{\rm *}}-T_{\rm model}}{T_{\rm *}}\right),
\end{equation}
where $\alpha,$ $\beta$, and $\eta$ are the coefficients we must solve for, $e_{\rm *}$ is the PSF ellipticity measured directly from stars, $T_{\rm model}$ is the modeled PSF size and $T_{\rm *}$ is the PSF size measured from stars. The first term on the r.h.s is proportional to the PSF model ellipticity (sometimes this term is referred to as PSF leakage). Non-null $\alpha$ could arise from errors in the deconvolution of the PSF model from the galaxy image. The second and third terms describe the impact of PSF model ellipticity and size errors. As PSF model errors produce an error in the shear estimate of similar order of magnitude (\citealt{Paulin-Henriksson2008},\Zuntz), the coefficients $\beta$ and $\eta$ are expected to be of the order of unity, although their exact value will depend on the detailed properties of both PSF and galaxy profiles. In general, we think the formalism outlined by Eq.~\ref{eq:new} is a good effective model to capture additive biases due to PSF modelling errors in our measurements, although we note that slightly different models exist in literature (e.g., \citealt{Giblin2020}).

For simplicity of notation, we rename the terms in Eq.~\ref{eq:new} as $\pp \equiv \textbf{e}_{\rm model}, \,\, \vqst\equiv \vest_{\rm *}-\vest_{\textrm{model}}$, and $\vwst\equiv \vest_{\textrm{*}}\left(T_{\rm *}-T_{\textrm{model}}\right)/T_{\textrm{*}}$, and rewrite it as 
\begin{equation}
\label{eq: simple}
\delta \vest^{\textrm{model}}_{\textrm{PSF}}=\alpha \pp+\beta \vqst+\eta \vwst.
\end{equation}
To solve for the three unknown coefficients $\alpha$, $\beta$, and $\eta$, we correlated all the observed shears  $\vecgest$ (Eq.~\ref{eq: observed}) in the {\mcal}  catalogue with the quantities $\pp$, $\vqst$, $\vwst$ measured for a catalogue of `reserved' stars that have not been used to constrain the model of the PSF\footnote{The reserved stars constitute ~20\% of all the stars selected as explained in \S~\ref{sect:selection_stars}.}. Assuming that the true shear signal $\vecg$ does not correlate with PSF modeling errors, we obtain
\begin{eqnarray}
\left\langle \vecgest \pp\right\rangle  & = & \alpha\left\langle \pp \pp \right\rangle+\beta \left\langle \vqst \pp \right\rangle +\eta \left\langle \vwst \pp \right\rangle \label{eq: sys1},\\
\left\langle \vecgest \vqst \right\rangle  & = & \alpha\left\langle \pp \vqst \right\rangle+\beta \left\langle \vqst \vqst \right\rangle +\eta \left\langle \vwst \vqst \right\rangle\label{eq: sys2},\\
\left\langle \vecgest \vwst\right\rangle  & = & \alpha\left\langle \pp \vwst \right\rangle+\beta \left\langle \vqst \vwst \right\rangle +\eta \left\langle \vwst \vwst \right\rangle \label{eq: sys3}.
\end{eqnarray}
All quantities in the above equations are mean-subtracted. The resulting correlations can be re-written in terms of the $\rho$-statistics (\citealt{Rowe2010,Jarvis2016}; \Zuntz): $\rho_{0}=\left\langle \pp\pp \right\rangle $, $\rho_1= \left\langle \vqst \vqst \right\rangle$,  $\rho_2= \left\langle \vqst \pp \right\rangle$, $\rho_3= \left\langle \vwst \vwst \right\rangle$, $\rho_4= \left\langle \vqst \vwst \right\rangle$, and $\rho_5= \left\langle  \pp \vwst\right\rangle$. To make notation even more compact we define: $\tau_{0}=\left\langle \vecgest \pp\right\rangle$, $\tau_{2}=\left\langle \vecgest \vqst \right\rangle$, and $\tau_5= \left\langle \vecgest \vwst \right\rangle$:
\begin{eqnarray}
\tau_{0}  & = & \alpha\rho_{0}+\beta\rho_{2}+\eta\rho_{5}\label{eq: sys1},\\
\tau_2 & = & \alpha\rho_{2}+\beta\rho_{1}+\eta\rho_{4}\label{eq: sys2},\\
\tau_{5} & = & \alpha\rho_{5}+\beta\rho_{4}+\eta\rho_{3}.\label{eq: sys3}
\label{eq:systaurho}
\end{eqnarray}

Fig.~\ref{fig:rho_stats} shows the $\rho$-statistics measured from the catalogue of reserved stars for DES Y3. For comparison purposes, we also show the DES Y1 $\rho$-statistics; due to better PSF modeling, the DES Y3 $\rho$-statistics have a substantially smaller amplitude compared to Y1.  In DES Y1, some of the $\rho$-statistics were affected by large-scales constant contributions, which were partially responsible for a non-negligible mean shear measured at the catalogue level. For DES Y3, no evident large-scale constant contribution is measured. {We note that a few $\rho$-statistics are characterised by a steep change in their amplitude around  $\sim$ 3 arcmin. This feature is due to variations in the accuracy of the PSF solution as predicted by the PSF interpolation scheme. It is reasonable to assume these variations to happen at scales $\sim$ CCD size/order of the polynomial used for the interpolation, i.e., 1/3 of the CCD size, which corresponds to  $\sim$ 3 arcmin for the shorter side of the DES CCDs}.

It is important to recall that the idea here is not to solve the system of equations in each scale, but instead to find the best scalar parameters $\alpha,$ $\beta$, and $\eta$, that match the $\tau$ and $\rho$ measurements within our model. To sample the posteriors of our parameters, we generated Markov chain Monte Carlo (MCMC) samples that map out the posterior space, leading to parameter constraints. To this end, we used the public software package \textsc{EMCEE} \citep{Foreman-Mackey2013}, which is an affine-invariant ensemble sampler for MCMC. The $\tau$ measurement covariance was estimated using multiple \textsc{FLASK} realisations \footnote{In particular, we generated with \textsc{FLASK} different realisations of the DES Y3 shape catalogue. We then cross-correlated these simulated catalogues with the catalogue of reserved stars in data, and used the measured $\tau$ to infer the covariance matrix.} \citep{Xavier2016}, but we also checked that estimating it from jackknife resampling did not change the results. 
We considered the angular range between 0.1 and 250 arcmin for all measurements. This range also includes scales smaller than the ones used in the main cosmological analysis. We nonetheless checked that including only small scales (<10 arcmin) or only large scales (>10 arcmin) provided consistent best-fit values for the $\alpha,$ $\beta$, and $\eta$ parameters. Finally, we note that before measuring $\rho$ and $\tau$ from the catalogs, we assigned weights to stars to balance the ratio of the number densities of stars and galaxies across the footprint at large scales. In this test, we are using the PSF model and residual values of stars under the assumption that stars spatially sample PSF effects the same way galaxies do: therefore, large-scale differences in the number densities might alter the interpretation of the results. Nonetheless, we found these weights to have little impact on the best fitting values of $\alpha$, $\beta$, and $\eta$.

The best fitting values for $\alpha$, $\beta$, and $\eta$ are reported in Table~\ref{table:ab_fp}. The best fitting model to the measured $\tau$ is shown in Fig.~\ref{fig:rhoandtaustats}. The $\chi^2$ of the best fitting values is {$\chi^2/n = 95/120$}. We note that the coherent offset of the measured $\tau_{0+}$ with respect to the best fitting model is due to data points at scales larger than 1 arcmin being highly correlated. We further checked that the best fitting values were robust against dropping the $\tau_-$ components from the system of equations described by Eqs.~\ref{eq: sys1}, \ref{eq: sys2}, \ref{eq: sys3}, or against dropping the PSF size residuals from the modelling (i.e., $\eta=0.$). {Last, we checked that computing $\alpha$, $\beta$, $\eta$ values at every angular scale and then fitting for a constant value across all the scales produced values compatible with the ones reported in Table~\ref{table:ab_fp}.}

We defer the assessment of the impact of PSF modelling uncertainties on our cosmological constraints constraints to \cite{Amon2020}. We note though that we expect a smaller impact compared to DES Y1. The $\alpha$ and $\beta$ values have the same order of magnitude of those measured in the DES Y1 analysis \citep{Troxel2018}, but the $\rho$ statistics now have a substantially smaller amplitude, as may be expected from relatively minor updates to shape measurement, but more substantial improvements \citep{Jarvis19} to PSF modeling since then.

\begin{table}
\centering
\caption{Values of the parameters $\alpha$, $\beta$, and $\eta$ as estimated from $\rho$ statistics in sky coordinates (left column) and from focal-plane averaged PSF ellipticity, PSF ellipticity residuals and size residual (right column).}
\begin{adjustbox}{width=0.45\textwidth}
\begin{tabular}{|c|c|c|c|}
 \hline

{parameter }&{sky} & {parameter }&{focal plane} \\
 \hline
 $\alpha$ &  {$0.001\pm0.005$} & $\alpha_{1}$ &  {$ -0.028\pm0.013$}\\
- &-                       & $\alpha_{2}$ &  {$ -0.025\pm0.013$}\\
 $\beta$ & {$1.09\pm0.07$}     & $\beta_{1}$ &   {$0.93\pm0.2      $}\\
 -& -                      & $\beta_{2}$ &   {$1.0\pm0.2      $}\\
 $\eta$ & {$-0.5\pm0.5$}         & $\eta_{1}$ &    {$-5\pm6          $}\\
- &-                       & $\eta_{2}$ &    {$-0.5\pm6          $}\\
 \hline
\label{table:ab_fp}
\end{tabular}
\end{adjustbox}
\end{table}

\subsubsection{$\rho$ statistics from focal plane-averaged quantities }\label{section:psf_modeling_error2}

The values of $\alpha$, $\beta$, and $\eta$ estimated in the previous section can also be estimated from the correlation with reserved stars in focal plane coordinates. To proceed with this test, we first computed the mean $\pp$, $\vqst$, and $\vwst$ in a grid in focal plane coordinates using the reserved stars catalogue; then, we assigned the quantities $\pp$, $\vqst$, and $\vwst$ to each galaxy based on the position in focal plane coordinates (i.e, the pixel of the grid they fall into). Values from differing exposures were averaged. Last, we estimated $\alpha$, $\beta$,  and $\eta$ coefficients performing a linear fit of the mean shear with respect to the focal plane-averaged $\pp$, $\vqst$, and $\vwst$:
\begin{equation}
\label{eq:dedp}
\frac{\partial \vecgest}{\partial \pp} = \alpha + \beta \frac{\partial \vqst}{\partial \pp} + \eta \frac{\partial \vwst}{\partial \pp},
\end{equation}
\begin{equation}
\label{eq:dedq}
\frac{\partial \vecgest}{\partial \vqst} = \alpha \frac{\partial \pp}{\partial \vqst} + \beta + \eta \frac{\partial \vwst}{\partial \vqst},
\end{equation}
\begin{equation}
\frac{\partial \vecgest}{\partial \vwst} = \alpha \frac{\partial \pp}{\partial \vwst} + \beta \frac{\partial \vqst}{\partial \vwst} + \eta.
\end{equation}
All the derivatives on the l.h.s of the above equations were also estimated from the data using a linear fit. This method provided two different estimates of the parameters $\alpha$,$\beta$, $\eta$, one for each component of the shear, {although our model for the PSF errors assumes there should be no difference between the two components}. The values are shown in Table \ref{table:ab_fp}, showing a good agreement between the two components. We also note that the values of the $\alpha$, $\beta$  and $\eta$ parameters estimated in such a way are generally compatible with the parameters estimated in the previous section, {although uncertainties are generally larger (especially for $\eta$). In general, the focal plane analysis is based on averaged quantities and it is sub-optimal with respect to measuring $\alpha$, $\beta$  and $\eta$ from 2-point correlation functions in sky-coordinates. This is due to the fact that different scales contribute differently to the constraints on $\alpha$, $\beta$  and $\eta$, but the focal plane analysis does not take this into account.
 }


\subsubsection{Mean shear-PSF correlation}\label{section:mean_e_psf}

The DES Y1 shape catalogue (\Zuntz) showed a linear dependence of the two components of mean shear $\langle e_i \rangle$ with the input PSF ellipticity at the galaxy position, caused mostly by PSF model ellipticity residuals. As shown in Fig. \ref{fig:e1e2psf}, for the DES Y3 shape catalogue this dependence vanished: the measured slopes for the two components are {$\partial \gamma_1^{\rm est}/\partial {\rm PSF_1}= -0.001 \pm 0.002$, $\partial \gamma_2^{\rm est}/\partial {\rm PSF_2}= -0.003 \pm 0.002$}. The measured slope (black solid line) is compared to that inferred from the $\rho$ statistics obtained from focal-plane-averaged quantities (red dashed lines).


The lower panels of Fig. \ref{fig:e1e2psf} show the correlation between the mean ellipticity and the PSF size. As for DES Y1 shape catalogue, no noticeable trend is observed.

\begin{figure*}
\label{fig:e1e2psf}
\begin{center}
\includegraphics[width=0.8
\textwidth]{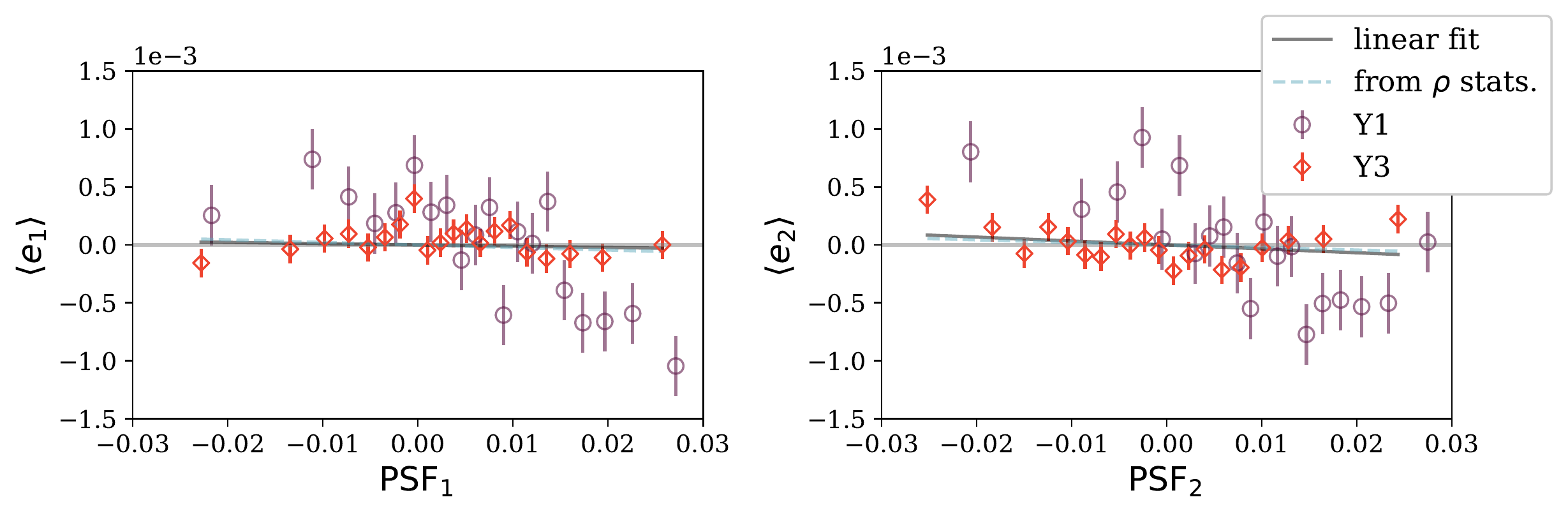}
\includegraphics[width=0.8
\textwidth]{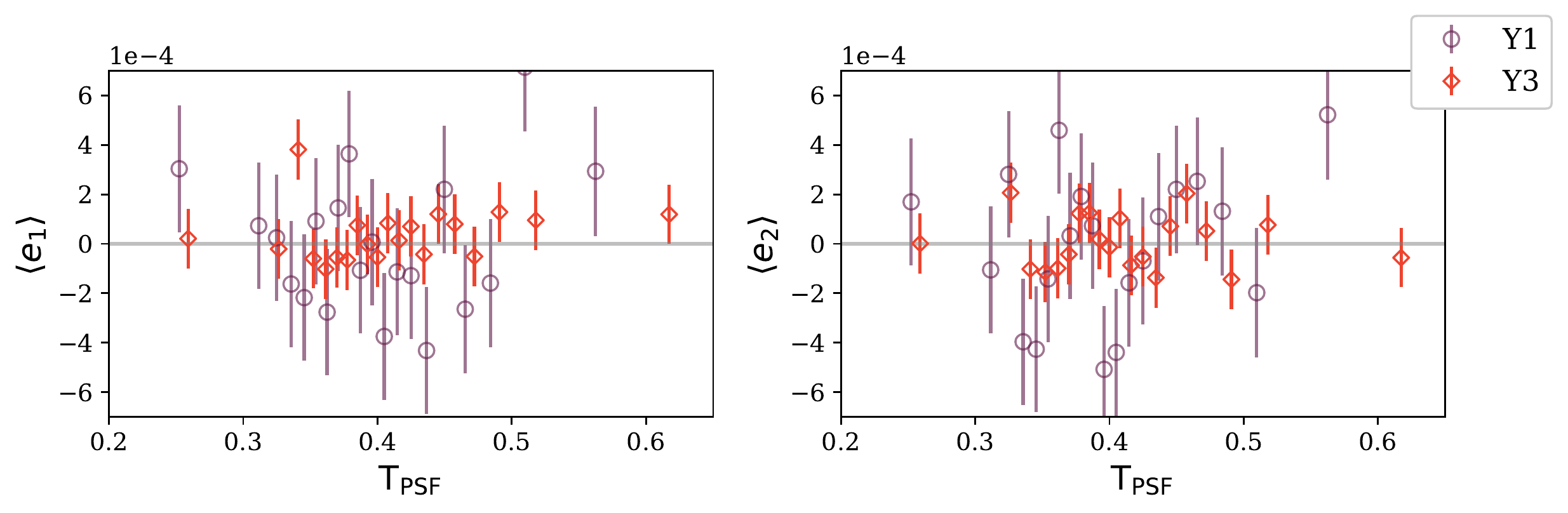}
\end{center}
\caption{\textit{Upper panels}: the mean shear $\langle e_i \rangle$ as a function of input PSF model ellipticity, for the two components. Solid lines are the linear best fit estimated using the PSF interpolated in real-space coordinates, dashed red lines are obtained using the $\alpha$, $\beta$ and $\eta$ parameters estimated using focal-plane-averaged quantities \textit{Lower panels}: the mean shear $\langle e_i \rangle$ as a function of input PSF size.}
\label{fig:e1e2psf}
\end{figure*}

\subsection{Tangential shear around stars}\label{sec:gammat_stars}


We discuss in this section the measurement of tangential shear around stars. We perform two different measurements: first, we measure the tangential shear around bright stars not used for the PSF modelling. {This measurement can be revealing of problems related to the light contamination from the outer light halos around bright stars. Second, we measure the tangential shear around faint stars. In principle, PSF modelling errors could generate a non null signal. E.g. we do find a non null signal when measuring the tangential shear of PSF residuals around PSF stars, indicating that an additive bias of the form $\sim \beta\vqst$ could contribute to the tangential shear of our shape catalog galaxies around PSF stars. We note, however, that we also expect this measurement to have a different sensitivity to PSF modelling errors compared to the $\rho$ statistics test presented in \S~\ref{sec:PSF_modeling_error}, due to the azimuthal averaging of the signal. If PSF modelling errors were negligible - or if the sensitivity of this measurement to PSF modelling errors were not good enough - we should expect a null detection.}

\begin{figure*}
\begin{center}
\includegraphics[width=0.8
\textwidth]{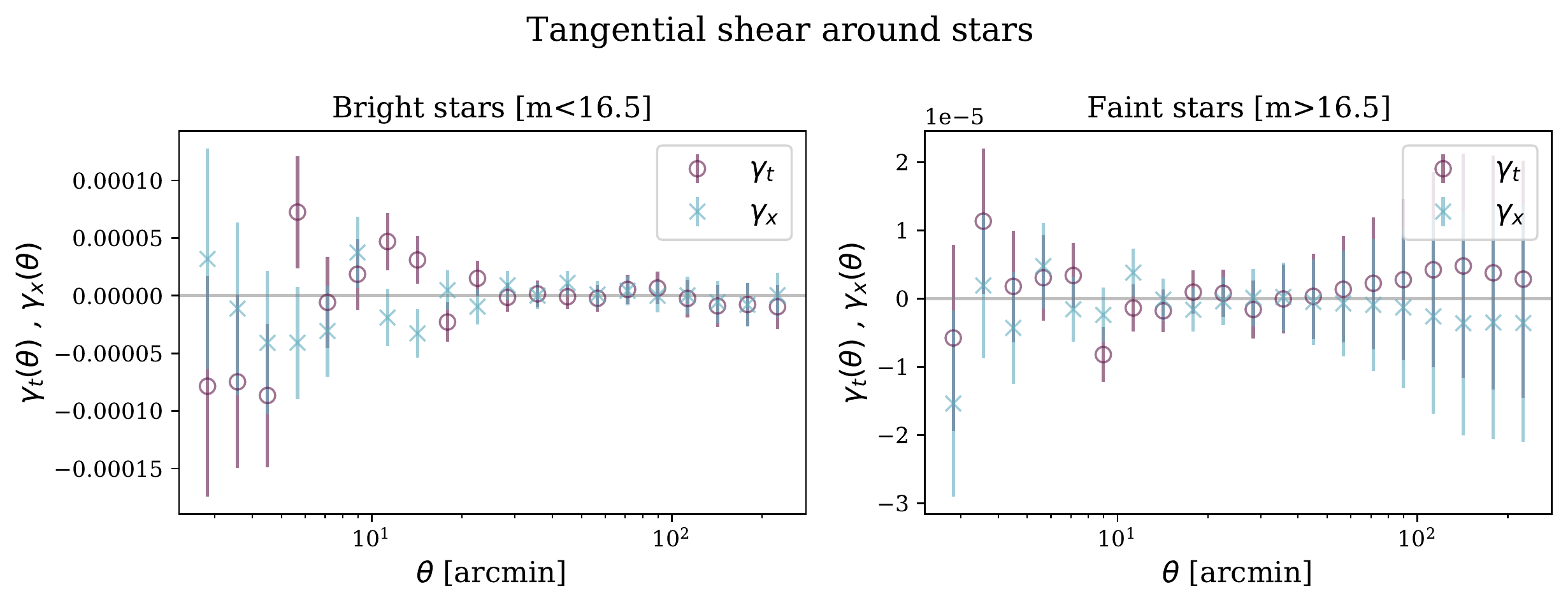}
\end{center}
\caption{Tangential shear around stars, which have been divided into a bright (m<16.5, left panel) and faint (m<16.5, right panel) sample.  The error bars are estimated form the jackknife method.}
\label{fig:gammat_stars}
\end{figure*}

{We show the measured signal in Fig.~\ref{fig:gammat_stars}. For this test, we divided the stars catalogue in two parts: bright stars with magnitude $m<16.5$, faint stars with magnitude $m>16.5$. We note that the faint stars sample is basically representative of the sample used for PSF modelling. Furthermore, we assigned weights to stars such that their distribution was uniform across the footprint. This step was needed as we noted that our star-finding algorithm tends to select slightly less stars in crowded regions, as stars could be contaminated or blended. This resulted in a distribution of stars slightly anti-correlated with the matter distribution. If not corrected, this would have generated a negative tangential signal, making harder to interpret the outcome of this test.}

{After applying the weights, we found the measured signals to be compatible with a null signal ({$\chi^2/n = 18/20$} and {$\chi^2/n = 11/20$} for bright and faint stars, respectively). As for the measurement involving faint stars, we checked that this test had actually not enough statistical power to detect a signal related to PSF modelling errors; based on the best-fit values of the $\alpha$, $\beta$ and $\eta$ parameters from Table~\ref{table:ab_fp}, the expected tangential shear signal due to PSF modelling errors is roughly one order of magnitude smaller than the statistical uncertainty of our measurement.}

\section{Shape catalogue tests}\label{sect:sherartests}
 Empirical tests lack an absolute calibration, therefore are more suited to test additive biases rather than multiplicative biases. They usually take the form of a `null' test. Deviations from a null signal might indicate the presence of additive biases. The tests included in this section are the following:
\begin{itemize}
    \item \textbf{shear variations in focal plane coordinates} (\S~\ref{sec:shearvarfocal});
    \item \textbf{tangential shear around field centres} (\S~\ref{sec:field_centers});
    \item \textbf{stellar contamination} (\S~\ref{sec:starcontam});
    \item \textbf{B-modes} (\S~\ref{sec:bmodes});
    \item \textbf{galaxy and survey properties} (\S~\ref{sec:additiveother}).
\end{itemize}
Note that unlike the other tests, the stellar contamination test is used to estimate a potential multiplicative shear bias. 

\subsection{Mean shear in focal plane coordinates}\label{sec:shearvarfocal}
\begin{figure*}
\begin{center}
\includegraphics[width=1 \textwidth]{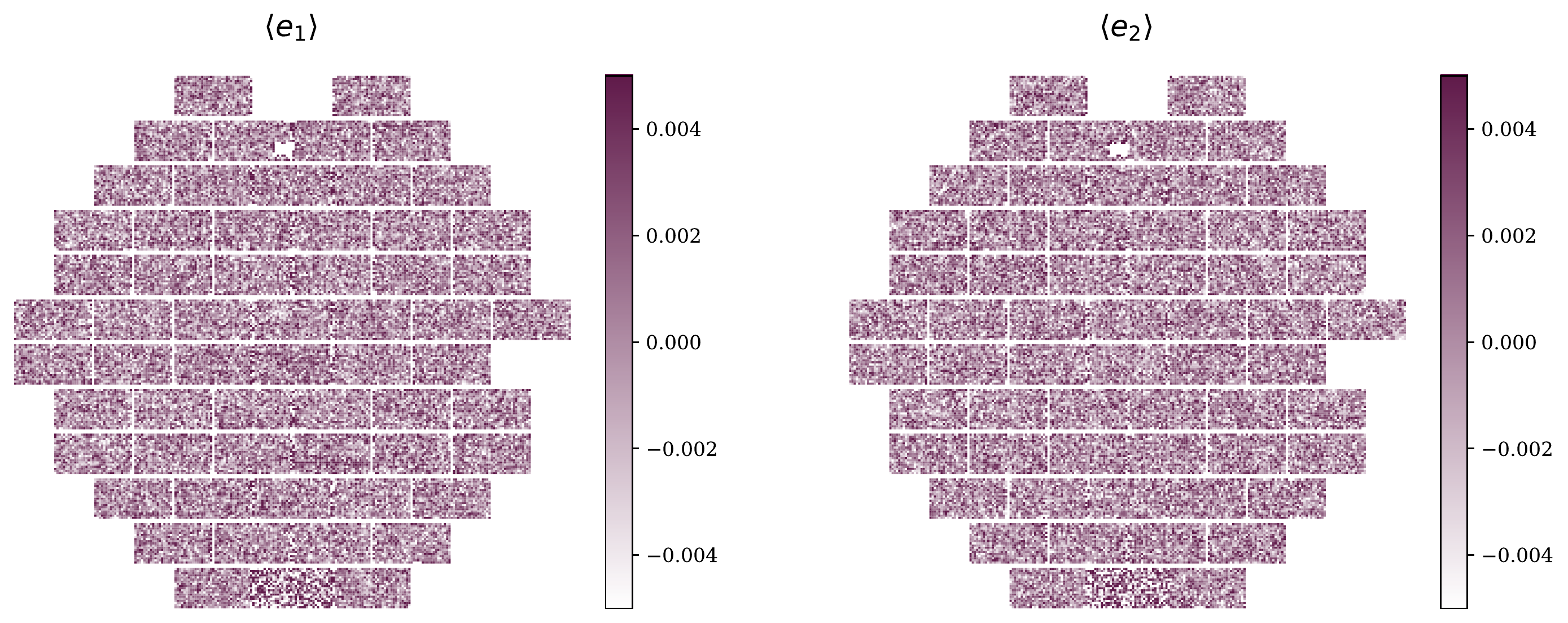}
\end{center}
\caption{Mean shear in focal plane coordinates, for the two component of the shear. For an explanation concerning the little hole in the upper part of the focal plane or other details concerning the DECam CCDs see \citealt{Flaugher2015}.}
\label{fig:focal_plane_meane}
\end{figure*}

\begin{figure*}
\begin{center}
\includegraphics[width=0.9 \textwidth]{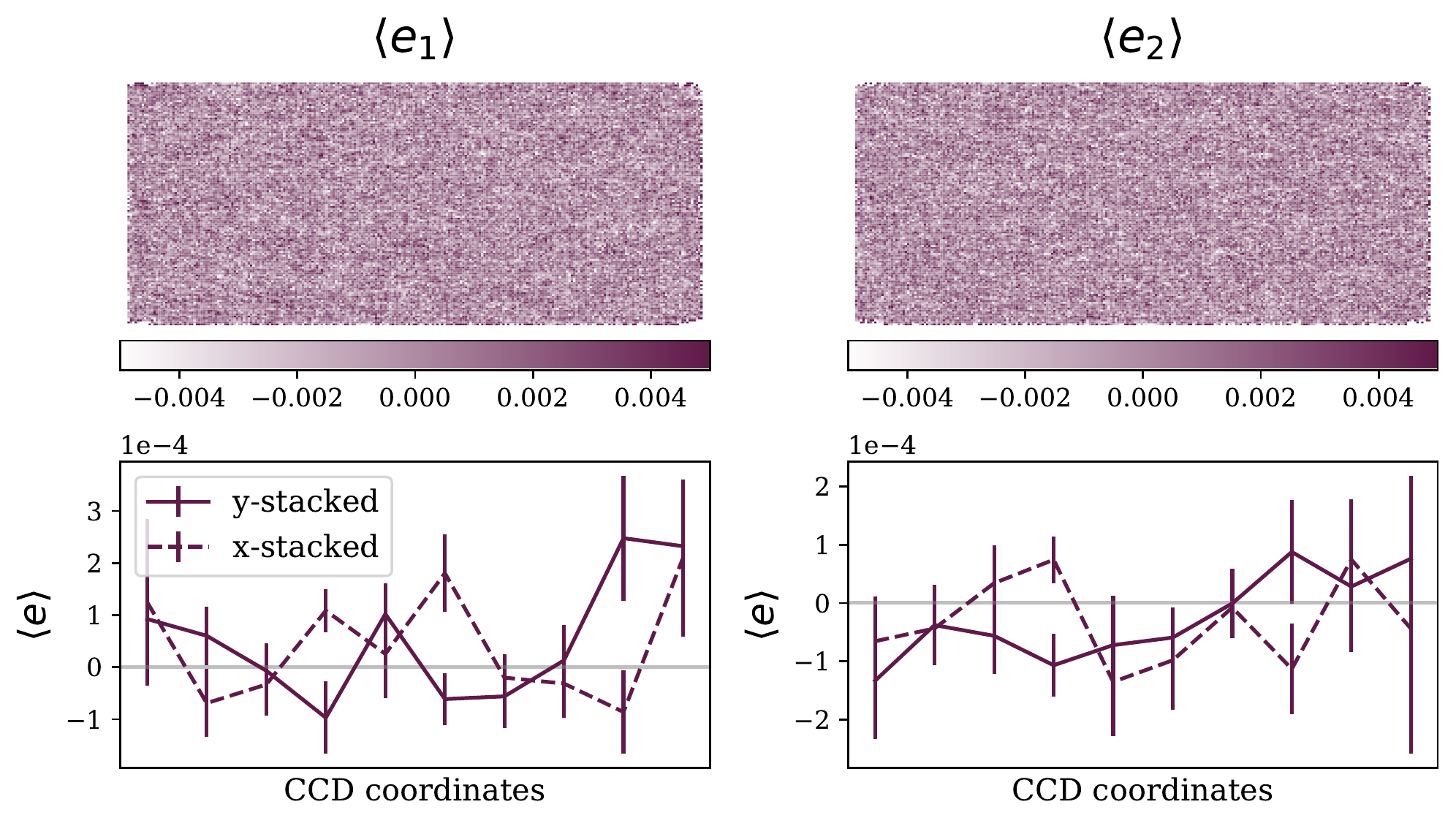}
\end{center}
\caption{Mean shear in CCD coordinates, obtained stacking all the CCDs signal. \textit{Upper panels}: the signal is stacked in a grid of 125 $\times$ 250 bins. \textit{Lower panels}: the signal is further stacked in 10 bins along the x or y directions.}
\label{fig:focal_plane_meane_ccd}
\end{figure*}

Fig.~\ref{fig:focal_plane_meane} shows the two components of the shear binned in focal plane coordinates.  The mean shear obtained stacking together all the CCDs is shown in Fig.~\ref{fig:focal_plane_meane_ccd}. It is possible for patterns to arise due to masking of bad columns in some of the CCDs, CCD pixels defects etc. In both Figs.~\ref{fig:focal_plane_meane} and \ref{fig:focal_plane_meane_ccd}, we observe no clear trends beyond variations due to shape noise and number count variations. This visual test is not stringent: given the bin size used to plot the two components of the shear, noise variations are much larger in amplitude than the mean shear measured at the catalogue level (\S~\ref{sect:number_density}). However, reducing the resolution of the plot did not show any significant pattern.

\subsection{Tangential shear around field centre}\label{sec:field_centers}
\begin{figure}
\begin{center}
\includegraphics[width=0.45 \textwidth]{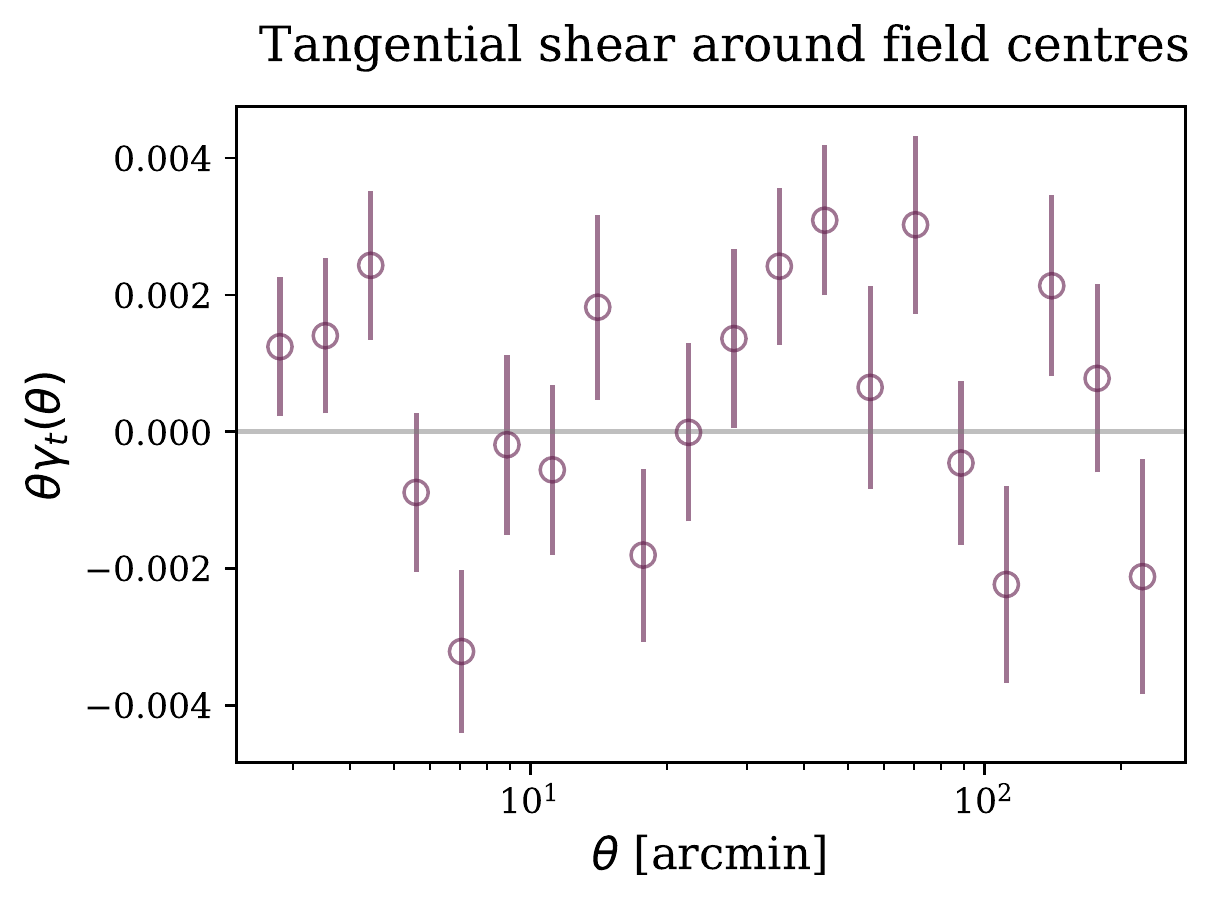}
\end{center}
\caption{Tangential shear around field centres, as a function of angular distance. Field centres in the $riz$ bands have been considered.}
\label{fig:field_centers}
\end{figure}
{We show in Fig.~\ref{fig:field_centers} the tangential shear binned by radius around field centres (the set of points where the centre of the focal plane is pointing over all exposures). A measurement of the tangential shear around a set of random points has been subtracted to the measurement. The measurement has been performed in sky coordinates; the field centres considered were 22331. A spurious signal might indicate residual systematics related to the position of the galaxies in the focal plane (due to, e.g., errors in the calibration of the focal plane distortions). The measured {$\chi^2/n = 46.1/20$} is too high ($p$-value of 0.0008) to neglect this signal. We first verified that the measurement could not have been explained by PSF modelling errors. We then proceeded assessing the impact of such spurious signal on the cosmic shear analysis. In particular, we interpolated the $\gamma_t$ measurement and converted it into a ${\mbox{\boldmath $\gamma(r)$}}$ signal for each exposure, where $r$ is the distance from the focal plane centre. We then assigned to each galaxy a new shape depending on its position in focal plane coordinates. Values from differing exposures were averaged. We last proceeded measuring the cosmic shear signal for the whole catalogue using these new shapes in sky coordinates. The resulting shear two-point measurement was 4 orders of magnitude smaller than the expected weakest cosmic shear signal (the lowest redshift tomographic bin), and therefore deemed negligible.}

\subsection{Stellar Contamination Test} \label{sec:starcontam}

The shape catalogue should contain only distant galaxies from which a cosmic shear signal may be measured. However, stars within our own galaxy may be detected in the images and erroneously pass the galaxy selection. Separating stars and galaxies at faint magnitudes is known to be a difficult problem.  It must be ensured that any stars which are mis-classified as galaxies and are included in the shape catalogue will not significantly dilute the measured shear.

Where stars are point-like and the PSF is accurately known, it is expected that their measured mean {ellipticity} and response should be zero $\langle  \boldsymbol{e} \rangle =\langle  \boldsymbol{R} \rangle = 0$. This will not be the case, however, if stars are included in the sample preferentially when their size is overestimated due to noise. Also, a mean non-zero response can result from even a small bias in the estimated PSF (see Fig.~11 of \Zuntz).

\subsubsection{Shear bias from stellar contamination}
\begin{figure*}
\begin{center}
\includegraphics[width=1.0\textwidth]{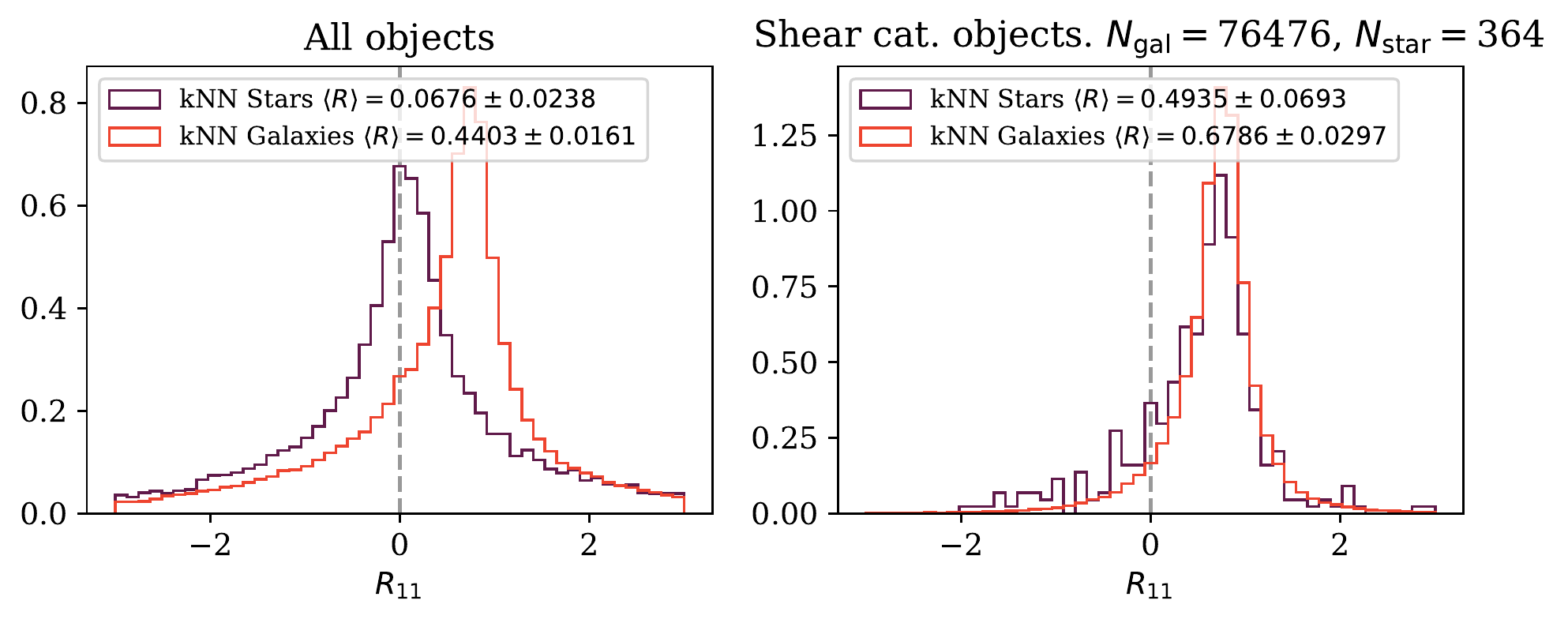}
\end{center}
\caption{\Mcal\ shear responses $R_{11}$ for star and galaxy objects in the C3, X3, and E2 fields as classified by the kNN classifier. The non-zero response of the stars which make it into the \mcal\ shape catalogue (\emph{right} panel) is the origin of the shear bias calculated in Eq.~\ref{eqn:stellar:m_bias}. The vertical dashed line represents zero on the $x$-axis.}
\label{fig:stellar:r11}
\end{figure*}

We assume the ellipticity distribution of the \mcal\ catalogue $P(\boldsymbol{e})$ is a weighted sum of galaxies and stars with ellipticity distributions $P_{G}(\boldsymbol{e})$ and $P_{\ast}(\boldsymbol{e})$ making up fractions $f_{G}$ and $f_{\ast}$ of the catalogue respectively:
\begin{equation}
    P = f_{G}P_{G} + f_{\ast}P_{\ast}.
    \label{eqn:stellar:ellip_distn}
\end{equation}
Assuming stars are not sheared, then the mean ellipticity of the \mcal\ catalogue is given by:
\begin{equation}
    \langle{\boldsymbol{e}}\rangle = f_{G}\langle\boldsymbol{R}_{G}\boldsymbol{\gamma}\rangle.
\end{equation}
Measuring the mean response for the full catalogue $\langle\boldsymbol{R}\rangle$ we can estimate the biased mean shear of the catalogue:
\begin{equation}
    \langle\boldsymbol{\gamma}^{\rm est, biased}\rangle = f_{G}\langle\boldsymbol{R}\rangle^{-1}\langle\boldsymbol{R}_{G}\boldsymbol{\gamma}\rangle.
    \label{eqn:stellar:biased_mean_shear}
\end{equation}
The mean response of the \mcal\ catalogue is given by:
\begin{equation}
    \langle\boldsymbol{R}\rangle = f_{G}\langle\boldsymbol{R}_{G}\rangle + f_{\ast}\langle\boldsymbol{R}_{\ast}\rangle.
\end{equation}
We can approximate its reciprocal using a Taylor series given that there are many more galaxies than stars:
\begin{equation}
    \langle\boldsymbol{R}\rangle^{-1} \approx \frac{1}{f_{G}}\left(1 - \frac{f_{\ast}}{f_{G}}\langle\boldsymbol{R}_{\ast}\rangle{\langle\boldsymbol{R}_{G}\rangle}^{-1}\right)\langle\boldsymbol{R}_{G}\rangle^{-1}.
    \label{eqn:stellar:total_response_taylor}
\end{equation}
Substituting Eq.~\ref{eqn:stellar:total_response_taylor} into Eq.~\ref{eqn:stellar:biased_mean_shear} and using $\langle\boldsymbol{\gamma}\rangle = \langle\boldsymbol{R}_{G}\rangle^{-1}\langle\boldsymbol{R}_{G}\boldsymbol{\gamma}\rangle$ for the mean shear gives:
\begin{equation}
    \langle\boldsymbol{\gamma}^{\rm est, biased}\rangle = \left(1 - \frac{f_{\ast}}{f_{G}}\langle\boldsymbol{R}_{\ast}\rangle{\langle\boldsymbol{R}_{G}\rangle}^{-1}\right)\langle\boldsymbol{\gamma}\rangle,
\end{equation}
and as such we identify the multiplicative bias $\boldsymbol{m}$ as a result of stellar contamination to be:
\begin{equation}
    \boldsymbol{m} = - \frac{f_{\ast}}{f_{G}}\langle\boldsymbol{R}_{\ast}\rangle{\langle\boldsymbol{R}_{G}\rangle}^{-1}.\
    \label{eqn:stellar:m_bias}
\end{equation}
In order to assess the level of contamination of the shape catalogue by stars $\frac{f_{\ast}}{f_{G}}$ and the stellar response $\langle\boldsymbol{R}_{\ast}\rangle$ we took advantage of the DES Deep Fields \citep{deepfields} to construct a star-galaxy separation algorithm which is expected to work at the faint magnitudes relevant for objects in the shape catalogue (see Appendix~\ref{sec:stargal_appendix} for more details). Within the COSMOS field we matched DES Deep Fields objects to the HST-ACS catalogue of \cite{2007ApJS..172..219L}, which covers the full range of magnitudes for the DES Deep Fields data and also includes the \textsc{mu\_class} morphological star-galaxy classification. Using the \textsc{mu\_class} as truth labels we trained a k-nearest neighbors (kNN) classifier in the available $ugrizJHK$ color space. {No DES-Y3 wide field depth reduction of the COSMOS field exists, meaning Y3 \mcal\ shear responses are not available for sources with the \textsc{mu\_class} classifications}. We then applied this classifier in the other DES Deep Fields, which have both Y3 \mcal\ measurements and optical and near-infrared colors available, but not HST-ACS \textsc{mu\_class}. The \mcal\ responses for objects in the DES C3, X3, and E2 Deep Fields classified in this way are shown in Fig.~\ref{fig:stellar:r11}. The left panel shows responses for all DES deep objects in the field. {Stars show a shear response consistent with zero, but within $2.5 \sigma$. This slight discrepancy is probably a consequence of PSF errors, since a non-zero response for stars is possible if the PSF model is biased}. {We have verified that the seeing distributions within the three Deep Fields used are representative of those in the full wide field}. The right panel shows shear responses for objects which pass the fiducial shape catalogue cuts described in \S~\ref{sec:mcalselect} and shows the expected behaviour (the non-zero response caused by cosmic shear we are trying to measure) for galaxies. Objects which are classified as stars but also pass the fiducial \mcal\ cuts and make it into the shape catalogue are $0.5\%$ of the shape catalogue, and have a response which peaks away from zero. Errors are jackknife resampling errors containing $66\%$ of the distribution.

\begin{figure}
\begin{center}
\includegraphics[width=0.5\textwidth]{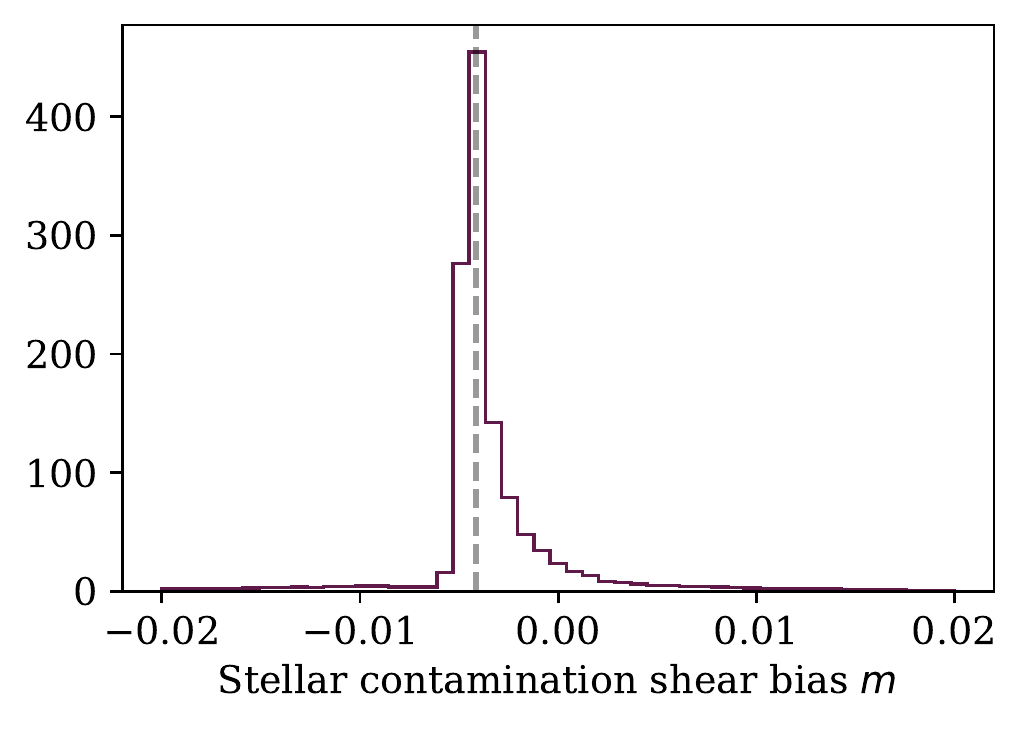}
\end{center}
\caption{Shear bias from contamination by objects classified as stars by the kNN classifier in the C3, X3, and E2 Deep Fields regions, calculated using Eq.~\ref{eqn:stellar:m_bias}. Distributions over $m$ are created by Monte Carlo sampling values of $\langle R_{\ast} \rangle$, $\langle R_{G} \rangle$, and $f_{\ast}/f_{G}$ from the revelant distributions in Fig.~\ref{fig:stellar:r11}; and the vertical dashed lines represent the median of each distribution.
}
\label{fig:stellar:bias}
\end{figure}

If we assume all of these objects are indeed contaminating stars and use their mean shear response $\langle\boldsymbol{R}_{\ast}\rangle$ then we may use Eq.~\ref{eqn:stellar:m_bias} to find the resultant shear bias. Fig.~\ref{fig:stellar:bias} shows this measured bias for the objects across three Deep Fields regions, C3, X3, and E2. {The median value for the inferred stellar contamination bias is {$m = -0.004_{-0.002}^{+0.001}$}.}  We note that we also performed this test on the DES Y3 image simulations \citep{MacCrannSims2019}, which include a catalogue of stars from the \textsc{TRILEGAL} \citep{trilegal2019} model. In versions of the simulations where objects (both stars and galaxies) were placed randomly, the stellar contamination fraction of the shape catalogue, histogram of responses of stellar objects in the shape catalogue, and histogram of resulting stellar contamination shear bias were all reproduced to a high degree of accuracy. In simulations where the sources (both stars and galaxies) were instead placed on a grid, the stellar contamination of the shape catalogue reduced by $2/3$, and the non-zero response of stellar objects in the shape catalogue disappeared, with the distribution peaking at zero. This indicates that the stellar contamination is an effect stemming from blending of star and galaxy sources, and one which is already included in the shear bias modelled by the image simulations. 

\subsection{E/B-modes decomposition and null tests with systematics} \label{sec:bmodes}

In this section, we show the measured B-mode signals obtained using both the pseudo-$C_\ell$ \citep{Hikage2011} and Complete Orthogonal Sets of E/B-Integrals \citep[COSEBIs; ][]{Schneider2010A&A...520A.116S} statistics. As the cosmic shear field to first-order predicts no B-modes, any detection in the shape catalogue could indicate a contamination by systematic effects, in particular by the PSF generating an additive bias. Note that a small B-mode power spectrum can be sourced by higher-order physical effects, including intrinsic alignments, clustering of sources and higher order contribution to the shear signal. If detectable at significant levels, these contributions should be included when modelling the two-point correlation functions \citep{Secco2020,y3methods,MassMapY3}.

\subsubsection{Pseudo-$C_\ell$ B-modes}
\begin{figure}
    \centering
    \includegraphics[width=0.44\textwidth]{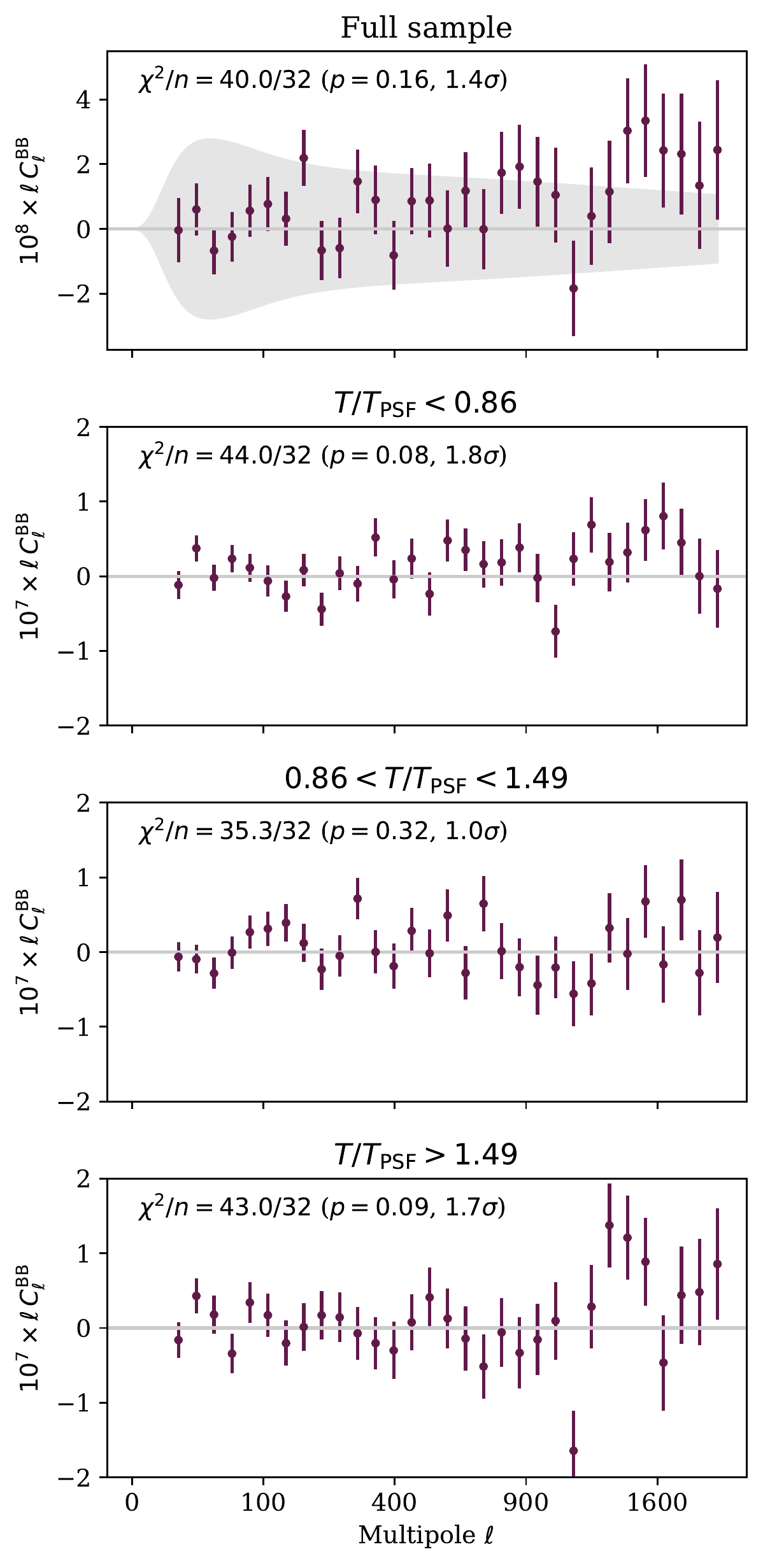}
    \caption{Pseudo-$C_\ell$ B-mode power spectrum measured from the DES Y3 shape catalogue in the multipole range 8-2048. The signal from the full catalogue is shown in the top panel; the other panels show the measured spectrum from the catalogue divided into three bins of equal weighted effective number density based on the galaxies' size ratio. The noise power spectrum bias has been estimated from Gaussian simulations (see text) and subtracted. {As an order of magnitude comparison, in the top panel we show as a grey shaded region 10 per cent of the expected E-mode power spectrum of the weakest cosmic shear signal (the lowest redshift tomographic bin)}.}
    \label{fig:bmode}
\end{figure}
We built two Healpix maps \citep{GORSKI2005} (with resolution ${n_{\rm side}=1024}$) of the cosmic shear signal by computing the weighted average of (response-corrected) ellipticities of galaxies within each pixel. We estimated the E- and B-mode power spectra of these maps with pseudo-$C_\ell$ using \texttt{NaMaster} \citep{2019MNRAS.484.4127A}, an open-source code that deconvolves the effects of masked regions from the harmonic space coefficients. We used the inverse-variance weight masks, given by the weighted count maps. We measured spectra for multipoles in the range $\ell=[8-2048]$ in 32 bins evenly separated on a square-root scale (spreading signal-to-noise more evenly than linear or logarithmic binning).

The measured power spectra receive an additive bias from the shape-noise power spectrum $N_\ell$, which may diverge from the approximation $N_\ell=\sigma_e^2/{n_{\rm eff}}$ due to mask effects and properties of the pseudo-$C_\ell$ estimator. Moreover, the mask induces a leakage between E- and B-modes (especially at large scales) which increases the variance of affected multipole bins. Therefore, we generated 2000 mock catalogs to both obtain an accurate measurement of the noise power spectrum and the covariance matrix of the pseudo-$C_\ell$ B-modes\footnote{We also applied the standard technique consisting in simply applying random rotations and obtained noise power spectra in agreement to better than $10^{-3}$. The covariance matrix obtained with this method, however, failed at capturing the contribution due to E-mode leakage, which is particularly relevant at large scales.}.

The procedure to generate the mock catalogs is as follows. Given a fiducial E-mode power spectrum, we generated 2000 full-sky healpix maps of the cosmic shear field. The cosmic shear field is assumed to be Gaussian.  For each galaxy in the catalog, we applied a random rotation to its measured ellipticity. Each rotated ellipticity is used as intrinsic ellipticity. We then sampled the shear field at the positions of galaxies and applied the shear addition formula (see, e.g, \citealt{Seitz1997}) to the mock intrinsic ellipticity. This method preserves both geometric properties of the catalogue and the ellipticity distribution over the DES Y3 footprint. We then applied the pseudo-$C_\ell$ estimator and obtained the noise power spectrum and an empirical covariance of the mock catalogue B-mode spectra. Finally, we excluded the first multipole bin which includes scales larger than the survey footprint and showed mild E-mode leakage reproduced by simulations. Note, however, that compared to the DES Y1 analysis, we extended the measurement to smaller scales, $\ell_{\rm max}=2048$, corresponding to angular scales of ${\sim 5.3^{\prime}}$. The measurement is shown in Fig.~\ref{fig:bmode}. Overall, we found a {$\chi^2/n=40./32$} for multipole bins in the range $\ell=[8-2048]$, corresponding to a $p$-value of {0.16}, or a {1.4$\sigma$} deviation from the null hypothesis of no B-mode, suggesting that B-modes in the data are consistent with pure shape-noise. 
Last, we note that in \cite{Amon2020} we further performed the B-mode measurements for our catalogue divided into the four fiducial tomographic bins used in the main cosmological analysis, and the overall measurement resulted in a null detection.

\subsubsection{COSEBIs}
\begin{figure}
    \centering
    \includegraphics[width=0.45\textwidth]{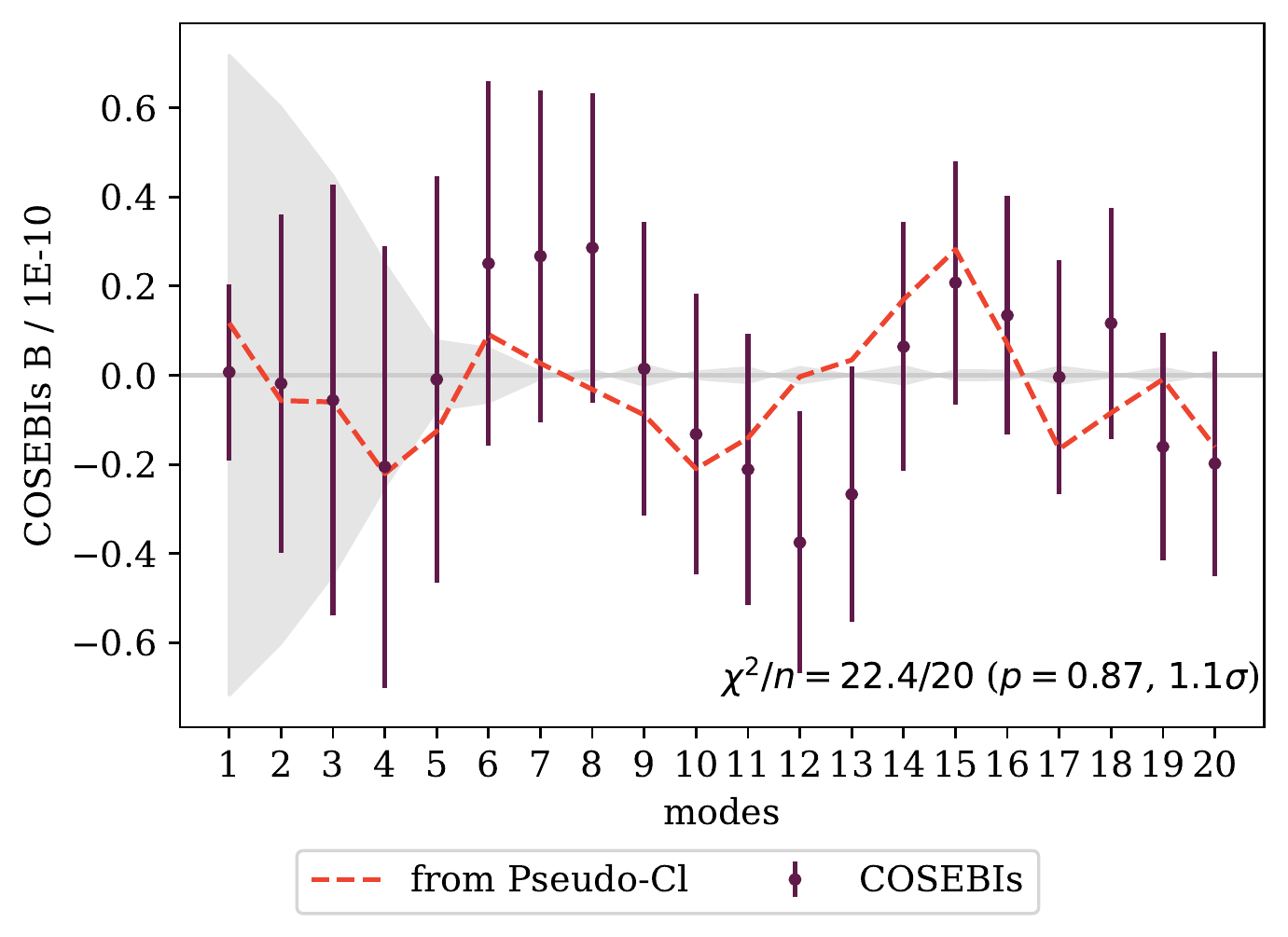}
    \caption{COSEBIs B-modes from the DES Y3 shape catalogue are shown as data points with uncertainties given by a noise-only analytical covariance matrix.  The dashed line shows the pseudo-$C_\ell$ shown in Figure~\ref{fig:bmode} converted to COSEBIs B-mode as described in the text.  Note that the COSEBIs are always discrete measurements for individual modes, but the dashed line is connected for clarity. {As an order of magnitude comparison, we show as a grey shaded region 10 per cent of the expected COSEBIs E-mode of the weakest cosmic shear signal (the lowest redshift tomographic bin).}}
    \label{fig:cosebis}
\end{figure}
We used Complete Orthogonal Sets of E/B-Integrals \citep[COSEBIs; ][]{Schneider2010A&A...520A.116S}, which is an estimator designed to separate E- and B-modes and has been measured for previous releases of DES as well as for KiDS \citep{AsgariKiDS4502019A&A...624A.134A,AsgariDESY12019MNRAS.484L..59A,AsgariKV4502020A&A...634A.127A,Giblin2020}.  To calculate the B-mode, we first computed the real space shear-shear correlations, $\xi_{+}$ and  $\xi_{-}$ with \treecorr~\citep{TreeCorr2015ascl.soft08007J} in 10000 logarithmically-spaced bins from $\theta_{\rm min}=2.5^{\prime}$ to $\theta_{\rm max}=250^{\prime}$ with a brute force calculation (\texttt{bin\_slop}=0).  These were converted to COSEBIs using filter functions that are described in \S~2 of \citet{AsgariKV4502020A&A...634A.127A} (see their Eq. 7).  We also calculated a noise-only covariance matrix following Appendix A of \citet{AsgariKV4502020A&A...634A.127A} (see their Eqs A.5 and A.6).  The resulting B-mode signal and corresponding square root of the diagonal of the covariance are plotted in Fig.~\ref{fig:cosebis}.

The {$\chi^{2}/n=22.4/20=1.1$} corresponds to a $p$-value of {0.87}, which indicates that the null hypothesis fits the data, and the DES Y3 shear catalogue is consistent with zero B-modes.  This result can be qualitatively compared to the DES Y1 values (also consistent with zero for \mcal\ shape measurements, with a $p$-value of 0.325) shown in the right panel of Fig. 1 from \citet{AsgariDESY12019MNRAS.484L..59A}, although note that different $\theta$-ranges were used. We also compared with the pseudo-$C_\ell$ shown in Fig.~\ref{fig:bmode} by converting those measurements to COSEBIs.  We did this by approximating the pseudo-$C_\ell$ as a piece-wise constant function and integrating it with the Hankel transform of the COSEBIs filter functions \citep[see Eq. 8 of][]{AsgariKV4502020A&A...634A.127A}. This converted measurement is displayed as the dashed line in Fig.~\ref{fig:cosebis}, and we see that the two different B-mode estimations agree well. {We note however that the two estimators generally probe scales differently, with 
COSEBIs being less sensitive to small scales \citep{AsgariKiDS4502019A&A...624A.134A,AsgariDESY12019MNRAS.484L..59A,AsgariKV4502020A&A...634A.127A,Asgari2020}. We verified this by zeroing the pseudo-$C_\ell$ measurement at $\ell>1500$ and then converting it to COSEBIs. The converted measurement little differed from the one obtained with no hard cut at $\ell \sim 1500$, demonstrating the insensitivity of COSEBIs to these small scales. }

\subsection{Galaxy and survey properties tests}\label{sec:additiveother}

Our shear catalogue  is characterized by a non-null mean shear in one of the two components, whose origin is unknown. The values for the two components are respectively {$\langle e_1 \rangle = 3.5$ $10^{-4}$} and {$\langle e_2 \rangle = 0.6$ $10^{-4}$}; for the first component, this value is larger than the one expected from cosmic variance ($\sim 0.5$ $10^{-4}$, as estimated from \texttt{FLASK} log-normal mocks). The mean shear is measured and subtracted at the catalogue  level as it would have an impact on the measured cosmic shear signal. We further investigated the possibility that the mean shear could vary across the footprint, and that these variations could be non-cosmological but could be related to other galaxy or observing properties. To this aim, we assumed $\langle e_1 \rangle$, $\langle e_2 \rangle$ to depend linearly on a number of different galaxy and observational properties: depth, S/N, size ratio $T/T_{\rm PSF}$ (i.e., the ratio between galaxy size and PSF size),  exposure time, brightness, and airmass. When applicable, these quantities were considered in the $i$ band. We did not explicitly include PSF ellipticity, ellipticity or size residuals as these had already been investigated in $\S$ \ref{section:psf_modeling_error2}.

\begin{figure}
\begin{center}
\includegraphics[width=0.45
\textwidth]{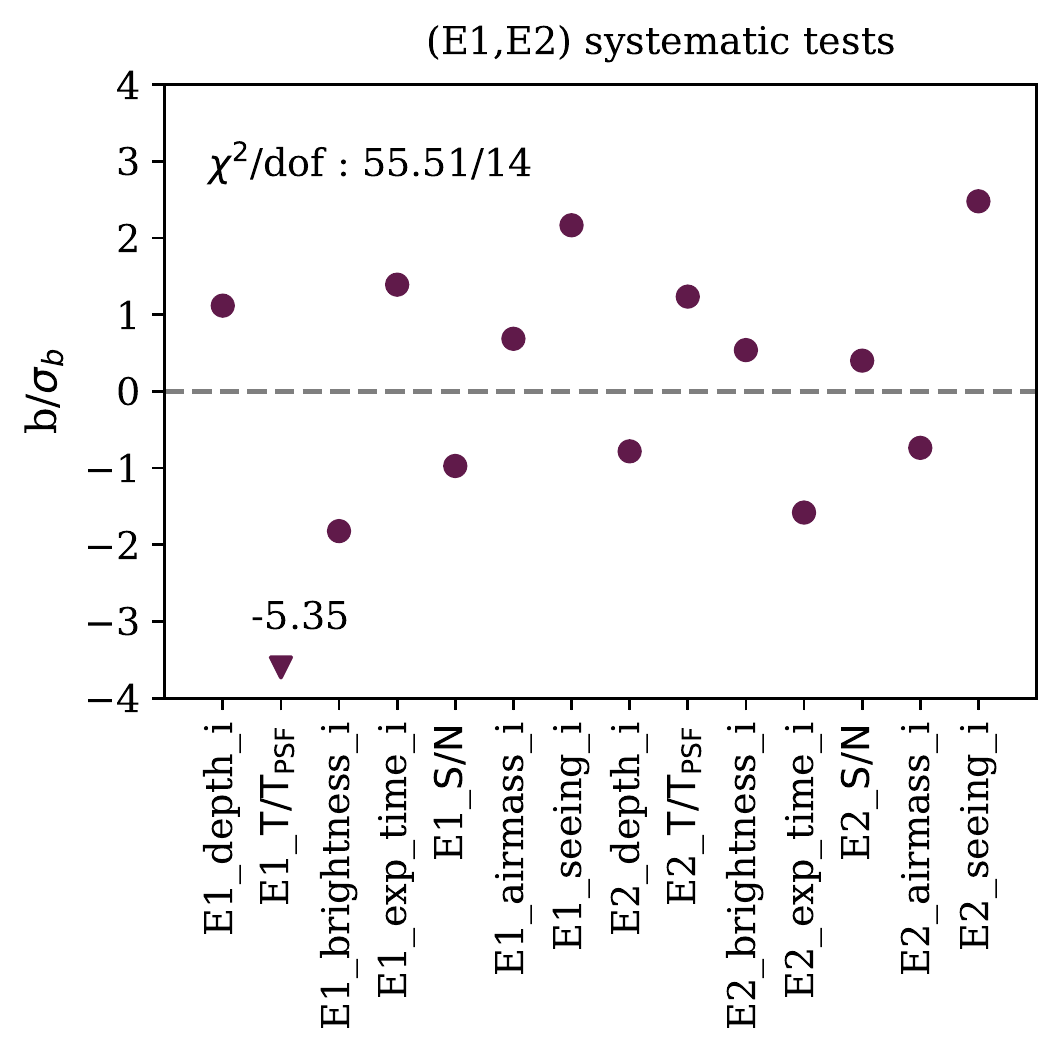}
\end{center}
    \caption{Best fit values for the coefficient of the relation $\langle e_i \rangle = b~ \mathrm{syst} + c$ with $\mathrm{syst}$ a given systematic map. The values of the slopes are shown for different tomographic bins, and the uncertainties are estimated through log-normal mocks. We also tested using uncertainties estimated by jackknife resampling, with no sensible difference. The reported $\chi^2$ takes into account correlations among different systematic maps.}
\label{fig:e1e2_linear}
\end{figure}

We then performed a linear fit (using \texttt{numpy polyfit}) for the two shear components as a function of the different properties across the footprint. When performing the fit, we included the weights of each galaxy and we accounted for varying selection effects by correcting the shear by a piecewise shear response. In principle, this is a null test, as we do not expect \textit{a priori} to detect any correlation. Any significant deviation from a null signal, however, could help shed light on the origin of the mean shear signal measured in the catalogue. We show the measured coefficient for each of these systematic maps in Fig.~\ref{fig:e1e2_linear}; uncertainties were estimated using 300 \texttt{FLASK} log-normal mocks. We also checked that using uncertainties estimated with jackknife resamples caused no significant difference in the results. We find a clear correlation between $\langle e_1 \rangle$ and the ratio between the galaxy size and the PSF size, while none of the other correlations are significant (removing the $\langle e_1 \rangle - $size ratio correlation from the analysis reduces the $\chi^2$ for the null hypothesis to 20 for 14 $d.o.f.$).

\begin{figure*}
\centering
\includegraphics[width=0.9\textwidth]{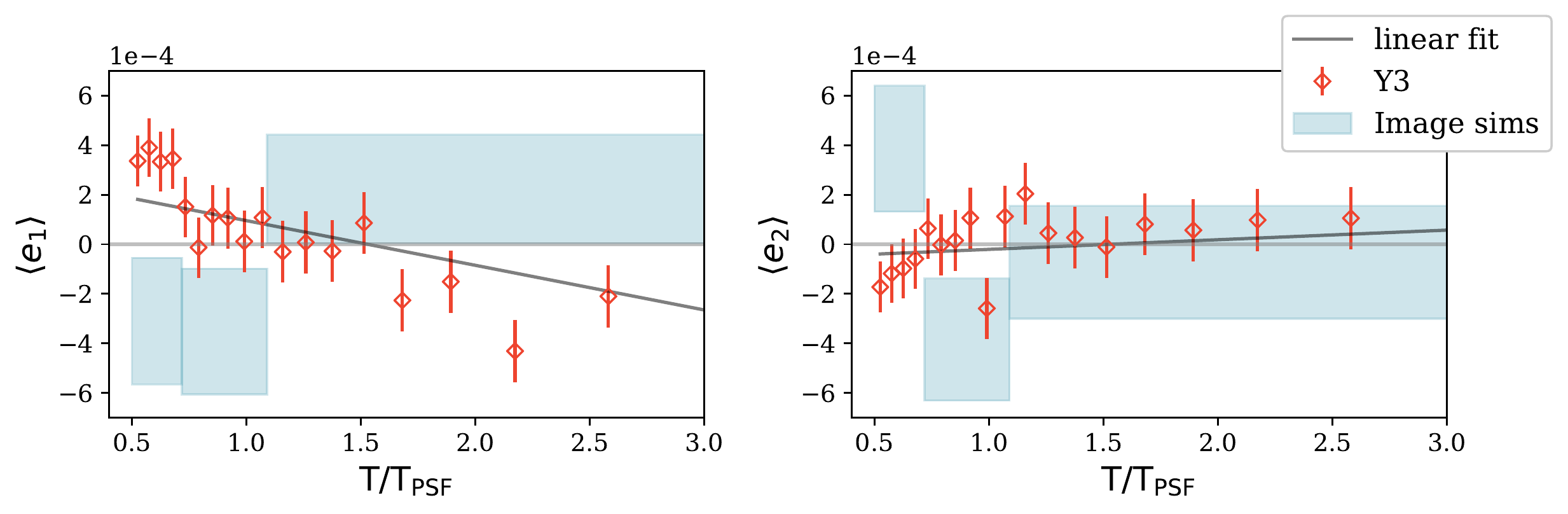}
\caption{The mean shear $\langle e_i \rangle$ as a function of size ratio $T/T_{\rm PSF}$, defined as the ratio between galaxy size and PSF size. A statistically significant trend with $T/T_{\rm PSF}$ is measured for the first component of the mean shear, while for the second component we measure no significant dependence. {The light blue boxes represent the mean shear as computed in the DES Y3 image simulations suite \citep{MacCrannSims2019}.}}
\label{fig:meanesizeratio}
\end{figure*}

We plot the mean shear as a function of size ratio in Fig.~\ref{fig:meanesizeratio}. As we showed in Fig.~\ref{fig:e1e2psf} that the mean shear has no dependence on the PSF size $T_{\rm PSF}$, this test mostly highlights a dependence of the mean shear with respect to the galaxy size: this implies that smaller galaxies are associated with a positive, spurious, mean shear signal. The origin of this signal is currently unknown. {In Fig.~\ref{fig:meanesizeratio} we also show the mean shear as a function of $T / T_{\rm PSF}$ as computed in the fiducial DES Y3 image simulations \citep{MacCrannSims2019}. The comparison is inconclusive, as the simulated tiles available do not allow us to measure with statistical significance a signal with an amplitude as the one measured in data.} Therefore, we cannot rule out whether the root cause of this trend is modelled and included in our image simulations or not. {We nonetheless checked that the scale-dependent part of this additive bias is sufficiently small to not bias the cosmological analysis. This has been achieved by the following procedure: first, we assigned to each galaxy of the catalogue an additive bias equal to $\delta e_i =b_i T / T_{\rm PSF}$, where $b_i$ are the measured per-component best fit to the linear dependence of the mean shear with respect to size ratio. We only care about the scale-dependent part of this bias (as the mean shear is always subtracted from our catalogue), so we made sure $\langle \delta e_i \rangle =0$.  Then we computed the shear two-point correlation function associated to these ``fake'' additive biases. The measured correlation function resulted to be three order of magnitude smaller than the weakest expected cosmic shear signal, at every scale; therefore, we considered the scale-dependent part of this additive bias negligible.}

{Last, we checked if this trend with size ratio could be responsible for any B-mode signal. Although we did not measure any statistically significant B-mode for the full catalogue, a spurious B-mode signal could be associated to a subset of it, e.g. galaxies with small size ratio. We divided the catalogue in three bins of equal weighted effective number density  as a function of size ratio and measured the B-mode power spectrum for each of them. The measurements are reported in Fig.~\ref{fig:bmode}; no statistically significant signal has been detected.}

\begin{figure*}
\centering
\includegraphics[width=0.9\textwidth]{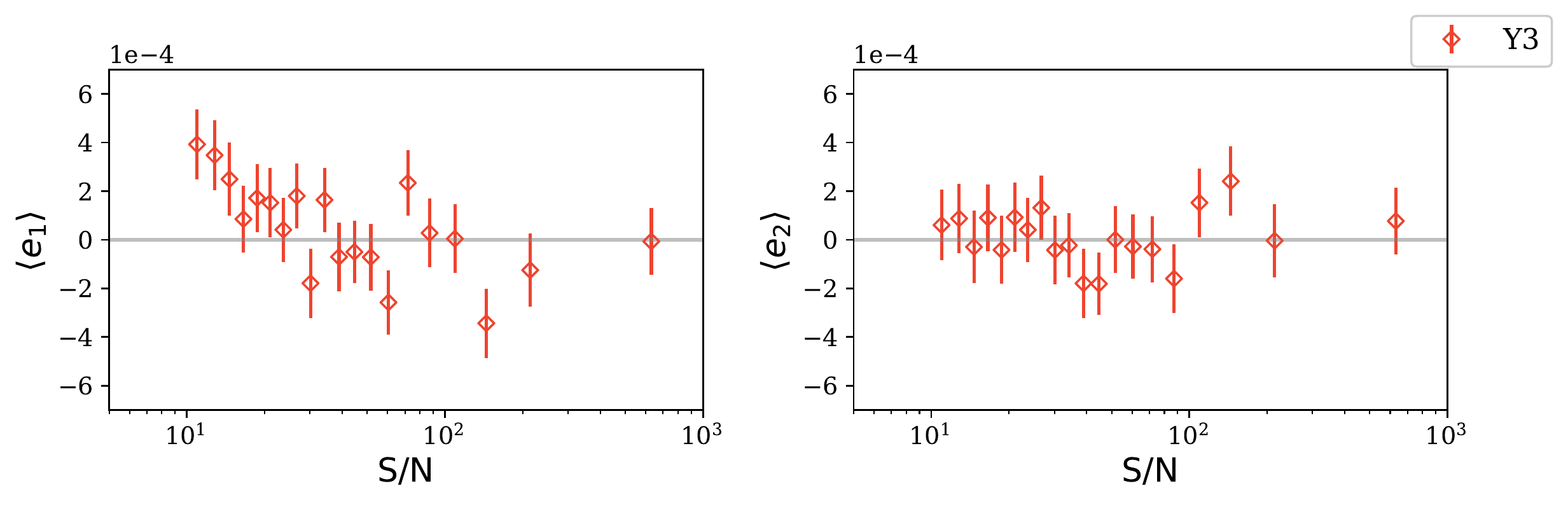}
\caption{The mean shear $\langle e_i \rangle$ as a function of S/N. No statistically significant trend is detected by the fit (Fig.~\ref{fig:e1e2_linear}), as the outcome of the fit is mostly determined by galaxies with \snr $>20$, which represent $\sim 75$ per cent of the total weight of the catalogue.}
\label{fig:meanesnr}
\end{figure*}

{We also show in Fig.~\ref{fig:meanesnr} the measured mean shear as a function of S/N, as \snr\ is a relevant quantity used to select the DES Y3 weak lensing sample. No statistically significant trend is detected by the fit, as shown in Fig.~\ref{fig:e1e2_linear}. Note that as galaxy weights are a strong function of \snr, the outcome of the fit is mostly determined by galaxies with \snr $>20$, as they represent $\sim 75$ per cent of the total weight of the catalogue.}

\bigskip

\section{Summary} \label{sec:summary}
This paper presented the weak lensing shape catalogue from the DES Y3 imaging data, covering $\sim 4143$ deg$^2$ of the southern hemisphere and comprising $\sim100$ million objects, resulting in a weighted source number density of {$n_{\rm eff} = 5.59$} gal/arcmin$
^{2}$ and corresponding shape noise {$\sigma_e = 0.261$}. We described the shape measurement pipeline used for the DES Y3 analysis, \mcal , which is based upon the pipeline used in the DES Y1 analysis (\Zuntz), but with the following improvements:
\begin{itemize}
    \item improved PSF solutions (\textsc{Piff}, \citealt{Jarvis19}) were used for the \mcal\ deconvolutions rather than the \psfex\ solutions that were used for Y1;
    \item improved astrometric solutions, based on \cite{Bernstein2017};
    \item inverse variance weighting for the galaxies.
\end{itemize}
We further discussed the sample selection adopted for the DES Y3 analysis and the changes compared to DES Y1, which improved the reliability of the weak lensing sample. The \mcal\ pipeline is capable of self-calibrating biases in the shear estimation by correcting for the response of the shear estimator and selection biases. The current \mcal\ implementation, however, does not correct for a shear-dependent detection bias, which is calibrated using a dedicated suite of image simulations in \cite{MacCrannSims2019}. It is expected that the DES Y6 release will implement an updated version of \mcal\ \citep{SheldonMetadetect2019}, which accounts for the aforementioned effect. {We note that we also expect to implement in the future a second shear measurement pipeline, following the BFD method outlined by \cite{BernBFD2016}, although more investigation is needed to see to what extent the BFD algorithm can cope with such shear-dependent detection bias.}

In this paper we performed a variety of empirical null tests, mostly aimed at identifying additive biases in our shape catalogue. We tested potential systematic errors connected to PSF corrections, demonstrating that the improved PSF solutions reduced additive biases due to PSF misestimation to negligible levels for the current analysis. PSF tests have been performed both in real space and in focal plane coordinates, showing agreement. We also tested that PSF chromatic effects (which are currently not modelled) were negligible. 

We further checked biases due to the erroneous inclusion of stars in the DES Y3 catalogue, estimating the stellar contamination bias, which we found to be in agreement with the results from image simulations. We looked at the signature of systematic effects by measuring the catalogue B-mode signals using both the COSEBIs and pseudo-$C_\ell$ estimators, which consistently revealed in a null detection. We checked the dependency of the two components of the shear with respect to a number of galaxy or survey properties, finding no significant correlations, except for a linear dependence between $\langle e_1 \rangle$ and the ratio between the galaxy size and PSF size. The origin of this trend is unknown, but we verified that it can be safely neglected in the main DES cosmological analysis. Finally, we tested the validity of using the mean response to also calibrate shear two-point correlation functions (see Appendix \ref{sect:2pt_response}), finding that a two-point response correction is not needed for the current DES Y3 analysis. 

We remind that this work is complemented by two other papers; the first one describes in more depth the DES Y3 PSF modeling \citep{Jarvis19}, whereas the second one describes the overall calibration of the catalogue using image simulations \citep{MacCrannSims2019}. In particular, the latter provides the multiplicative shear bias calibration for the catalogue, which needs to be applied before using the catalogue for any scientific purposes. We also note that the DES Y3 analysis relies on only one shape catalogue, contrary to the DES Y1 analysis where two different shape catalogues were produced with two different pipelines. While having two different catalogues in the DES Y1 analysis increased our confidence in the robustness of the catalogues' calibration, the DES Y3 shape catalogue is backed up by a much more powerful and accurate suite of image simulations \citep{MacCrannSims2019} compared to the DES Y1 analysis, which supports the overall calibration and robustness of the catalogue.

\section*{Data Availability}
The full \mcal\ catalogue will be made publicly available following publication, at the URL \url{https://des.ncsa.illinois.edu/releases}. The code used to perform the tests in this manuscript will be made available upon reasonable request to the authors.

\section*{Acknowledgements}
This paper has gone through internal review by the DES collaboration. ES is supported by DOE grant DE-AC02-98CH10886. AC acknowledges support from NASA grant 15-WFIRST15-0008.

Funding for the DES Projects has been provided by the U.S. Department of Energy, the U.S. National Science Foundation, the Ministry of Science and Education of Spain, 
the Science and Technology Facilities Council of the United Kingdom, the Higher Education Funding Council for England, the National Center for Supercomputing 
Applications at the University of Illinois at Urbana-Champaign, the Kavli Institute of Cosmological Physics at the University of Chicago, 
the Center for Cosmology and Astro-Particle Physics at the Ohio State University,
the Mitchell Institute for Fundamental Physics and Astronomy at Texas A\&M University, Financiadora de Estudos e Projetos, 
Funda{\c c}{\~a}o Carlos Chagas Filho de Amparo {\`a} Pesquisa do Estado do Rio de Janeiro, Conselho Nacional de Desenvolvimento Cient{\'i}fico e Tecnol{\'o}gico and 
the Minist{\'e}rio da Ci{\^e}ncia, Tecnologia e Inova{\c c}{\~a}o, the Deutsche Forschungsgemeinschaft and the Collaborating Institutions in the Dark Energy Survey. 

The Collaborating Institutions are Argonne National Laboratory, the University of California at Santa Cruz, the University of Cambridge, Centro de Investigaciones Energ{\'e}ticas, 
Medioambientales y Tecnol{\'o}gicas-Madrid, the University of Chicago, University College London, the DES-Brazil Consortium, the University of Edinburgh, 
the Eidgen{\"o}ssische Technische Hochschule (ETH) Z{\"u}rich, 
Fermi National Accelerator Laboratory, the University of Illinois at Urbana-Champaign, the Institut de Ci{\`e}ncies de l'Espai (IEEC/CSIC), 
the Institut de F{\'i}sica d'Altes Energies, Lawrence Berkeley National Laboratory, the Ludwig-Maximilians Universit{\"a}t M{\"u}nchen and the associated Excellence Cluster Universe, 
the University of Michigan, the National Optical Astronomy Observatory, the University of Nottingham, The Ohio State University, the University of Pennsylvania, the University of Portsmouth, 
SLAC National Accelerator Laboratory, Stanford University, the University of Sussex, Texas A\&M University, and the OzDES Membership Consortium.

Based in part on observations at Cerro Tololo Inter-American Observatory at NSF's NOIRLab (NOIRLab Prop. ID 2012B-0001; PI: J. Frieman), which is managed by the Association of Universities for Research in Astronomy (AURA) under a cooperative agreement with the National Science Foundation.

The DES data management system is supported by the National Science Foundation under Grant Numbers AST-1138766 and AST-1536171.
The DES participants from Spanish institutions are partially supported by MINECO under grants AYA2015-71825, ESP2015-66861, FPA2015-68048, SEV-2016-0588, SEV-2016-0597, and MDM-2015-0509, 
some of which include ERDF funds from the European Union. IFAE is partially funded by the CERCA program of the Generalitat de Catalunya.
Research leading to these results has received funding from the European Research
Council under the European Union's Seventh Framework Program (FP7/2007-2013) including ERC grant agreements 240672, 291329, and 306478.
We  acknowledge support from the Brazilian Instituto Nacional de Ci\^encia
e Tecnologia (INCT) e-Universe (CNPq grant 465376/2014-2).

This manuscript has been authored by Fermi Research Alliance, LLC under Contract No. DE-AC02-07CH11359 with the U.S. Department of Energy, Office of Science, Office of High Energy Physics.







\appendix
\section{The shear two-point correlation \MCAL\ Response }\label{sect:2pt_response}

We derive in this Appendix the response for the shear two-point correlation function, following \cite{SheldonMcal2017}. We can write the two-point correlation function as follows

\begin{equation}
\label{eq1}
\xi = \int d\mbox{\vesta}d\mbox{\vestb} S_{\alpha} S_{\beta} P(\mbox{\vesta},\mbox{\vestb}) \mbox{\vesta} \mbox{\vestb} 
\end{equation}
where $S_{\alpha}$ and $S_{\beta}$ are selection functions, and $P(\mbox{\vesta},\mbox{\vestb})$ the joint probability distribution of $\mbox{\vesta}$ and $\mbox{\vestb}$. In the DES Y1 analysis, we assumed that the shapes of galaxies were not correlated in the absence of lensing, i.e. $\left.P(\mbox{\vesta},\mbox{\vestb})\right|_{\gamma=0} = \left. P(\mbox{\vesta})\right|_{\gamma=0} \left.P(\mbox{\vestb})\right|_{\gamma=0}$. Under this hypothesis, the response of the shear two point function is equal to the mean response squared: $\langle  R^{\rm 2pt} \rangle \approx \langle  {R} \rangle^2$ (also assuming the response matrix is diagonal and that it does not vary across the footprint).
In what follows we drop the assumption of zero correlation in absence of lensing. The response at leading order can be written as:

\begin{equation}
\label{eq2}
\langle  R^{\rm 2pt} \rangle = \int d\mbox{\vesta} d\mbox{\vestb} \frac{\partial^2 \left(S_{\alpha}S_{\beta} P( \mbox{\vesta}, \mbox{\vestb}) \mbox{\vesta} \mbox{\vestb}\right)}{\partial \gamma_{\alpha} \partial \gamma_{\beta}}
\end{equation}

\begin{multline}
\label{eq3}
\langle  R^{\rm 2pt} \rangle = \int d\mbox{\vesta} d\mbox{\vestb} \frac{\partial }{\partial \gamma_{\alpha}} S_{\alpha} \times \\ \left[P(\mbox{\vesta},\mbox{\vestb}) \mbox{\vesta}\mbox{\vestb} \frac{\partial S_{\beta}}{\partial \gamma_{\beta}} + S_{\beta}\mbox{\vesta} \frac{\partial (P(\mbox{\vesta},\mbox{\vestb}) \mbox{\vestb})}{\partial \gamma_{\beta}}\right]
\end{multline}

\begin{multline}
\label{eq4}
\langle  R^{\rm 2pt} \rangle = \int d\mbox{\vesta} d\mbox{\vestb} \left[ P(\mbox{\vesta},\mbox{\vestb}) \mbox{\vesta}\mbox{\vestb} \frac{\partial S_{\alpha}}{\partial \gamma_{\alpha}}\frac{\partial S_{\beta}}{\partial \gamma_{\beta}}\right] +   \\  \int d\mbox{\vesta} d\mbox{\vestb}\left[S_{\alpha}\mbox{\vestb} \frac{\partial(  P(\mbox{\vesta},\mbox{\vestb}) \mbox{\vesta})}{\partial \gamma_{\alpha}}\frac{\partial S_{\beta}}{\partial \gamma_{\beta}}\right] +\\       \int d\mbox{\vesta} d\mbox{\vestb}\left[S_{\beta}\mbox{\vesta} \frac{\partial(  P(\mbox{\vesta},\mbox{\vestb}) \mbox{\vestb})}{\partial \gamma_{\beta}}\frac{\partial S_{\alpha}}{\partial \gamma_{\alpha}}\right] +\\ \int d\mbox{\vesta} d\mbox{\vestb} \left\{ S_{\alpha}S_{\beta} \frac{\partial}{\delta \gamma_{\alpha}} \left[ \mbox{\vesta} \frac{\partial(  P(\mbox{\vesta},\mbox{\vestb}) \mbox{\vestb})}{\partial \gamma_{\beta}}\right]\right\}  
\end{multline}

For the first term of Eq.~\ref{eq4}:

\begin{multline}
\label{eq5}
 \int d\mbox{\vesta} d\mbox{\vestb} \left[ P(\mbox{\vesta},\mbox{\vestb}) \mbox{\vesta}\mbox{\vestb}\frac{\left(S^{+}_{\alpha}-S^{-}_{\alpha}\right)}{\Delta \gamma}\frac{\left(S^{+}_{\beta}-S^{-}_{\beta}\right)}{\Delta \gamma}\right] = \\ \frac{1}{(\Delta \gamma)^2} \left[\xi ^ {++}(\alpha,\beta) - \xi ^ {-+}(\alpha,\beta) - \xi ^ {+-}(\alpha,\beta) +\xi ^ {--}(\alpha,\beta)\right]
\end{multline}
where derivatives have been approximated using finite differences. The notation $\xi ^ {-+}(\alpha,\beta)$ indicates that the shear two-point correlation function has been computed applying the negatively sheared selection on the sample $\alpha$ and the positively sheared selection on the sample $\beta$.

The second term of Eq.~\ref{eq4} reads:

\begin{multline}
\label{eq6}
 \int d\mbox{\vesta} d\mbox{\vestb} \left[S_{\alpha} \mbox{\vestb}\frac{\left(P(\mbox{\vesta}^+,\mbox{\vestb}) \mbox{\vesta}^+ - P(\mbox{\vesta}^-,\mbox{\vestb}) \mbox{\vesta}^- \right)}{\Delta \gamma}\frac{\left(S^{+}_{\beta}-S^{-}_{\beta}\right)}{\Delta \gamma}\right] = \\ \frac{1}{(\Delta \gamma)^2} \left[\xi ^ {0+}(\alpha^+,\beta) - \xi ^ {0-}(\alpha^+,\beta) - \xi ^ {0+}(\alpha^-,\beta) +\xi ^ {0-}(\alpha^-,\beta)\right]
\end{multline}
where the notation $\xi ^ {0-}(\alpha^+,\beta)$ indicates that the shear two-point correlation function has been computed applying the normal selection to the positively sheared sample $\alpha$, and applying the negatively sheared selection to the sample $\beta$.

The third term of Eq.~\ref{eq4} reads:

\begin{multline}
\label{eq7}
 \int d\mbox{\vesta} d\mbox{\vestb} \left[S_{\beta} \mbox{\vesta}\frac{\left(P(\mbox{\vesta},\mbox{\vestb}^+) \mbox{\vestb}^+ - P(\mbox{\vesta},\mbox{\vestb}^-) \mbox{\vestb}^- \right)}{\Delta \gamma}\frac{\left(S^{+}_{\alpha}-S^{-}_{\alpha}\right)}{\Delta \gamma}\right] = \\ \frac{1}{(\Delta \gamma)^2} \left[\xi ^ {+0}(\alpha,\beta^+) - \xi ^ {-0}(\alpha,\beta^+) - \xi ^ {+0}(\alpha,\beta^-) +\xi ^ {-0}(\alpha,\beta^-)\right]
\end{multline}

Lastly, the fourth term of Eq.~\ref{eq4} is: 

\begin{multline}
\label{eq8}
 \int d\mbox{\vesta} d\mbox{\vestb} \left\{S_{\beta} S_{\alpha} \frac{\partial}{\partial \gamma_{\alpha}} \left[ \mbox{\vesta}\frac{\left(P(\mbox{\vesta},\mbox{\vestb}^+) \mbox{\vestb}^+ - P(\mbox{\vesta},\mbox{\vestb}^-)\mbox{\vestb}^- \right)}{\Delta \gamma} \right]\right\} = \\ \int d\mbox{\vesta} d\mbox{\vestb} S_{\beta} S_{\alpha} \left[ \frac{\left(P(\mbox{\vesta}^+,\mbox{\vestb}^+) \mbox{\vesta}^+\mbox{\vestb}^+ - P(\mbox{\vesta}^-,\mbox{\vestb}^+) \mbox{\vesta}^-\mbox{\vestb}^+ \right)}{(\Delta \gamma)^2} \right]+\\-\int d\mbox{\vesta} d\mbox{\vestb} S_{\beta} S_{\alpha} \left[ \frac{\left(
 P(\mbox{\vesta}^+,\mbox{\vestb}^-) \mbox{\vesta}^+\mbox{\vestb}^- + 
 P(\mbox{\vesta}^-,\mbox{\vestb}^-) \mbox{\vesta}^-\mbox{\vestb}^- 
 \right)}{(\Delta \gamma)^2} \right] = \\ \frac{1}{(\Delta \gamma)^2} \left[\xi ^ {00}(\alpha^+,\beta^+) - \xi ^ {00}(\alpha^-,\beta^+) - \xi ^ {00}(\alpha^+,\beta^-) +\xi ^{00}(\alpha^-,\beta^-)\right]
\end{multline}

Putting together Eq.~\ref{eq5}, \ref{eq6}, \ref{eq7} and \ref{eq8}:

\begin{multline}
\label{eq9}
\langle  R^{\rm 2pt} \rangle = \frac{1}{(\Delta \gamma)^2} \left[ \xi ^ {++}(\alpha,\beta) - \xi ^ {-+}(\alpha,\beta) - \xi ^ {+-}(\alpha,\beta) +\xi ^ {--}(\alpha,\beta)\right] + \\ \frac{1}{(\Delta \gamma)^2} \left[ \xi ^ {0+}(\alpha^+,\beta) - \xi ^ {0-}(\alpha^+,\beta) - \xi ^ {0+}(\alpha^-,\beta) +\xi ^ {0-}(\alpha^-,\beta)\right]+\\ \frac{1}{(\Delta \gamma)^2} \left[ \xi ^ {+0}(\alpha,\beta^+) - \xi ^ {-0}(\alpha,\beta^+) - \xi ^ {+0}(\alpha,\beta^-) +\xi ^ {-0}(\alpha,\beta^-)\right] + \\ \frac{1}{(\Delta \gamma)^2} \left[ \xi ^ {00}(\alpha^+,\beta^+) - \xi ^ {00}(\alpha^-,\beta^+) - \xi ^ {00}(\alpha^+,\beta^-) +\xi ^ {00}(\alpha^-,\beta^-) \right].
\end{multline}

Ideally, the response would need to be computed shearing $e_t$ or $e_{\times}$. These are the tangential and cross components of the shear along the line connecting two galaxies. We cannot do this because the shear would depend on the pair of galaxies considered. We can just shear $e_1$ and $e_2$ (which are the tangential and cross components along two arbitrary fixed axes). Let us define:

\begin{equation}
\label{eq10}
\langle R^{\rm 2pt}_{tt,tt}\rangle = \int de_{\alpha,t} de_{\beta,t} \frac{\partial^2 \left(S_{\alpha}S_{\beta} P(e_{\alpha,t}, e_{\beta,t}) e_{\alpha,t} e_{\beta,t}\right)}{\partial \gamma_{\alpha,t} \partial \gamma_{\beta,t}}
\end{equation}

Analogously we can define $\langle R^{\rm 2pt}_{\times \times,\times \times}\rangle$. Under the hypothesis of isotropy,  $\langle R^{\rm 2pt}_{\times \times,\times \times}\rangle = \langle R^{\rm 2pt}_{tt,tt}\rangle  \equiv \langle R^{\rm 2pt} \rangle$, which would be the response needed to correct $\xi_{+}$ and $\xi_{-}$. However, these two responses are not directly accessible. Using Eq.~\ref{eq9} and shearing $e_1$ and $e_2$, the estimator $\langle \hat{R}^{\pm}\rangle $ we can measure from the data is:

\begin{multline}
\label{eq11}
 \langle \hat{R}^{\pm}\rangle =\langle R^{\rm 2pt}_{tt,11}\rangle + 2\langle R^{\rm 2pt}_{tt,12}\rangle + \langle R^{\rm 2pt}_{tt,22}\rangle \pm\\ \left( \langle R^{\rm 2pt}_{\times \times,11}\rangle + 2\langle R^{\rm 2pt}_{\times \times,12}\rangle + \langle R^{\rm 2pt}_{\times \times,22}\rangle\right)
\end{multline}
where now the derivatives are with respect to $e_1$ and $e_2$. The $\pm $ depends on whether we chose $\xi_{+}$ or $\xi_{-}$ as statistics to infer the response. Changing variables, Eq.~\ref{eq11} becomes:
\begin{multline}
\label{eq12}
 \langle \hat{R}^{\pm}\rangle =\langle R^{\rm 2pt}\rangle \left[\langle \left(\frac{\partial \gamma_t}{\partial e_1}\right)^2\rangle + 2\langle \frac{\partial \gamma_t}{\partial e_1} \frac{\partial \gamma_t}{\partial e_2}\rangle + \langle \left(\frac{\partial \gamma_t}{\partial e_2}\right)^2\rangle  \right] \pm\\ \langle R^{\rm 2pt}\rangle  \left[ \langle \left(\frac{\partial \gamma_{\times}}{\partial e_1}\right)^2\rangle + 2\langle \frac{\partial \gamma_{\times}}{\partial e_1} \frac{\partial \gamma_{\times}}{\partial e_2}\rangle + \langle \left(\frac{\partial \gamma_{\times}}{\partial e_2}\right)^2\rangle  \right]
\end{multline}
$\gamma_t$ and $\gamma_{\times}$ are related to $e_1$ and $e_2$ by a rotation matrix; we can assume for instance 
\begin{equation}
\label{eq13}
\begin{split}
\frac{\partial \gamma_t}{\partial e_1} = -{\rm cos}(2\phi) \quad
\frac{\partial \gamma_t}{\partial e_2} = -{\rm sin}(2\phi) \\
\frac{\partial \gamma_{\times}}{\partial e_1} = {\rm sin}(2\phi) \quad
\frac{\partial \gamma_{\times}}{\partial e_2} = -{\rm cos}(2\phi)
\end{split}
\end{equation}
Eq.~\ref{eq12} leads to 

\begin{multline}
\label{eq14}
 \langle \hat{R}^{\pm}\rangle =\langle R^{\rm 2pt}\rangle \left[\langle  {\rm cos}^2(2\phi)\rangle +  2\langle  {\rm cos}(2\phi) {\rm sin} (2\phi)\rangle +\langle  {\rm sin}^2(2\phi)\rangle  \right] \pm\\ \langle R^{\rm 2pt}\rangle \left[\langle  {\rm sin}^2(2\phi)\rangle -  2\langle  {\rm cos}(2\phi) {\rm sin} (2\phi) \rangle +\langle  {\rm cos}^2(2\phi)\rangle  \right] = \\\langle R^{\rm 2pt}\rangle  \pm \langle R^{\rm 2pt}\rangle 
\end{multline}

\begin{figure*}
\begin{center}
\includegraphics[width=0.9 \textwidth]{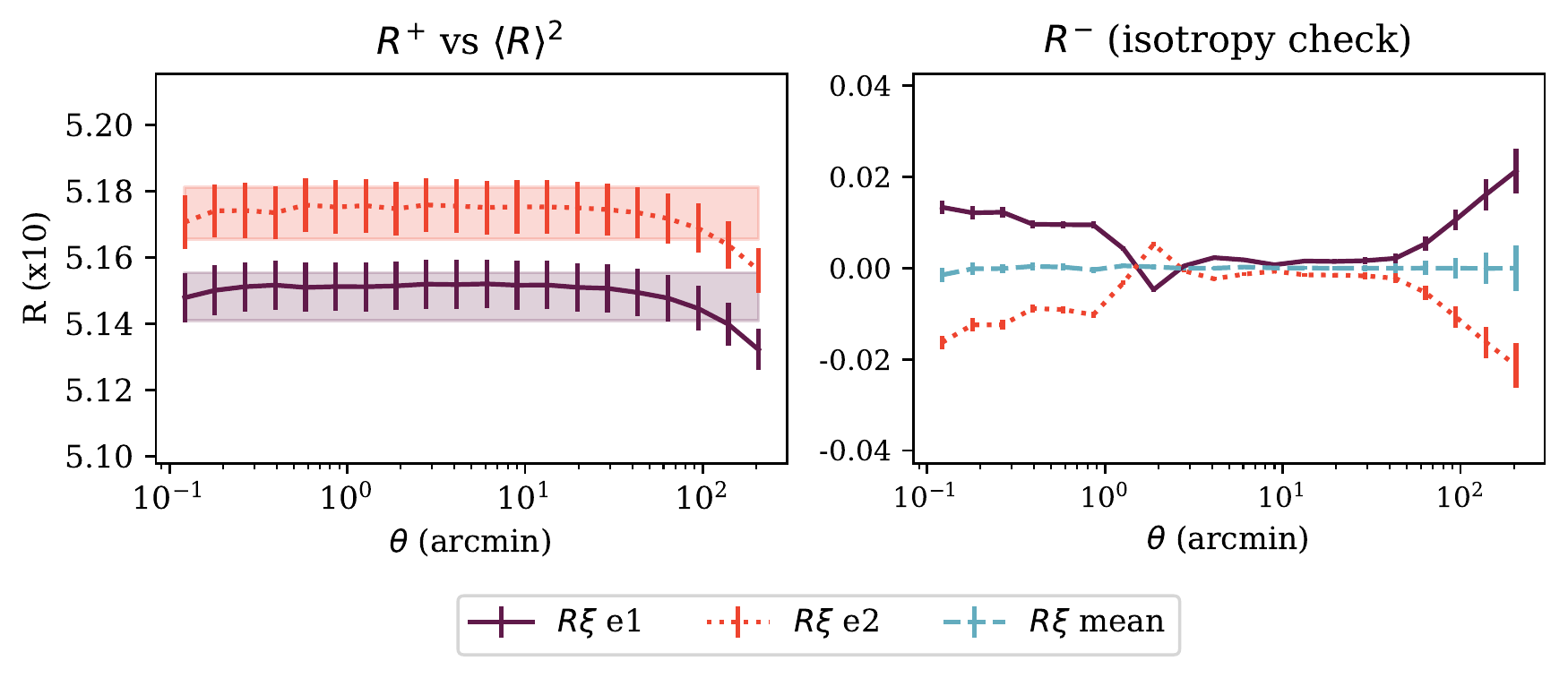}
\end{center}
\caption{\textit{Left:} response inferred using $\xi_+$, i.e. $\langle \hat{R}^{+}\rangle/2$ (from Eq.~\ref{eq14} and represented by the lines in plot) compared with the standard mean response squared used in the DES Y1 analysis to calibrate the shear two-point statistics (horizontal bands). \textit{Right:} Response inferred using $\xi_-$, i.e. $\langle \hat{R}^{-}\rangle/2$ (from Eq.~\ref{eq14}). If the hypothesis of isotropy holds, this should be compatible with 0.}
\label{fig:xi+}
\end{figure*}

Fig.~\ref{fig:xi+} shows the response obtained from Eq.~\ref{eq14}. We sheared $e_1$ and $e_2$ separately so as to better compare with the standard procedure implemented in the DES Y1 analysis. The shear two-point measurement has been computed in 20 bins from 2.5 to 250 arcminutes. Error bars were obtained from 100 jackknives. We note that the values obtained from $e_1$ and $e_2$ separately show differences of the order of $\sim$ 0.4 per cent on $\langle R^{\rm 2pt} \rangle$. {This corresponds to a difference of $\sim$ 0.2 per cent on $\langle R \rangle$, indicating that the hypothesis of isotropy holds down to a 0.2 per cent level.  The two diagonal components of the response matrix are expected to be identical if there was no preferred direction in the measurement process. In practice, this is not true, due to PSF anisotropies or mask effects with distinct orientation with respect to the two shear axes (as found by  \citealt{SheldonMcal2017}). We do not expect this level of bias to impact the DES Y3 analysis, and its amplitude is within the overall calibration error budget from the image simulations. We also note that in the fiducial methodology for the DES Y1 and Y3 analyses the responses from the two components are averaged, which should mitigate this effect (see below).}

\begin{figure}
\begin{center}
\includegraphics[width=0.45\textwidth]{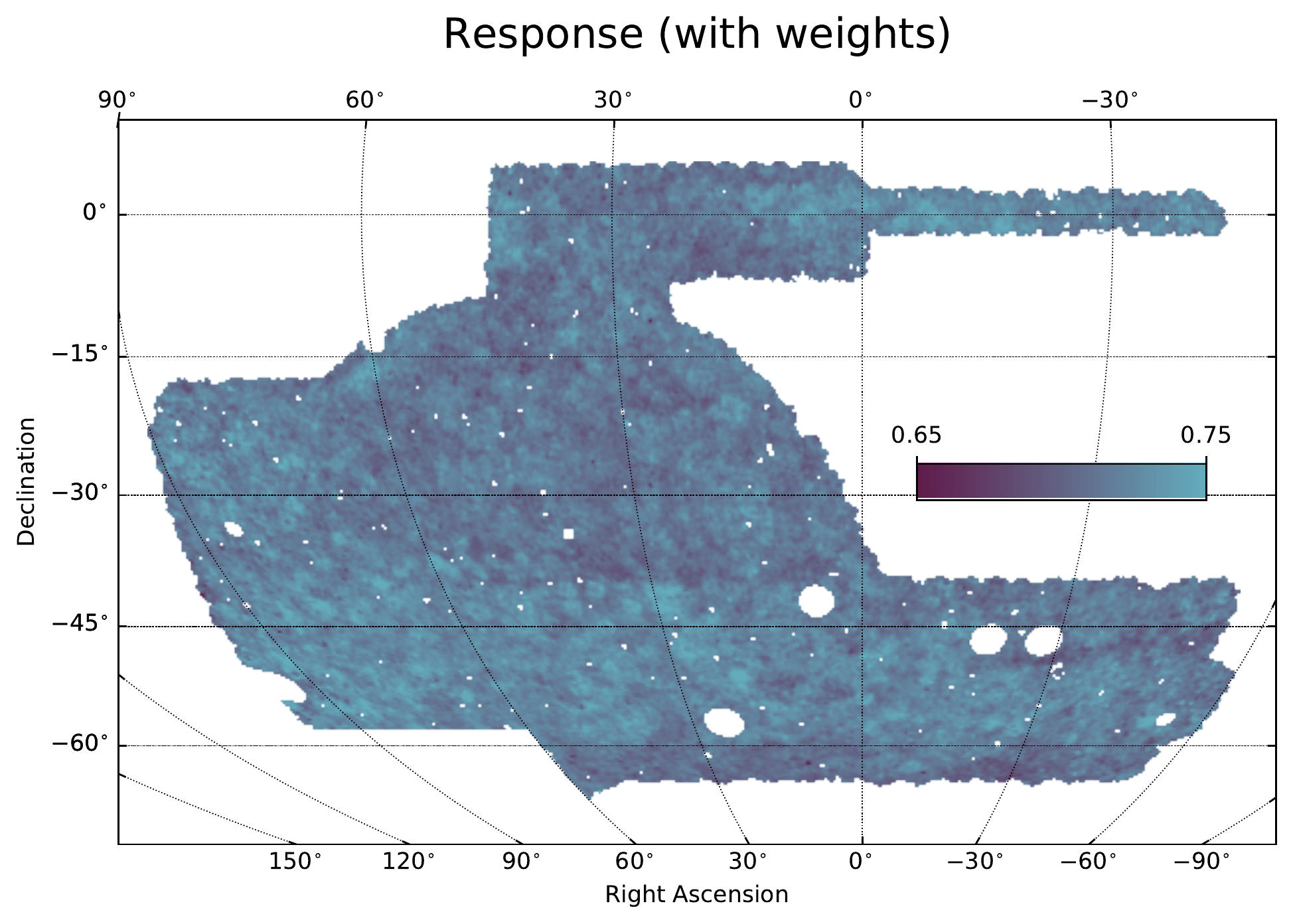}

\end{center}
\caption{Weighted mean response across the survey footprint.}
\label{fig:response_map}
\end{figure}

Fig.~\ref{fig:xi+} also shows the comparison with the mean response implemented in DES Y1 ($\langle  R \rangle^2$). The responses obtained with the two methods are in good agreement within errors for most of the angular scales probed here, except at large scales, where a small difference is measured. This large-scale discrepancy is expected to have a negligible impact for the DES Y3 analysis, given its amplitude. Such difference might be explained by the large-scale pattern of the response across the DES Y3 footprint (Fig.~\ref{fig:response_map}). This pattern cannot be captured by the mean response correction implemented in the DES Y1 analysis, since the mean response is computed over the full sample, losing any spatial/angular information. The presence of a pattern in the mean response is not unexpected and can be caused by a variety of factors: e.g., the mean response is expected to be correlated with imaging depth.

Finally, the right panel of Fig.~\ref{fig:xi+} tests our assumption of isotropy made before Eq.~\ref{eq11}: if isotropy holds for $e_1$ and $e_2$, it is not possible to estimate the response using $\xi_-$, since in Eq.~\ref{eq14} the two terms cancel out. The response obtained for the two components separately are not compatible with zero. This again suggests that the assumption of isotropy is good only at the sub-percentage level with respect to the response computed with $\xi_+$, {in quantitative agreement with the results shown in the left panel of Fig.~\ref{fig:xi+}. The signal vanishes when the two components are averaged, effectively erasing the bias in the estimate of the response.}

\section{Colour based Star-Galaxy Separation}\label{sec:stargal_appendix}

We made use of the star-galaxy separation at faint magnitudes using the DECam observations in $ugriz$ made as part of the DES Deep Fields, combined with $JHKs$ bands as observed by the UltraVISTA survey, as detailed in \cite[][in particular Section 8]{deepfields}. {This star galaxy separation uses colours  as features for supervised machine learning classification. The training set for the classification comes from the HST-ACS \textsc{mu\_class} available within the COSMOS field \citep{2007ApJS..172..219L}.} In particular, we chose the Nearest Neighbors (kNN) star-galaxy classification, as it is shown to have the best performance in terms of stellar purity, and therefore is appropriate for assessing the contamination of stars in the shape catalogue.

\begin{figure*}
\begin{center}
\includegraphics[width=0.9\textwidth]{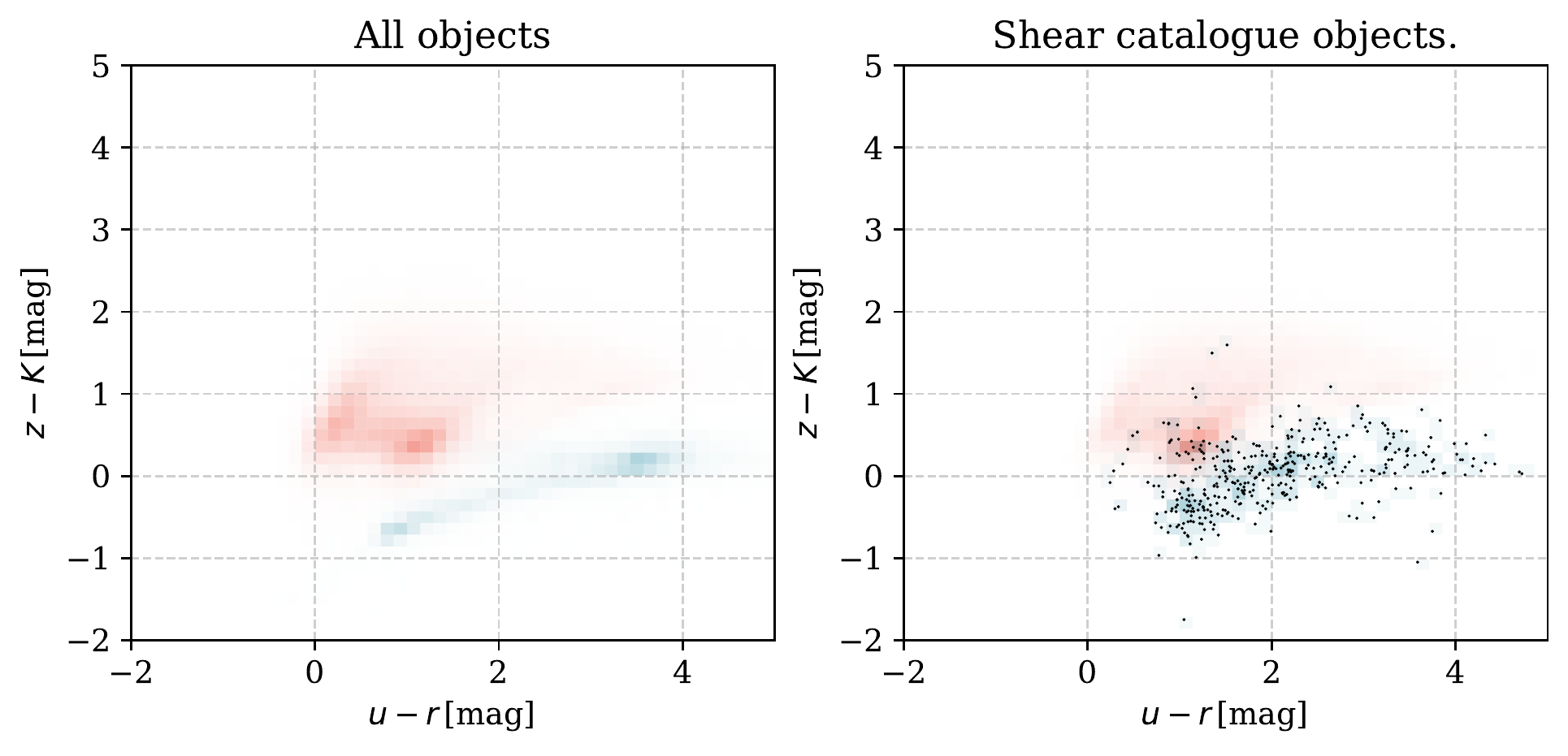}
\end{center}
\caption{Objects as separated by the kNN classifier. {Red} boxes represent objects identified as galaxies; f{blue} boxes stars. {The individual points in the right panel further highlight all of the objects classified as stars in the shear catalogue}. \emph{Left} shows the color distributions for all matched objects in the DES Deep Fields C3, E2 and X3, \emph{right} shows color distributions for objects in this set which pass the fiducial cuts and make it into the shape catalogue. {The apparent broadening of the stellar locus in the right panel is likely due to the fact typical stellar objects close to the locus are removed preferentially by the shear catalogue cuts. Additional broadening may also be due to mis-classifications of some true galaxies.}}
\label{fig:stargal:color-sep}
\end{figure*}

\begin{figure}
\begin{center}
\includegraphics[width=0.45\textwidth]{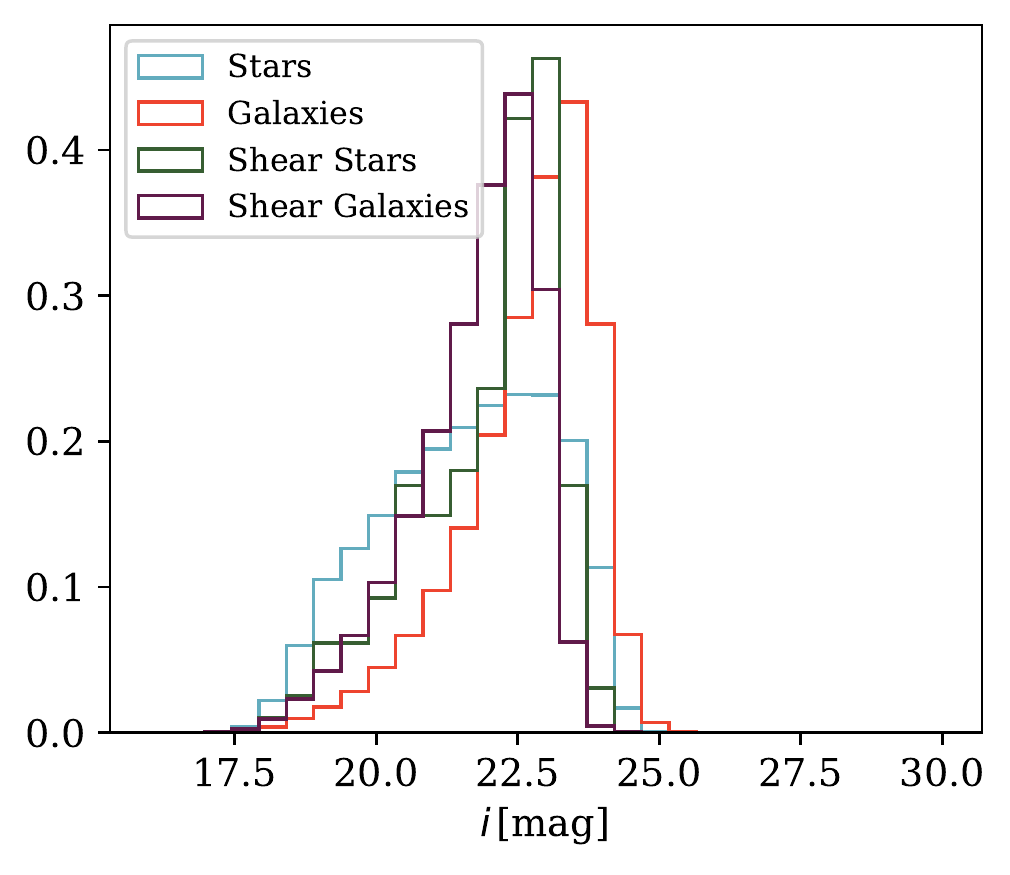}
\end{center}
\caption{$i$ band magnitude distributions for objects as separated by the kNN classifier in the DES Deep Fields C3, E2 and X3.}
\label{fig:stargal:magi}
\end{figure}

The colour-colour plots in Fig.~\ref{fig:stargal:color-sep} show the results of this classifier when applied to all objects in the DES catalogues in the Deep Fields C3, X3 and E2 regions for which both $ugrizJHK$ colours and \mcal\ shape measurements are available. As can be seen, the colour-based classification of shape catalogue objects (which are not selected by colour) shows a small fraction of contaminating objects which have colours highly consistent with those of the stellar population. Fig.~\ref{fig:stargal:magi} also shows the $i$ band magnitudes of the objects in the C3, X3 and E2 regions, as classified by the colour-based kNN method.

\bsp	
\label{lastpage}

\bibliography{references}
\bibliographystyle{mn2e_2author_arxiv_amp.bst}

\section*{Affiliations}
$^{1}$ Institut de F\'{\i}sica d'Altes Energies (IFAE), The Barcelona Institute of Science and Technology, Campus UAB, 08193 Bellaterra (Barcelona) Spain\\	
$^{2}$ Brookhaven National Laboratory, Bldg 510, Upton, NY 11973, USA\\
$^{3}$ Kavli Institute for Particle Astrophysics \& Cosmology, P. O. Box 2450, Stanford University, Stanford, CA 94305, USA\\
$^{4}$ Argonne National Laboratory, 9700 South Cass Avenue, Lemont, IL 60439, USA	\\
$^{5}$ Department of Physics, Duke University Durham, NC 27708, USA\\
$^{6}$ Center for Cosmology and Astro-Particle Physics, The Ohio State University, Columbus, OH 43210, USA	\\
$^{7}$ Department of Physics and Astronomy, University of Pennsylvania, Philadelphia, PA 19104, USA	\\
$^{8}$ Department of Physics, The Ohio State University, Columbus, OH 43210, USA	\\
$^{9}$ 	Instituto de F\'isica Gleb Wataghin, Universidade Estadual de Campinas, 13083-859, Campinas, SP, Brazil	\\
$^{10}$ Jodrell Bank Center for Astrophysics, School of Physics and Astronomy, University of Manchester, Oxford Road, Manchester, M13 9PL, UK	\\
$^{11}$ Department of Physics, Stanford University, 382 Via Pueblo Mall, Stanford, CA 94305, USA	\\
$^{12}$ SLAC National Accelerator Laboratory, Menlo Park, CA 94025, USA	\\
$^{13}$ Jet Propulsion Laboratory, California Institute of Technology, 4800 Oak Grove Dr., Pasadena, CA 91109, USA\\
$^{14}$ Department of Astronomy and Astrophysics, University of Chicago, Chicago, IL 60637, USA	\\
$^{15}$ Kavli Institute for Cosmological Physics, University of Chicago, Chicago, IL 60637, USA\\
$^{16}$ Institut d'Estudis Espacials de Catalunya (IEEC), 08034 Barcelona, Spain\\
$^{17}$ Institute of Space Sciences (ICE, CSIC),  Campus UAB, Carrer de Can Magrans, s/n,  08193 Barcelona, Spain\\
$^{18}$ Department of Astrophysical Sciences, Princeton University, Peyton Hall, Princeton, NJ 08544, USA\\
$^{19}$ Institute for Astronomy, University of Edinburgh, Edinburgh EH9 3HJ, UK\\
$^{20}$ Cerro Tololo Inter-American Observatory, NSF's National Optical-Infrared Astronomy Research Laboratory, Casilla 603, La Serena, Chile \\
$^{21}$ Departamento de F\'isica Matem\'atica, Instituto de F\'isica, Universidade de S\~ao Paulo, CP 66318, S\~ao Paulo, SP, 05314-970, Brazil\\
$^{22}$ Laborat\'orio Interinstitucional de e-Astronomia - LIneA, Rua Gal. Jos\'e Cristino 77, Rio de Janeiro, RJ - 20921-400, Brazil\\
$^{23}$ Fermi National Accelerator Laboratory, P. O. Box 500, Batavia, IL 60510, USA\\
$^{24}$ Instituto de Fisica Teorica UAM/CSIC, Universidad Autonoma de Madrid, 28049 Madrid, Spain\\
$^{25}$ Institute of Cosmology and Gravitation, University of Portsmouth, Portsmouth, PO1 3FX, UK\\
$^{26}$ CNRS, UMR 7095, Institut d'Astrophysique de Paris, F-75014, Paris, France\\
$^{27}$ Sorbonne Universit\'es, UPMC Univ Paris 06, UMR 7095, Institut d'Astrophysique de Paris, F-75014, Paris, France\\
$^{28}$Department of Physics and Astronomy, Pevensey Building, University of Sussex, Brighton, BN1 9QH, UK\\
$^{29}$ Department of Physics \& Astronomy, University College London, Gower Street, London, WC1E 6BT, UK\\
$^{30}$ Kavli Institute for Particle Astrophysics \& Cosmology, P. O. Box 2450, Stanford University, Stanford, CA 94305, USA\\
$^{31}$ Instituto de Astrofisica de Canarias, E-38205 La Laguna, Tenerife, Spain \\
$^{32}$ Universidad de La Laguna, Dpto. Astrofísica, E-38206 La Laguna, Tenerife, Spain\\
$^{33}$ Department of Astronomy, University of Illinois at Urbana-Champaign, 1002 W. Green Street, Urbana, IL 61801, USA\\
$^{34}$ National Center for Supercomputing Applications, 1205 West Clark St., Urbana, IL 61801, USA\\
$^{35}$ University of Nottingham, School of Physics and Astronomy, Nottingham NG7 2RD, UK\\
$^{36}$ INAF-Osservatorio Astronomico di Trieste, via G. B. Tiepolo 11, I-34143 Trieste, Italy \\
$^{37}$ Institute for Fundamental Physics of the Universe, Via Beirut 2, 34014 Trieste, Italy\\
$^{38}$ Laborat\'orio Interinstitucional de e-Astronomia - LIneA, Rua Gal. Jos\'e Cristino 77, Rio de Janeiro, RJ - 20921-400, Brazil \\
$^{39}$ Observat\'orio Nacional, Rua Gal. Jos\'e Cristino 77, Rio de Janeiro, RJ - 20921-400, Brazil\\
$^{40}$ School of Mathematics and Physics, University of Queensland,  Brisbane, QLD 4072, Australia\\
$^{41}$ Centro de Investigaciones Energ\'eticas, Medioambientales y Tecnol\'ogicas (CIEMAT), Madrid, Spain\\
$^{42}$ Department of Physics, IIT Hyderabad, Kandi, Telangana 502285, India\\
$^{43}$ Faculty of Physics, Ludwig-Maximilians-Universit\"at, Scheinerstr. 1, 81679 Munich, Germany\\
$^{44}$ Santa Cruz Institute for Particle Physics, Santa Cruz, CA 95064, USA,\\
$^{45}$ Institute of Theoretical Astrophysics, University of Oslo. P.O. Box 1029 Blindern, NO-0315 Oslo, Norway\\
$^{46}$ Department of Astronomy, University of Michigan, Ann Arbor, MI 48109, USA\\
$^{47}$ Department of Physics, University of Michigan, Ann Arbor, MI 48109, USA \\
$^{48}$ Institute of Astronomy, University of Cambridge, Madingley Road, Cambridge CB3 0HA, UK\\
$^{49}$ Kavli Institute for Cosmology, University of Cambridge, Madingley Road, Cambridge CB3 0HA, UK\\
$^{50}$ D\'{e}partement de Physique Th\'{e}orique and Center for Astroparticle Physics, Universit\'{e} de Gen\`{e}ve, 24 quai Ernest Ansermet, CH-1211 Geneva, Switzerland\\
$^{51}$ Max Planck Institute for Extraterrestrial Physics, Giessenbachstrasse, 85748 Garching, Germany\\
$^{52}$ Universit\"ats-Sternwarte, Fakult\"at f\"ur Physik, Ludwig-Maximilians Universit\"at M\"unchen, Scheinerstr. 1, 81679 M\"unchen, Germany\\
$^{53}$ Center for Astrophysics $\vert$ Harvard \& Smithsonian, 60 Garden Street, Cambridge, MA 02138, USA\\
$^{54}$ Department of Astronomy/Steward Observatory, University of Arizona, 933 North Cherry Avenue, Tucson, AZ 85721-0065, USA\\
$^{55}$ George P. and Cynthia Woods Mitchell Institute for Fundamental Physics and Astronomy, and Department of Physics and Astronomy, Texas A\&M University, College Station, TX 77843,  USA\\
$^{56}$ Instituci\'o Catalana de Recerca i Estudis Avan\c{c}ats, E-08010 Barcelona, Spain\\
$^{57}$ Physics Department, 2320 Chamberlin Hall, University of Wisconsin-Madison, 1150 University Avenue Madison, WI  53706-1390\\
$^{58}$ Department of Physics, Carnegie Mellon University, Pittsburgh, Pennsylvania 15312, USA\\
$^{59}$ School of Physics and Astronomy, University of Southampton,  Southampton, SO17 1BJ, UK \\
$^{60}$ Computer Science and Mathematics Division, Oak Ridge National Laboratory, Oak Ridge, TN 37831\\

\end{document}